\documentclass[prd,aps,a4paper,twocolumn,eqsecnum,nofootinbib,floatfix]{revtex4}  %

\usepackage{graphicx} 
\usepackage{amsmath,amsfonts,amssymb}
\usepackage{color}

\newcommand{\beq}{\begin{equation}}
\newcommand{\eeq}{\end{equation}}
\newcommand{\bea}{\begin{eqnarray}}
\newcommand{\eea}{\end{eqnarray}}
\newcommand{\pinf}{p_\infty}
\newcommand{\g}{\gamma}

\begin{document}

\title{Radiated momentum and radiation-reaction in gravitational two-body scattering \\
including time-asymmetric effects}

\author{Donato Bini$^{1,2}$, Thibault Damour$^3$, Andrea Geralico$^1$}
  \affiliation{
$^1$Istituto per le Applicazioni del Calcolo ``M. Picone,'' CNR, I-00185 Rome, Italy\\
$^2$INFN, Sezione di Roma Tre, I-00146 Rome, Italy\\
$^3$Institut des Hautes Etudes Scientifiques, 91440 Bures-sur-Yvette, France
}

\date{\today}

\begin{abstract}
We compute to high post-Newtonian accuracy the 4-momentum (linear momentum, and energy), radiated as gravitational waves  in a two-body system undergoing gravitational scattering. We include, for the first time, all the relevant {\it time-asymmetric}  effects that arise
when consistently going three post Newtonian orders beyond the leading post Newtonian order.  
We find that the inclusion of time-asymmetric radiative effects (both in tails and in the radiation-reacted hyperbolic motion)
 is crucial to ensure the  mass-polynomiality of the post-Minkowskian  expansion
 ($G$  expansion) of the radiated 4-momentum. 
Imposing the mass-polynomiality of the corresponding individual impulses determines the 
 conservativelike radiative contributions at the fourth post-Minkowskian order, and
 strongly constrains them at the fifth  post-Minkowskian order.
 \end{abstract}

\maketitle

\section{Introduction}

Gravitational scattering has attracted a renewed interest in recent years, both for conservative 
and dissipative (i.e., gravitational-radiation-related) effects. Various approximation methods (post-Newtonian,
post-Minkowskian, quantum perturbation theory, effective field theory, string theory) 
have been applied to this problem. For a sample of results on (classical or quantum) post-Minkowskian (PM)
gravitational scattering, see, e.g., 
Refs \cite{Amati:1990xe,Damour:2017zjx,
Bjerrum-Bohr:2018xdl,Kosower:2018adc,Cheung:2018wkq,Bjerrum-Bohr:2019kec,
Bern:2019crd,Bern:2019nnu,Mogull:2020sak,Kalin:2020mvi,Kalin:2020fhe,
Bern:2021dqo,Bern:2021yeh,Dlapa:2021npj,Dlapa:2021vgp,Bjerrum-Bohr:2021din,
Saketh:2021sri,Kalin:2022hph,Khalil:2022ylj}. 
For recent PM results on radiative losses during gravitational scattering and related results, see, e.g., 
Refs. \cite{Herrmann:2021tct,Mougiakakos:2021ckm,Jakobsen:2021smu,Riva:2021vnj,Manohar:2022dea,Dlapa:2022lmu}.

The state of the art for the PM scattering  of spinless bodies  is  $O(G^3)$ for radiation-reacted scattering 
\cite{DiVecchia:2020ymx,Damour:2020tta,DiVecchia:2021ndb,Herrmann:2021tct}, 
and $O(G^4)$ for the {\it conservative} case \cite{Bern:2021yeh,Dlapa:2021vgp}. The state
of the art for radiative losses during gravitational scattering is $O(G^3)$ for radiated angular momentum \cite{Manohar:2022dea},
and $O(G^3)$ for radiated 4-momentum
\cite{Bern:2021dqo,Herrmann:2021tct,Mougiakakos:2021ckm,Riva:2021vnj}. 
While finalizing this work a  $O(G^4)$-accurate computation of  the (radiation-reacted) 
individual 4-momentum changes (or ``impulses"),
$\Delta p_{a \mu}$, and of the loss of 4-momentum of the system, appeared on arXiv  \cite{Dlapa:2022lmu}.

The relation between radiative losses of energy, linear momentum, and angular momentum, and the radiation-reaction
contribution to scattering has been worked out, {\it to linear order in radiation-reaction}, in Refs. \cite{Bini:2012ji,Bini:2021gat}.
One of the aims of the present work is to go beyond the purely linear-in-radiation-reaction treatment of 
Refs. \cite{Bini:2012ji,Bini:2021gat}. This will be done by focussing on the various {\it time-asymmetric}  effects arising
in the radiative losses of energy and linear momentum during hyperbolic encounters.

The post-Newtonian (PN) approximation method has also recently played a useful role in tackling
gravitational scattering. The state of the art for the PN scattering (of spinless bodies) in the {\it conservative}
case is the fourth post-Newtonian (4PN) accuracy \cite{Bini:2017wfr}. This was generalized in Refs. 
\cite{Bini:2020wpo,Bini:2020hmy,Bini:2020rzn} to the 5PN, and 6PN, accuracies (modulo
the knowledge of a few, yet undetermined, Hamiltonian coefficients).  
The state of the art for the PN-expanded computation of the radiative losses 
(to gravitational waves) of energy, angular momentum and linear momentum\footnote{We recall that 
the leading PN orders of radiative losses is the  2.5PN order for energy and angular momentum, while it is
the 3.5PN order for linear momentum.} is as follows:
the radiated energy and angular momentum (for spinless bodies) have been computed at the absolute 4.5PN order
(corresponding to a 2PN fractional accuracy) in Refs.  \cite{Bini:2021gat,Bini:2021jmj,Bini:2021qvf}.  
Higher-order terms (corresponding to, at least, the 3PN fractional accuracy) have been computed in
 Refs. \cite{Cho:2021onr,Cho:2022pqy,Bini:2022xpp}.  
  The radiated linear momentum is currently known to the (absolute)  5.5PN order 
  \cite{Bini:2021gat,Bini:2022yrk}  (corresponding to a 2PN fractional accuracy).
The state of the art for the PN-expanded computation of the scattering of spinless bodies
is the 5PN level, at which an inconsistency with the mass-polynomiality of the conservative $G^4$ (4PM) contribution was
highlighted in  \cite{Bini:2021gat}, and remains puzzling despite recent work
on the additional radiative contributions \cite{Almeida:2022jrv}.

The aims of the present paper are:
\begin{enumerate} 

\item  to complete the PN knowledge of the radiated energy  
by including both the fractional 2.5PN contribution (linked to the 2.5PN radiation-reaction modification of the hyperbolic motion)
which was incorrectly argued to vanish in Ref. \cite{Cho:2021onr},
and  the ``instantaneous" 3PN-level contribution first derived in Ref. \cite{Cho:2021onr}, and rederived here;

\item to improve the knowledge of the radiated angular momentum   
by including both the fractional 2.5PN contribution (computed here for the first time) and the 3PN-level
contribution (obtained here by adding instantaneous 3PN terms \cite{Cho:2021onr} and higher-order tails \cite{Bini:2021qvf});  

\item
 to raise the knowledge of the radiated linear momentum to the  fractional  3PN accuracy (corresponding to the
absolute 6.5PN order). 

\item
 to bring new light on the  mass-polynomiality structure of the scattering at the 4PM and 5PM orders. 
 
\end{enumerate}

The  accuracy increase (from 2PN to 3PN fractional accuracy) 
in the radiated linear momentum requires that many new physical effects 
be taken into account: indeed, we will need to take into account:
(i) 2.5PN radiation-reaction effects in the hyperbolic motion; (ii)  2.5PN ``instantaneous" contributions
to the radiative multipole moments \cite{Blanchet:2013haa,Mishra:2011qz}; (iii) 
the 1PN fractional correction to the leading-order tail\footnote{We recall that tail contributions
 to gravitational radiation start at the fractional 1.5PN order  \cite{Blanchet:1992br,Blanchet:2013haa}.}
 contribution to the radiated linear momentum
(which was first computed in Ref.  \cite{Bini:2021gat}); (iv) 3PN accuracy in several multipoles
and in the hyperboliclike motion; and (v) higher-order tails in the momentum loss \cite{Bini:2022yrk}.

To complete the so-obtained increased PN-expanded knowledge of the radiated 4-momentum 
$P^\mu_{\rm rad}=(E_{\rm rad}, P^i_{\rm rad})$ we will re-express it in terms of Lorentz-invariant
form factors by decomposing it  on the basis $u_{1-}^\mu$, $u_{2-}^\mu$,  $\hat {b}_{12}^\mu$,
defined by the initial four velocities of the bodies and the direction of the vectorial impact parameter
 $\hat { b}_{12}^\mu=b_{12}^\mu/b$, with $b_{12}^\mu= b_{1}^\mu-b_{2}^\mu$. More precisely,
 it will be useful to decompose it as
 \beq \label{decompPrad0}
 P^\mu_{\rm rad}= P^{\rm rad}_{1+2} (u_{1-}^\mu + u_{2-}^\mu) + P^{\rm rad}_{1-2} (u_{1-}^\mu - u_{2-}^\mu) +  P^{\rm rad}_{b_{12}} \hat { b}_{12}^\mu.
 \eeq
We will show below (generalizing considerations introduced in Refs \cite{Damour:2019lcq,Bini:2021gat}) that, at each
order in $G$, the 
PM expansion of the form factors, $  P^{\rm rad}_{1+2}, P^{\rm rad}_{1-2}, P^{\rm rad}_{b_{12}}$
(expressed as functions of $b$ and of the relative Lorentz factor 
$\gamma \equiv - u_{1-}^\mu u_{2- \mu}$)\footnote{We use a mostly plus signature.},
 have a {\it polynomial structure} in the two masses
$m_1, m_2$, e.g.
\beq
 P^{\rm rad}_{1+2}= \frac{G^3}{b^3} m_1^2 m_2^2 \, \hat P^{\rm rad}_{1+2}\,,
\eeq
with
\beq
 \hat P^{\rm rad}_{1+2}=\sum_{n \geq 3}  \frac{G^{n-3}}{b^{n-3}} SP^{1+2}_{n-3}(m_1,m_2;\g)\,.
\eeq
Here, and in the following, the notation $ SP^{1+2}_{N}(m_1,m_2;\g)$ denotes a homogeneous {\it symmetric polynomial} of
order $N$ in the two masses, with coefficients depending on the Lorentz factor $\gamma$.
 
At the 3PM level ($O(G^3)$),  only one form factor of $ P^\mu_{\rm rad}$
 is non vanishing, namely $ {P^{{\rm rad}\, G^3}_{1+2}}$, with
\beq \label{PradG3}
{P^{{\rm rad}\, G^3}_{1+2}}= \frac{G^3}{b^3} m_1^2 m_2^2  \frac{{\mathcal E} (\g)}{\g+1}.
\eeq
The exact value of the function ${\mathcal E} (\g)$ has been computed in Refs. 
\cite{Bern:2021dqo,Herrmann:2021lqe,Herrmann:2021tct,Manohar:2022dea,Jakobsen:2022psy},
while its PN expansion was computed to order $v^{15}$ included in \cite{Bini:2021gat}, see Eq. (5.19) there. For illustration,
let us display the beginning of the PN expansion of   ${\mathcal E}(\g)$, when expressed in terms of $\pinf \equiv \sqrt{\g^2-1}$,
\bea
{\mathcal E} (\g) &=& \pi \left(\frac{37}{15} \pinf + \frac{1357}{840}\pinf^3 + \frac{27953}{10080}\pinf^5\right.\nonumber\\
&&\left.
-\frac{676273}{354816}\pinf^7 +O(\pinf^9) \right)\,.
\eea
Using our newly acquired PN-expanded knowledge on the values of $E_{\rm rad}$, and $P^i_{\rm rad}$ [computed
in the center-of-mass (c.m.) frame], we will be able both to check the mass-polynomiality structure of the form factors
 $  P^{\rm rad}_{1+2}, P^{\rm rad}_{1-2}, P^{\rm rad}_{b_{12}}$ entering the decomposition \eqref{decompPrad0},
 and to compute their expansions in powers of $\pinf$ at the   fractional 3PN accuracy.

Finally, we will use the so-acquired improved knowledge of $P^\mu_{\rm rad}$ to constrain the radiation-reaction induced
contributions to  the individual changes $\Delta p_a^\mu$ (also called ``impulses") of the 4-momenta of the two bodies. 
As we will recall in more detail below, Refs. \cite{Bini:2012ji,Bini:2021gat}
have derived the effect of radiation reaction on the individual momentum changes  $\Delta p_a^\mu$ only to linear order
in radiation reaction, and within a restricted set of assumptions. Namely, writing 
 the equations of motion of each particle as a perturbed ``conservative" (Hamiltonian) system involving an additional
``radiation-reaction force" $\cal F^\mu_{\rm rr}$, Refs. \cite{Bini:2012ji,Bini:2021gat} worked only to
{\it linear order} in  $\cal F^\mu_{\rm rr}$, and, furthermore, often assumed that
 the latter radiation-reaction force  was  time-antisymmetric\footnote{The time reversal operation
is taken around the moment of closest approach of the time-symmetric unperturbed
conservative dynamics, considered in the center-of-mass  frame.}. Under these assumptions, 
Refs. \cite{Bini:2012ji,Bini:2021gat} derived an expression for  $\Delta p_a^\mu$ of the form
\beq \label{Dpa0}
\Delta p_{ a \mu}= \Delta p_{ a \mu}^{\rm cons} + \Delta p_{ a \mu}^{\rm rr \, lin } + \Delta p_{ a \mu}^{\rm rr \, nonlin}\,.
\eeq
Here  the term $\Delta p_{ a \mu}^{\rm rr \, lin }$ denotes the contribution  linear in the  radiation
reaction derived in \cite{Bini:2021gat}, while the term $ \Delta p_{ a \mu}^{\rm rr \, nonlin}$ denotes the
missing remainder, due  to non-linear effects in  $\cal F^\mu_{\rm rr}$.  
Ref.  \cite{Bini:2021gat} had illustrated the existence of   non-linear effects in  $\cal F^\mu_{\rm rr}$
by computing (within the standard PN approach) a contribution to  $\Delta p_{ a \mu}$ quadratic in $\cal F^\mu_{\rm rr}$.
It has been known for a long time \cite{Blanchet:1987wq,Damour:2014jta} that there are {\it hereditary},
tail-related contributions to the equations of motion. These contributions are {\it time-asymmetric}, i.e. neither
time-even, nor time-odd. At the 4PN level, they can be uniquely decomposed 
in a time-even conservative piece (contributing to the Hamiltonian), and a time-odd piece giving
a nonlocal-in-time contribution to  $\cal F^\mu_{\rm rr}$ (see, Section VI of \cite{Damour:2014jta}).
However,  this simple decomposition becomes more tricky at the 5PN level.
This is indeed the PN level where quadratic effects in  $\cal F^\mu_{\rm rr}$ enter, and where
past-related tail effects contribute to the linear-response results of  \cite{Bini:2021gat} (via
the presence of a ``conservativelike", 5PN-level, past-tail contribution to $ P^\mu_{\rm rad}$;
see Eq. (H3) there). [These 5PN-level subtleties arise at the 4PM level ($O(G^4)$).]
The new results presented here will complete the results of \cite{Bini:2021gat}
by fully taking into account time-asymmetric effects in various observables.
First, in $ P^\mu_{\rm rad}$ (which we compute here with higher PN accuracy than before, 
including all needed hereditary tail effects), and second, in the radiative contributions to the impulses
 $\Delta p_{ a \mu}^{\rm rr }$. We will improve below
the results of \cite{Bini:2021gat} by completing the linear-response term  $\Delta p_{ a \mu}^{\rm rr \, lin }$
with the effect of the time-even part of   $\cal F^\mu_{\rm rr}$ on the relative scattering angle. In addition,
 our strategy to scrutinize the mass-polynomiality of the impulses will allow us to obtain valuable
 information on the remainder term  $\Delta p_{ a \mu}^{\rm rr \, nonlin}$ in Eq. \eqref{Dpa0}. This information is enough to uniquely determine 
$\Delta p_{ a \mu}^{\rm rr \, nonlin}$ at order $G^4$ at to strongly constrain its value at order $G^5$.

\section{Framework}

To set up the stage for our computations below, let us recall that 
 the general expressions for the radiative fluxes (at infinity) of energy, linear momentum and angular momentum  in terms of the 
{\it radiative} multipole moments $U_L$ and $V_L$ (defined at future null infinity) read \cite{Blanchet:1985sp,Blanchet:1987wq,Blanchet:1989ki,Damour:1990ji,Blanchet:1998in,Poujade:2001ie}

\bea \label{FEUV}
\frac{dE^{\rm rad}}{dt_{\rm ret}}&\equiv&{\mathcal F}_{E} 
=\sum_{l=2}^\infty\frac{G}{c^{2l+1}} \left[\frac{(l+1)(l+2)}{(l-1)ll!(2l+1)!!}U_{L}^{(1)}U_L^{(1)}\right.\nonumber\\
&+&\left.\frac{4l(l+2)}{c^2(l-1)(l+1)!(2l+1)!!}V_{L}^{(1)}V_{L}^{(1)} 
 \right]\,,
\eea

\bea \label{FPUV}
\frac{dP^{\rm rad}_i}{dt_{\rm ret}}&\equiv&{\mathcal F}_{P}{}_i
=\sum_{l=2}^\infty \left[\frac{G}{c^{2l+3}}\frac{2(l+2)(l+3)}{l(l+1)!(2l+3)!!}U_{iL}^{(2)}U_L^{(1)}\right.\nonumber\\
&+&
\frac{G}{c^{2l+5}}\frac{8(l+3)}{(l+1)!(2l+3)!!}V_{iL}^{(2)}V_{L}^{(1)}\nonumber\\
&+&\left. 
\frac{G}{c^{2l+3}}\frac{8(l+2)}{(l-1)(l+1)!(2l+1)!!}\epsilon_{iab}U_{aL-1}^{(1)}V_{bL-1}^{(1)}
\right]\,,\nonumber\\
\eea
and
\bea \label{FJUV}
\frac{dJ^{\rm rad}_i}{dt_{\rm ret}}&\equiv&{\mathcal F}_{J_i} \nonumber\\
&=&\epsilon_{iab}\frac{G}{c^{2l+1}}\sum_{l=2}^\infty \left[\frac{(l+1)(l+2)}{(l-1)l!(2l+1)!!}U_{aL-1}U_{bL-1}^{(1)}\right.\nonumber\\
&+&\left.
\frac{4l^2(l+2)}{c^2(l-1)(l+1)!(2l+1)!!}V_{aL-1}V_{bL-1}^{(1)} 
\right]\,.\nonumber\\
\eea

Here, $t_{\rm ret}= t - \frac{r}{c}- \frac{2 G \cal{M} }{c^3} \ln \left( \frac{r}{r_0} \right)+O(G^2)$ is the retarded time
(with $\cal{M}$ denoting the total Arnowitt-Deser-Misner (ADM) mass of the spacetime, and $r_0$ a constant length scale), while $U_L$ and $V_L$ 
are the mass-type and current-type radiative multipole moments, respectively (with $L=i_1i_2\cdots i_l$ being a multi-index consisting of $l$ spatial indices).
 They are related to the {\it source} multipole moments 
$I_L$ and $J_L$ by relations having the structure\footnote{Henceforth we replace the argument  $t_{\rm ret}$
of the radiative multipole moments simply by the dynamical time variable $t = t_{\rm ret}+ {\rm cst}$ describing
the binary motion (in the center-of-mass system).} \cite{Blanchet:2013haa}
\bea \label{UvsI}
U_L(t)&=& I_L^{(l)}(t) \nonumber\\
&+& \frac{G}{c^3} (\hbox{tail + semi-hered. + instantaneous})  
\nonumber\\
&+& \frac{G}{c^5} (\hbox{semi-hered. + instantaneous})\nonumber\\
&+&  \hbox{higher-order tails}
\,,\nonumber\\
V_L(t)&=& J_L^{(l)}(t) \nonumber\\
&+& \frac{G}{c^3} (\hbox{tail + instantaneous})  
\nonumber\\
&+& \frac{G}{c^5} (\hbox{instantaneous})\nonumber\\
&+&  \hbox{higher-order tails}\,.
\eea
Here the ``tail terms" are given by integrals extending over the full past history of the source of the type
\beq
 + \frac{2 G \cal{M} }{c^3} \int_0^{\infty} d\tau I_L^{(l+2)}(t-\tau )  \left( \ln \frac{\tau}{\tau_0} + {\rm const.}\right)  
\,,
\eeq
with $\tau_0=r_0/c$.
The semi-hereditary (semi-hered.) terms (also known as memory terms) are time anti-derivatives of products of multipole moments, whereas  the instantaneous terms  are polynomials in (time-derivatives) of the source multipole moments.
Notice that there are no semi-hereditary contributions to the radiative current moments.
In the case of radiative mass moments, instead, the $O(\frac{G}{c^3})$ semi-hereditary terms first appear for $l=4$, while at the next order $O(\frac{G}{c^5})$ they are already present for $l=2$.
Furthermore, both the energy and the linear momentum fluxes \eqref{FPUV} only contain  time derivatives of the radiative moments \eqref{UvsI}, so that all the semi-hereditary terms give instantaneous contributions to both ${\mathcal F}_E(t)$ and ${\mathcal F}_{P}{}_i(t)$. 

The higher-order tail contributions (tail-squared, tails-of-tails, etc.) start at fractional order $(\frac{2 G \cal{M} }{c^3})^2$, i.e., 3PN. We will take into account these fractional 3PN contributions  in all radiated quantities: energy, angular momentum and linear momentum. 
 To reach the 3PN accuracy,   we also
need to take into account all semi-hereditary and instantaneous terms that contribute at the fractional 2.5PN level. 
Among the 2.5PN effects, an important, and subtle one, comes from the 2.5PN-level correction to the hyperbolic
motion induced by the leading-order radiation-reaction force. It is the subject of the next Section.

\section{2.5PN correction to the quasi-Keplerian parametrization for hyperboliclike orbits}
\label{2.5PNmotion}

In order to explicitly compute the 2.5PN correction to hyperbolic motion\footnote{For our
present purpose it is enough to study the relative 2-body planar motion, considered  in the center-of-mass
system, and in harmonic coordinates.} caused by the leading-order
radiation-reaction force (considered as a first-order perturbation of the 2PN equations of motion), 
it is convenient to follow Ref. \cite{Damour:2004bz} in using Lagrange's method of variation of constants. 
This is done by rewriting the  
{\it hyperbolic version}\footnote{A straightforward analytic continuation to positive binding energies of the
{\it ellipticlike} 2PN quasi-Keplerian parametrization would involve complex parameters.} \cite{Cho:2018upo} of the solution of the 2PN-level equations of motion  \cite{Damour:1988mr,SW1993,Memmesheimer:2004cv}
(which depends on four integration constants, say $c_1,c_2, c_l, c_\phi$)
in terms of four time-dependent versions of the integration constants, say $c_1(t),c_2(t), c_l(t), c_\phi(t)$.
Namely, one writes
\bea
\label{sys0}
r&=& S(l,c_1(t),c_2(t))\,,\nonumber\\
\dot r&=&\bar n(c_1(t),c_2(t)) \frac{\partial S(l,c_1(t),c_2(t))}{\partial l}
\,,\nonumber\\
\phi&=& c_\phi(t) +{\bar W}(l,c_1(t),c_2(t)) \,,\nonumber\\ 
\dot \phi&=&\bar n(c_1(t),c_2(t)) \frac{\partial {\bar W}(l,c_1(t),c_2(t))}{\partial l}
\,.
\eea
Here the functions $S(l,c_1,c_2)$ and  ${\bar W}(l,c_1,c_2)$ are defined by eliminating the
auxiliary variables $v$ and $V$ (by expressing them as functions of $l$, $c_1$ and $c_2$) from the four equations
\bea \label{SW}
S&=&\bar a_r (e_r \cosh v-1)
\,,\nonumber\\
{\bar W}&=&K[V+f_\phi \sin 2V+g_\phi \sin 3V]\,, \nonumber\\
l&=&e_t \sinh v-v + f_t V+g_t \sin V\,,\nonumber\\
V&=&2\, {\rm arctan}\left[\sqrt{\frac{e_\phi+1}{e_\phi-1}}\tanh \frac{v}{2}  \right]\,.
\eea
In these equations, the Quasi-Keplerian orbital parameters $a_r, e_r, e_t, e_\phi, K\equiv 1+k, f_\phi ,g_\phi,f_t,  g_t$
are functions of the two (2PN) integrals of motion $c_1,c_2$. Similarly to  $S$ and $\bar W$, the auxiliary variable $v$
can  be considered as a function of $l,c_1,c_2$: $v=v(l,c_1,c_2)$. One could choose as
basic 2PN constants, $c_1,c_2$, the energy $E$ of the system (or the specific binding energy $\bar E \equiv (E-Mc^2)/(\mu c^2)$) and 
the angular momentum $J$  of the system (or the dimensionless angular momentum $j=c J/(GM\mu)$ 
(see, e.g.,  Table VIII of Ref. \cite{Bini:2020hmy}, for the  harmonic-coordinates-case expressions of the orbital parameters). 
In the following, we 
find more convenient to use $c_1=\bar a_r$ and $c_2=e_r$. 
The harmonic-coordinates expressions of the Quasi-Keplerian orbital parameters, as functions of $\bar a_r$ 
and $e_r$, will be presented below  
when discussing the generalization of this representation at the 3PN level.
The auxiliary variable\footnote{The variable $v$ is the hyperbolic analog of the usual Kepler eccentric anomaly
$u$ (solution of Kepler's equation $l=u - e_t \sin u + O(\frac1{c^2})$) used
in the description of elliptic motions. See Appendix B for a discussion of the complex analytic continuation relating 
elliptic and hyperbolic motions.}
 $v$ is then considered as a function of the form $v=v(l,\bar a_r,e_r)$, with the dependence on $\bar a_r$ entering only beyond the 
 leading order (LO).

The perturbed motion is then expressed, besides allowing $c_1$, $c_2$ and $c_\phi$ to be functions of time,
 by describing the time dependence of the basic angle $l$ of the hyperboliclike planar motion 
 in the following way:
\beq
\label{ldef}
l(t)=\int_{t_0}^t\bar n(c_1(t),c_2(t)) dt+c_l(t)\,.
\eeq
Here, $t_0$ is an arbitrary reference time, and the four  former ``constants" $c_1(t),c_2(t),c_l(t),c_\phi(t)$
are now time varying. Inserting Eqs. \eqref{sys0}, \eqref{ldef} in the perturbed equations of motion determines the
system of four first-order evolutions equations that must be satisfied by the four quantities $c_1(t),c_2(t),c_l(t),c_\phi(t)$, say
\beq
\frac{dc_\alpha}{dt}=F_\alpha(l,c_\beta)\,, \qquad  
\alpha,\beta=1\,,2\,,l\,,\phi \,,
\eeq
where the functions $F_\alpha$ are linear in the perturbing (relative) acceleration. 
They generally read
\bea \label{LagrangeHyp}
\frac{d c_1}{dt}&=&  \frac{\partial{c_1({\mathbf x},{\mathbf v})}}{\partial {\mathbf v}} \cdot {\mathbf A}_{\rm rr}
\,,\nonumber\\
\frac{d c_2}{dt}&=&  \frac{\partial{c_2({\mathbf x},{\mathbf v})}}{\partial {\mathbf v}} \cdot {\mathbf A}_{\rm rr}
\,,\nonumber\\
 \frac{d c_l}{dt} &=&-\left(\frac{\partial S}{\partial l}\right)^{-1}\left[\frac{\partial S}{\partial  c_1}\frac{d c_1}{dt}+\frac{\partial S}{\partial c_2}\frac{c_2}{dt}\right]  
\,,\nonumber\\
\frac{d c_\phi}{dt} &=&-\frac{\partial {\bar W}}{\partial l}\frac{d c_l}{dt}-\frac{\partial {\bar W}}{\partial c_1}\frac{d  c_1}{dt}-\frac{\partial {\bar W}}{\partial c_2}\frac{d  c_2}{dt}\, ,
\eea
where ${\mathbf A}^{\rm rr}$ denotes the (relative) radiation-reaction acceleration (which starts at 2.5 PN). 
When choosing $c_1=E$ and $c_2=J$, and when working in the Hamiltonian formalism,  the first two varying-constant
equations read:
\beq \label{EJevolution}
\frac{d  E}{dt}= \frac{\partial H}{\partial p_i} {\mathcal F}^{\rm rr}_i \,,\qquad 
\frac{d  J}{dt}= {\mathcal F}^{\rm rr}_\phi\,,
\eeq
where ${\mathcal F}^{\rm rr}_i$ denotes the relative radiation-reaction force.

When choosing $c_1=\bar a_r$ and $c_2=e_r$, and when working at the leading PN order, these two equations read
\bea \label{Lagrange}
\frac{d \bar a_r}{dt}&=&-2\bar a_r^2{\mathbf v}\cdot {\mathbf A}_{\rm rr}
\,,\nonumber\\
\frac{d e_r}{dt}&=&\frac{e_r^2-1}{e_r}\bar a_r{\mathbf v}\cdot {\mathbf A}_{\rm rr}
+\frac{\sqrt{e_r^2-1}}{e_r\sqrt{\bar a_r}} [{\mathbf x}\times {\mathbf A}_{\rm rr}]_z
\,.
\eea 
As we need to compute the time dependence of the source multipole moments expressed in {\it harmonic coordinates},
we shall use here the (leading order) value of  ${\mathbf A}^{\rm rr}$ in  harmonic coordinates, namely (denoting $\nu\equiv \mu/M$)
\bea
\label{Arr}
{\mathbf A}^{\rm rr}&=&-\frac85 \nu \frac{G^2}{c^5} \frac{M^2}{r^3}\left[
-\left(3v^2+\frac{17}3 \frac{GM}{r}\right)\dot r {\mathbf n}\right.\nonumber\\
&&\left.+\left(v^2+3\frac{GM}{r}\right){\mathbf v}\right]\,.
\eea

Working at the leading 2.5PN order, denoting
\beq
{\mathcal X} \equiv e_r\cosh v-1\,,
\eeq
(where the auxiliary variable $v$ is the same as in Eqs. \eqref{SW}), and decomposing the four varying constants
$c_\alpha(t)$ as
\beq
c_\alpha(t) = c_\alpha^0+ \delta^{\rm rr} c_\alpha(t),
\eeq
with constants $c_\alpha^0$,\footnote{Below, we ease the notation by denoting $c_\alpha^0$ simply as $c_\alpha$, 
while the full quantity $c_\alpha(t)$ is always indicated with its time dependence.} one finds the following explicit 
(2.5 PN-accurate) evolution system\footnote{Following usual practice, we often use scaled variables (factoring out some appropriate powers of $M$ or $\mu$) when studying the relative motion.} for the four $\delta^{\rm rr} c_\alpha(t)$'s:
\bea
\label{syshyp}
\frac{d\,\delta^{\rm rr}\bar a_r}{dt}&=&\frac{\nu}{\bar a_r^3}\left[-\frac{32}{5{\mathcal X}^3}- \frac{512}{15{\mathcal X}^4}+\frac{16(-49+9e_r^2)}{15{\mathcal X}^5}\right.\nonumber\\
&&\left. +\frac{112(e_r^2-1)}{3{\mathcal X}^6}\right] 
\,,\nonumber\\
\frac{d\,\delta^{\rm rr} e_r}{dt}&=&\frac{\nu(e_r^2-1)}{\bar a_r^4e_r}\left[
-\frac{56(e_r^2-1)}{3{\mathcal X}^6}-\frac{8(9e_r^2-49)}{15{\mathcal X}^5} \right.\nonumber\\
&& \left.
+\frac{136}{15{\mathcal X}^4}+\frac{8}{5{\mathcal X}^3}  
  \right] 
\,,\nonumber\\
\frac{d\,\delta^{\rm rr} c_l}{dt}&=&\frac{\nu\sinh v}{\bar a_r^4e_r} \left[
-\frac{56(e_r^2-1)^2}{3{\mathcal X}^6}
\right.\nonumber\\  
&&
-\frac{8 (e_r^2-1) (9 e_r^2-14)}{15{\mathcal X}^5}  \nonumber\\
&&\left.+\frac{8(43 e_r^2-3)}{15{\mathcal X}^4}
+\frac{32}{5}  \frac{e_r^2}{{\mathcal X}^3 } \right]
\,,\nonumber\\
\frac{d \delta^{\rm rr} c_\phi}{dt}&=& \frac{\nu \sinh v}{\bar a_r^4}\frac{\sqrt{e_r^2-1}}{e_r}\left[ \frac{8}{5{\mathcal X}^4}-\frac{8}{15}\frac{9e_r^2-14}{{\mathcal X}^5}\right. \nonumber\\
&&\left.-\frac{56}{3}\frac{e_r^2-1}{{\mathcal X}^6} \right]
\,.
\eea
Let us note in passing that we have checked these results on the 2.5PN-level variation of the 2PN 
 quasi-Keplerian parameters of {\it hyperboliclike} motions 
by relating them to the results of Ref. \cite{Damour:2004bz} on the 2.5PN-level, radiation-reaction correction to the quasi-Keplerian parametrization of {\it ellipticlike} motions. In order to relate the two types of results we 
used the fact that the latter 2.5PN-level, radiation-reaction correction only depends on the Newtonian-level 
Keplerian parametrization (which admits a smooth analytic continuation when changing the sign of the
binding energy).
We then had to go through two different steps:
(i) to relate the elliptic and hyperbolic quasi-Keplerian parametrizations by a simple analytic continuation
(as used, e.g., at the 1PN level in Ref. \cite{DD85}); and (ii) to take into account the fact that 
Ref. \cite{Damour:2004bz} worked in a different coordinate system [namely ADM coordinates], 
corresponding to a different explicit expression for the radiation-reaction force. Some partial results on the comparison to the
results of Ref. \cite{Damour:2004bz} are given in Appendix \ref{comparingDGI2004}.

It is convenient to integrate perturbed quantities with respect to the auxiliary variable $v$ by using the unperturbed relation $\frac{dt}{dv}=\bar a_r^{3/2}{\mathcal X}$.
The explicit solution of the above evolution system then reads
\bea \label{dadedcldclam}
\delta^{\rm rr}\bar a_r(t)&=&\frac{\nu}{\bar a_r^{3/2}}\left\{\frac{4(37 e_r^4+292 e_r^2+96)}{15(e_r^2-1)^{7/2}}At(v)\right.\nonumber\\
&+&\sinh v \left[\frac{28e_r }{3{\mathcal X}^4}+\frac{4e_r(36 e_r^2+49)}{45 (e_r^2-1){\mathcal X}^3}  
\right. \nonumber\\
&+&\left. \frac{2 e_r  (111 e_r^2+314)}{ 45(e_r^2-1)^2 {\mathcal X}^2} +\frac{2e_r (673 e_r^2+602)}{45 (e_r^2-1)^3 {\mathcal X}}\right] \nonumber\\
&+&\left. \frac{2(673 e_r^2+602)}{45 (e_r^2-1)^3}  
\right\}
\,,\nonumber\\
\delta^{\rm rr} e_r(t)&=& \frac{\nu}{\bar a_r^{5/2}}\left[
-\frac{2e_r (121 e_r^2+304)}{15(e_r^2-1)^{5/2} }At(v) \right.\nonumber\\
&+&\sinh v\left(
-\frac{14(e_r^2-1)}{3{\mathcal X}^4}
-\frac{2 (36 e_r ^2+49)}{45{\mathcal X}^3} \right.\nonumber\\
&-&\left.
\frac{(291 e_r ^2+134) }{ 45{\mathcal X}^2 (e_r^2-1)}  
-\frac{(72 e_r ^4+1069 e_r ^2+134)}{ 45 (e_r^2-1)^2{\mathcal X}}
\right)\nonumber\\
&-&\left.
\frac{(72 e_r ^4+1069 e_r ^2+134)}{45 (e_r^2-1)^2 e_r }
\right]
\,,\nonumber\\
\delta^{\rm rr} c_l(t)&=&\frac{\nu}{\bar a_r^{5/2}}\left[
\frac{14(e_r^2-1)^2}{3e_r^2 {\mathcal X}^4}  
+\frac{8(e_r^2-1) (9 e_r^2-14)}{45e_r^2{\mathcal X}^3}\right.\nonumber\\  
&-&\left. \frac{4(-3+43 e_r^2)}{15e_r ^2 {\mathcal X}^2}-\frac{32}{5{\mathcal X}}  
\right]
\,,\nonumber\\
\delta^{\rm rr} c_\phi(t) &=& \frac{\nu \sqrt{e_r^2-1}}{\bar a_r^{5/2}e_r^2}\left[ \frac{14(e_r^2-1)}{3{\mathcal X}^4}+\frac{8}{45}\frac{9e_r^2-14}{{\mathcal X}^3}\right.\nonumber\\
&&\left.-\frac{4}{5{\mathcal X}^2} \right]\,,
\eea
where 
\beq
At(v) \equiv  {\rm arctan}\left[\alpha\tanh\left(\frac{v}{2}\right)\right]
+{\rm arctan}\,\alpha\,, 
\eeq
with 
\beq
\alpha \equiv \sqrt{\frac{e_r+1}{e_r-1}}\,,
\eeq
and where the dependence of $v$ on $t$ is the unperturbed one.
Here we have assumed the boundary conditions $\lim_{t\to -\infty} \delta c_\alpha(t)=0$.

By looking at this solution, one sees that $\delta^{\rm rr} c_l(t)$ and $\delta^{\rm rr} c_\phi(t)$ 
are {\it even} functions of $t$,  so that they tend to the same value
(here chosen to be zero) both at $t= - \infty$ and at $t= + \infty$). By contrast, the two other quantities 
$\delta^{\rm rr} \bar a_r(t)$ and $\delta^{\rm rr}  e_r(t)$ vary between $t= - \infty$ and $t= + \infty$.
More precisely, one gets
total variations $[f] \equiv f(+ \infty)-  f(- \infty)$ given by
\bea
\left[\delta^{\rm rr} \bar a_r \right] &=& \frac{4}{15}\frac{\nu}{\bar a_r^{3/2}(e_r^2-1)^3}\left[
\frac{673e_r^2+602}{3}\right.\nonumber\\
&+&\left. 
\frac{37e_r^4+292e_r^2+96}{(e_r^2-1)^{1/2}}{\rm arccos}\left(-\frac{1}{e_r}\right)
\right]
\,,\nonumber\\
\left[\delta^{\rm rr}  e_r \right] &=&-\frac{2}{15}\frac{\nu}{\bar a_r^{5/2}(e_r^2-1)^2e_r}\left[
\frac{72 e_r^4+1069e_r^2+134}{3}\right. \nonumber\\
&+&\left.
\frac{e_r^2(121e_r ^2+304)}{(e_r^2-1)^{1/2}} {\rm arccos}\left(-\frac{1}{e_r}\right) \right] 
\,,\qquad
\eea
with $[\delta^{\rm rr} \bar a_r]= (2 \bar a_r^2/\nu)\,  \delta^{\rm rr} E^{\rm  N}$, as from Eqs. (C7)-(C9) of Ref. \cite{Bini:2021gat}.
These total variations agree with the total scattering changes in $\bar a_r$ and $e_r$
obtained in Eqs. (6.1) and (6.2) of Ref. \cite{Bini:2022xpp}  by assuming (to leading PN order) balance equations for  energy and angular momentum, between the system and radiation.

To complete the solution of the radiation-reacted  motion one needs to inject the 
results, Eqs. \eqref{dadedcldclam}, in the definitions of $l(t)$  and $\phi(t)$. In other words,
one must now evaluate the functions $l(t)= l^0(t) + \delta^{\rm rr} l(t)$ and
 $\phi(t)= \phi^0(t) + \delta^{\rm rr} \phi(t)$,
 where $ l^0(t)= \bar n^0 (t-t_0)$, $\phi^0(t) = c_\phi^0 +{\bar W}(l^0(t),c_1^0,c_2^0)$, and where  
\bea
 \delta^{\rm rr} l(t) &=& \int_{t_0}^t \delta^{\rm rr}\bar n(t) dt+ \delta^{\rm rr} c_l(t)\,, \nonumber\\
\delta^{\rm rr} \phi(t) &=& \delta^{\rm rr} c_\phi(t) +\frac{\partial \bar W}{\partial l} \delta^{\rm rr} l(t) \nonumber\\
&&+\frac{\partial \bar W}{\partial c_1} \delta^{\rm rr} c_1(t)
+\frac{\partial \bar W}{\partial c_2} \delta^{\rm rr} c_2(t)\,.
\eea
If we work only to the leading PN order (i.e. the 2.5PN order) we can (in the radiation-reacted contributions)
use the Newtonian-level approximation (notably $\bar n \approx (\bar a_r)^{-3/2}$, $K\approx 1$) so as to
get 
\bea
\delta^{\rm rr} l(t)&=&\frac{\nu}{15\bar a_r^{5/2}(e_r^2-1)^3} \left[-(673e_r^2+602)\times\right.\nonumber\\
&&[{\mathcal X}+1+e_r\sinh v-v]\nonumber\\
&-&
(111e_r^2+314)(e_r^2-1)\ln{\mathcal X}\nonumber\\
&+&
\frac{2(36e_r^2+49)(e_r^2-1)^2}{{\mathcal X}}
+\frac{105(e_r^2-1)^3}{{\mathcal X}^2}
\nonumber\\
&-&
\frac{6(37e_r^4+292e_r^2+96)}{\sqrt{e_r^2-1}}\times\nonumber\\
&&\left.
[{\mathcal L}+(e_r\sinh v-v)\arctan\alpha]
\right]\nonumber\\
&+&\hbox{\rm const.}
\,,\nonumber\\
\eea
where 
\beq
\label{calLdef}
{\mathcal L} \equiv \int^v dv\,{\mathcal X}(v)\, {\rm arctan}\left(\alpha\tanh\left(\frac{v}{2}\right)\right)\,,
\eeq
and where the integration constant can be chosen at will (e.g., to make $\delta^{\rm rr} l(t)$ vanish when $v=0$,
i.e., at the moment of closest approach in the hyperbolic motion).
Changing the integration variable as $v=2\,{\rm arctanh}\,T$ in the above integral yields
\bea
\label{calLdef2}
{\mathcal L}&=&\int dT\left[-\frac{1}{1+T}-\frac{1}{1-T}\right.\nonumber\\
&+&\left.
e_r\left(\frac{1}{(1+T)^2}+\frac{1}{(1-T)^2}\right)\right]
{\rm arctan}\left(\alpha T\right)\,,\qquad
\eea
which can be solved in terms of dilogarithms.
Explicitly, 
\bea
&&\int dT\,\frac{{\rm arctan}\left(\alpha T\right)}{(1\pm T)^2}=
\mp\frac{{\rm arctan}\left(\alpha T\right)}{1\pm T}
\nonumber\\
&&\qquad\,
+\frac{\alpha}{2(1+\alpha^2)}\left[
2\ln(1\pm T)-(1\pm i\alpha)\ln(1+i\alpha T)\right.\nonumber\\
&&\qquad\,\left.
-(1\mp i\alpha)\ln(1-i\alpha T)
\right]\,,
\eea
and
\bea
&&\int dT\,\frac{{\rm arctan}\left(\alpha T\right)}{1\pm T}=
\pm\frac{i}{2}\left[
\ln\left(\frac{\alpha(1\pm T)}{\mp i+\alpha}\right)\ln(1-i\alpha T)
\right.\nonumber\\
&&\qquad\,
+\ln\left(\frac{\alpha(1\pm T)}{\pm i+\alpha}\right)\ln(1+i\alpha T)\nonumber\\
&&\qquad\,\left.
-{\rm Li}_2\left(\frac{i-\alpha T}{i\pm\alpha}\right)
+{\rm Li}_2\left(\frac{i+\alpha T}{i\mp\alpha}\right)
\right]\,.
\eea

Finally, the solution for the orbit $x^i(t)=x^i_{\le{\rm 2PN}}(t)+\delta^{\rm rr} x^i(t)$, obtained by varying 
$l,  c_1, c_2, c_\phi$ in the functions $r(l,c_1,c_2)$ and $\phi(l,c_1,c_2, c_\phi)$ defined
by  Eqs. \eqref{sys0}--\eqref{SW}, reads
\bea
\label{deltarphi}
\delta^{\rm rr} r(t)&=&\frac1{{\mathcal X}}\bar a_re_r\sinh v\,\delta^{\rm rr} l(t)
+{\mathcal X}\delta^{\rm rr} \bar a_r(t)\nonumber\\
&+&
\frac{\bar a_r}{e_r}\left(-1+\frac{e_r^2-1}{{\mathcal X}}\right)\delta^{\rm rr} e_r(t)
\,,\nonumber\\
\delta^{\rm rr} \phi(t)&=&  \frac{\sqrt{e_r^2-1}}{{\mathcal X}^2}\delta^{\rm rr} l(t)
+\delta^{\rm rr} c_\phi(t)\nonumber\\
&-&
\frac{\sinh v}{{\mathcal X}}\left(\frac{\sqrt{e_r^2-1}}{{\mathcal X}}+\frac{1}{\sqrt{e_r^2-1}}\right)\delta^{\rm rr} e_r(t)
\,.\nonumber\\
\eea 
Taking into account the time-even character of $\delta^{\rm rr} c_l(t)$ and $\delta^{\rm rr} c_\phi(t)$,
the total change, $[\delta^{\rm rr}\phi]$, between $- \infty$ and $+ \infty$, of the value of $\delta^{\rm rr}\phi (t)$ is then 
easily seen\footnote{The term proportional to $\delta^{\rm rr} l(t)$ in $\delta^{\rm rr} \phi(t)$ vanishes at infinity.
Furthermore, $[\delta^{\rm rr} c_\phi]=0$ as from Eq. \eqref{dadedcldclam}.} to be
\beq
\label{deltaphivar}
[\delta^{\rm rr}\phi] = - \frac{[\delta^{\rm rr} e_r]}{e_r \sqrt{e_r^2-1}}\,.
\eeq
This agrees with the leading PN order result obtained in Ref. \cite{Bini:2012ji} for the radiation-reaction
contribution to the (relative) scattering angle: $\chi_{\rm rr}^{\rm 2.5PN} = [\delta^{\rm rr}\phi]$.
As already mentioned in Ref. \cite{Bini:2012ji} the general linear-response formula, Eq. (5.99) there, for 
$\chi_{\rm rr}(E,j)$ is generally valid (to linear order in radiation reaction) beyond the leading PN order,
under the two conditions that the unperturbed conservative motion be time-symmetric, and that the
radiation-reaction force be time-antisymmetric. [These two conditions generally ensure that 
$\frac{d\,\delta^{\rm rr}\bar a_r}{dt}$ and $\frac{d\,\delta^{\rm rr} e_r}{dt}$ will be time-symmetric,
while  $\frac{d\,\delta^{\rm rr} c_l}{dt}$ and $\frac{d\,\delta^{\rm rr} c_\phi}{dt}$ will be
time-antisymmetric, so that $[c_l]=0$ and $[c_\phi]=0$.]
For completeness, we present in Appendix A the explicit expressions of the 2.5PN, 3.5PN
and 4.5PN contributions to the function $\chi_{\rm rr}(E,j)$.

\section{Contribution to the radiated linear momentum coming from the radiation-reaction correction to hyperbolic motion}

Having in hands the radiation-reaction correction to hyperbolic motion we can now come back to the analytical determination
of the linear-momentum loss at the fractional 3PN accuracy.

Inserting in Eq. \eqref{FPUV} the expressions \eqref{UvsI} for the radiative moments in terms of the source moments,
and taking into account all instantaneous, semi-hereditary and hereditary terms contributing at the 3PN level, we get
a radiative linear-momentum flux of the form
\bea
{\mathcal F}_{P }{}_i&=& {\mathcal F}_{P }{}_i{}^{{\rm inst}\,I,J \, \leq {\rm 3PN}} +  \Delta{\mathcal F}_{P }{}_i^{{\rm inst}\,I,J}\nonumber\\
& +& {\mathcal F}_{P }{}_i^{\rm tail}+ {\mathcal F}_{P }{}_i{}^{\rm  higher-order \, tails}\,.
\eea
Here: the  ``leading-order instantaneous" term
$ {\mathcal F}_{P }{}_i{}^{{\rm inst}\,I,J \leq 3PN}$ is defined by replacing in Eq. \eqref{FPUV} the radiative moments
$U_L$ and $V_L$ by the source ones, $I_L$ and $J_L$; the ``supplementary instantaneous" contribution $\Delta{\mathcal F}_{P }{}_i^{{\rm inst}\,I,J}$ combines  contributions  bilinear in (the derivatives of) $I_L, J_L$ 
coming both from the instantaneous terms and the 
semi-hereditary ones in Eq. \eqref{UvsI}; finally the ``tail" terms  (both linear tails, and higher-order tails)   
${\mathcal F}_{P }{}_i^{\rm tail}+ {\mathcal F}_{P }{}_i{}^{\rm  higher-order \, tails}$ denote the contribution bilinear in $I_L, J_L$ and in the various hereditary contributions to $U_L, V_L$. 

The complete expression for the linear momentum flux at the 2.5PN fractional accuracy level is given in Eqs. (2.3)-(2.5) of Ref. \cite{Mishra:2011qz}. The notation used there is 
\bea
\left({\mathcal{F}_{P}^{i}}\right)_{\rm inst}&=&{\mathcal F}_{P }{}_i{}^{{\rm inst}\,I,J} 
+\Delta{\mathcal F}_{P }{}_i^{{\rm inst}\,I,J}
\,,\nonumber\\
\left({\mathcal{F}_{P}^{i}}\right)_{\rm hered}&=&{\mathcal F}_{P }{}_i^{\rm tail}\,.
\eea
In order to reach the 3PN accuracy, we need: (i) to insert in these
expressions the 3PN-accurate expressions of the source moments 
$I_L(t), J_L(t)$ considered as functions of dynamical time $t$; and (ii) to add the
higher-order tail contribution to the hereditary term $\left({\mathcal{F}_{P}^{i}}\right)_{\rm hered}$.
 When evaluating 3PN-accurate values
of the relevant $k$th time-derivatives, $I_L^{(k)}{}^{\rm  \leq 3PN}(t), J_L^{(k)}{}^{\rm \leq 3PN}(t)$, of the source moments one needs to use the 3PN-level equations of motion (including the 2.5PN 
radiation-reaction contribution), and then to express these time-differentiated moments along
 radiation-reacted  hyperboliclike solutions of the equations of motion. The latter are obtained by
 adding the 2.5PN-level, radiation-reaction effects discussed in the previous Section to the conservative 3PN
 hyperboliclike solutions (which will be discussed below).

 Let us symbolically write the motions as 
\bea
x^{\leq \rm 3PN}(t) &=& x^{\rm 3PN, cons}(t)+ \delta^{\rm rr} x(t)
\,,\nonumber\\
v^{\leq \rm 3PN}(t) &=&  v^{\rm 3PN, cons}(t)+ \delta^{\rm rr} v(t)\,.
\eea
As a consequence, the first contribution, ${\mathcal F}_{P }{}_i{}^{{\rm inst}\,I,J \, \leq 3PN}(t)$, to the linear-momentum flux
is naturally decomposed as a sum of two terms:
\bea
{\mathcal F}_{P }{}_i{}^{{\rm inst}\,I,J \,\leq {\rm 3PN}}(t)&=& {\mathcal F}_{P }{}_i{}^{{\rm inst}\,I,J\,{\rm 3PN, cons} }(t)\nonumber\\
& +& \delta^{\rm rr}{\mathcal F}_{P }{}_i{}^{{\rm inst}\,I,J}(t)\,.
\eea
In these expressions, and below, the symbol $ \delta^{\rm rr}$ will be used to denote the 2.5PN-level radiation-reaction-generated
contribution to some physical quantity, $Q(t)=Q(x(t),v(t))$, {\it considered as a function of dynamical time} $t$.
In the previous section, we obtained (at leading order) the various needed radiation-reaction
contributions, $\delta^{\rm rr} x(t), \delta^{\rm rr} v(t), \delta^{\rm rr} Q(t)$  by using Lagrange's
method of variation of constants.

Finally,  integrating ${\mathcal F}_{P}{}_i$ over $t$ (from $- \infty$ to $+ \infty$) we get the
total linear momentum radiated in gravitational waves during a full hyperbolic encounter:
\beq
P^{\rm rad}_i= \int_{- \infty}^{+ \infty}dt  \,{\mathcal F}_{P}{}_i(t)\,.
\eeq
The 6.5PN-accurate value of  $P^{\rm rad}_i$ is then obtained as a sum of various contributions,  say:
\bea \label{decompPrad}
P^{\rm rad}_i&=& P^{\rm rad}_i{}^{\,{\rm inst}\,I,J\, {\rm 3PN, cons}}+ \delta^{\rm rr}P^{\rm rad}_i{}^{\,{\rm inst}\,I,J}\nonumber\\
&+&
\Delta P^{\rm rad}_i{}^{\,{\rm inst}\,I,J}
+P^{\rm rad}_i{}^{\,\rm tail} +P^{\rm rad}_i{}^{\,\rm higher-order\, tails}\,.\nonumber\\
\eea
The resulting vectorial contributions will be projected on an orthonormal basis ${\bf e}_x$, ${\bf e}_y$ defined
in terms of the vectorial impact parameter
 $ b_{12}^\mu=b \hat { b}_{12}^\mu$,  of 
 the initial  four velocities $u_{1-}^\mu$ and $u_{2-}^\mu$ of the two bodies, and of the conservative\footnote{In our treatment below, the coordinate basis ${\bf e}_x$, ${\bf e}_y$, ${\bf e}_z$, enters via the
2PN-accurate quasi-Keplerian representation of the scattering motion.} part of the scattering angle, $\chi_{\rm cons}$ (see e.g., Table X of Ref. \cite{Bini:2021gat}, also recalled at 2PN  in Eqs. \eqref{chi_cons_2PN_formal}--\eqref{chi_cons_2PN_coeffs} below for convenience). The basis ${\bf e}_x$, ${\bf e}_y$ was already used in Ref. \cite{Bini:2021gat}
(see Eq. (3.49) there). Its definition is
recalled in Appendix \ref{appA}. Let us only mention here that  ${\bf e}_x$  is in the direction of the major axis of the
hyperboliclike relative orbit (direction of closest approach).

The 2PN-accurate value of the instantaneous contribution to linear-momentum loss
 has been evaluated in Ref. \cite{Bini:2021gat}, see Eqs. (G6)-(G9) there.
We have extended this result by including both  the higher-order tail effects (which
were computed in Ref. \cite{Bini:2022yrk}), and the 3PN-level conservative effects.
The technology (including a 3PN-accurate quasi-Keplerian representation of hyperboliclike motions)
needed for computing 3PN-level conservative instantaneous contribution will be discussed below.

 Let us discuss here the evaluation of the radiation-reaction-related contribution 
 $\delta^{\rm rr}{\mathcal F}_{P }{}_i{}^{{\rm inst},\,I,J}$.
 To obtain it, it is enough to evaluate the Newtonian flux
\bea
{\mathcal F}_{P }{}_i{}^{{\rm inst},\, I,J\,{\rm N}}&=& -\frac{64}{105}\frac{G^3M^4}{r^4c^7}\frac{m_2-m_1}{M}\nu^2 \nonumber\\
&&\times 
\left(A^{\rm N} n_i+B^{\rm N} v_i\right)\,,
\eea
with
\bea
\label{ABNdefs}
A^{\rm N}&=&\dot r\left(\frac{55}{8}v^2-\frac{45}{8}\dot r^2+\frac32\frac{GM}{r}\right)
\,,\nonumber\\
B^{\rm N}&=&-\left(\frac{25}{4}v^2-\frac{19}{4}\dot r^2+\frac{GM}{r}\right) 
\,,
\eea
along the radiation-reaction-perturbed orbit, i.e., by substituting in it
\bea
r(t)&=&r_{\rm N}(t)+\delta^{\rm rr} r (t)\,, \qquad
\dot r(t)=\dot r_{\rm N}(t)+\delta^{\rm rr} \dot r (t)\,, \nonumber\\
\phi(t)&=&\phi_{\rm N}(t)+\delta^{\rm rr}\phi (t)\,, \qquad
\dot \phi(t)=\dot \phi_{\rm N}(t)+\delta^{\rm rr} \dot \phi (t)\,, 
\nonumber\\
\eea
taking then the time integral, retaining only linear corrections.
[The magnitude of the relative velocity in Eq. \eqref{ABNdefs} above should not be confused with the auxiliary variable used to parametrize the orbit, denoted by the same letter $v$.]

The variations $\delta^{\rm rr} r(t)$ and $\delta^{\rm rr}\phi(t)$ are given by Eq. \eqref{deltarphi}.
The related variations $\delta^{\rm rr} \dot r (t)$ and $\delta^{\rm rr} \dot \phi (t)$ are obtained
either by taking the time derivatives of $\delta^{\rm rr} r(t)$ and $\delta^{\rm rr}\phi(t)$ or
 by varying the {\it functions}  $\dot r (l, c_1,c_2)$ and $\dot \phi (l,c_1,c_2,c_\phi)$ in
Eqs. \eqref{sys0}--\eqref{SW}.
This yields 
\bea
\delta^{\rm rr} \dot r(t) &=&
\frac{e_r}{\bar a_r^{1/2}{\mathcal X}^3}(e_r-\cosh v) \delta^{\rm rr} l(t)
-\frac12\frac{e_r\sinh v}{\bar a_r^{3/2}{\mathcal X}}\delta^{\rm rr} \bar a_r(t)\nonumber\\
&-&
\frac{(e_r^2-1)\sinh v}{\bar a_r^{1/2}{\mathcal X}^3}\delta^{\rm rr} e_r(t)
\,,\nonumber\\
\delta^{\rm rr} \dot \phi (t)&=&
-2\frac{e_r\sqrt{e_r^2-1}}{\bar a_r^{3/2}{\mathcal X}^4}\sinh v\,\delta^{\rm rr} l(t)
-\frac32 \frac{\sqrt{e_r^2-1}}{\bar a_r^{5/2}{\mathcal X}^2}\delta^{\rm rr} \bar a_r(t)\nonumber\\
&+&
\frac{e_r}{\bar a_r^{3/2}{\mathcal X}^2\sqrt{e_r^2-1}}\left[
1+\frac{2(e_r^2-1)}{e_r^2{\mathcal X}}\left(1-\frac{e_r^2-1}{{\mathcal X}}\right)
\right]\nonumber\\
&\times& 
\delta^{\rm rr} e_r(t)
\,.
\eea
We have  checked that they satisfy $\frac{d\delta^{\rm rr} r(t)}{dt}=\delta^{\rm rr} \dot r(t)$ and $\frac{d\delta^{\rm rr} \phi(t)}{dt}=\delta^{\rm rr} \dot \phi(t)$.

We finally get the 2.5PN correction to the Newtonian flux
\beq
{\mathcal F}_{P }{}_i{}^{{\rm inst},\,I,J\,{\rm N}}\big\vert_{x^\mu=x^\mu_{\rm N}}+\delta^{\rm rr}{\mathcal F}_{P }{}_i{}^{{\rm inst},\,I,J}
\,,
\eeq
which has to be integrated along the orbit to yield
\beq
 \delta^{\rm rr} P_i^{{\rm rad\,inst},\,I,J}= \int dt \, \delta^{\rm rr}{\mathcal F}_{P }{}_i{}^{{\rm inst},\,I,J}\,.
\eeq
The final exact results are given by the following functions of $\bar a_r$ and $e_r$ (here, and below, $\eta$ is a place holder
to indicate a half PN order $\frac1c$):
\begin{widetext}
\bea
\label{Pxy_rad_inst_25PNexact}
\delta^{\rm rr} P_x^{{\rm rad\,inst},\, I,J}&=&-(Mc)\frac{m_2-m_1}{M}\nu^3\eta^5\frac{1}{e_r[\bar a_r(e_r^2-1)]^{13/2}}\times\nonumber\\
&&
\left[{\rm arccos}\left(-\frac{1}{e_r}\right)\left(\frac{110416}{675}+\frac{132304}{135}e_r^2+\frac{5134544}{4725}e_r^4+\frac{1365802}{4725}e_r^6+\frac{30331}{1800}e_r^8\right)\right.\nonumber\\
&&\left. 
+\frac{(e_r^2-1)^{1/2}}{e_r^2}\left(\frac{8576}{2025}+\frac{11644714}{33075}e_r^2+\frac{22762729}{18375}e_r^4+\frac{1623094259}{1984500}e_r^6+\frac{159585499}{1323000}e_r^8+\frac{15872}{6125}e_r^{10}\right)
\right]
\,,\nonumber\\
\delta^{\rm rr} P_y^{{\rm rad\,inst},\, I,J}&=&(Mc)\frac{m_2-m_1}{M}\nu^3\eta^5\frac{1}{[\bar a_r(e_r^2-1)]^{13/2}}\left[
e_r{\rm arccos}^2\left(-\frac{1}{e_r}\right)\left( \frac{2479}{225} e_r^6+\frac{22616}{45} e_r^4+\frac{35416}{75} e_r^2+\frac{48256}{75}  \right)\right.\nonumber\\
&&
+\frac{(e_r^2-1)^{1/2}}{ e_r} {\rm arccos}\left(-\frac{1}{e_r}\right)\left(\frac{296}{25} e_r^8+\frac{277966}{675} e_r^6+\frac{752812}{675} e_r^4+\frac{75592}{45} e_r^2+\frac{1072}{27}\right)\nonumber\\
&&\left.
+\frac{e_r^2-1}{e_r^3}\left(\frac{9352}{75} e_r^8+\frac{8027}{45} e_r^6+\frac{2686964}{2025} e_r^4-\frac{10084}{2025} e_r^2+\frac{8576}{2025}\right)
 \right]
\,.
\eea

\end{widetext}
The first terms of their expansions in inverse powers of $j$ (equivalent,
remembering $j \propto G^{-1}$ to a PM expansion) read
\bea
\label{delta_rr_P_xy_rad_inst_IJ}
 \delta^{\rm rr} P_x^{{\rm rad\,inst},\, I,J}&=&
-(Mc)\frac{m_2-m_1}{M}\nu^3\eta^5\left[
\frac{15872}{6125}\frac{p_\infty^8}{j^5} \right.\nonumber\\
&+&
\frac{30331}{360}\pi\frac{p_\infty^7}{j^6} \nonumber\\
&+&\left.
\frac{24234752}{165375}\frac{p_\infty^6}{j^7}
+O\left(\frac{1}{j^8}\right)
\right]
\,,\nonumber\\
 \delta^{\rm rr} P_y^{{\rm rad\,inst},\, I,J}&=&
(Mc)\frac{m_2-m_1}{M}\nu^3\eta^5\left[
\frac{148}{25}\pi\frac{p_\infty^8}{j^5}\right.\nonumber\\
&+&
\left(\frac{2048}{15}+\frac{2479}{900}\pi^2\right)\frac{p_\infty^7}{j^6}\nonumber\\
&+&\left.
\frac{160406}{675}\pi\frac{p_\infty^6}{j^7}
+ O\left(\frac{1}{j^8}\right)
\right]
\,.
\eea

\section{2.5PN instantaneous contributions to the  radiated linear momentum}

Let us now evaluate the third contribution to $P^{\rm rad}_i$ denoted 
 $\Delta P^{\rm rad}_i{}^{\rm inst}$ in Eq. \eqref{decompPrad}. 
 This contribution is obtained by integrating over time
 the (2.5PN level) ``instantaneous" part of the linear momentum flux  in terms of the source multipole moments.
Using the results of Ref. \cite{Blanchet:2008je},  Ref. \cite{Mishra:2011qz} has explicated 
this instantaneous part as the $O(\frac1{c^5})$ term in  Eq. (2.3) there.
Recently Ref. \cite{Kastha:2021kyn} has provided an explicit expression for this 2.5PN instantaneous part of the linear momentum flux as a function of the (relative) position and velocity along the orbit,
see Eq. (4.1) there. For clarity, we reproduce here this explicit expression:
\bea
\Delta{\mathcal F}_{P }{}_i^{{\rm inst}\,I,J}&=& -\frac{64}{105}\frac{G^3M^4}{r^4c^7}\frac{m_2-m_1}{M}\nu^2  \nonumber\\
&&\times 
\left(A^{\rm 2.5PN} n_i+B^{\rm 2.5PN} v_i\right)\,,
\eea
where
\bea
A^{\rm 2.5PN}&=& \frac{GM}{rc^5}\nu \left[ \frac{701}{90}v^6-\frac{51137}{96}v^4 \dot r^2+\frac{41611}{40}v^2\dot r^4
\right.\nonumber\\
&-&
\frac{49219}{96}\dot r^6-\frac{4}{15}\frac{G^3M^3}{r^3}\nonumber\\
&+&
\frac{G^2M^2}{r^2}\left(\frac{1237}{90}v^2-\frac{6607}{180}\dot r^2\right)\nonumber\\
&-&\left.
\frac{GM}{r}\left(\frac{4261}{120}v^4 +\frac{8397}{40}v^2 \dot r^2-\frac{3778}{15}\dot r^4 \right) \right]\,, \nonumber\\
B^{\rm 2.5PN}&=& \frac{GM}{rc^5}\nu\dot r \left[\frac{157787}{480}v^4 -\frac{39869}{60}v^2 \dot r^2+\frac{31913}{96}\dot r^4\right.\nonumber\\
 &+&\left.
\frac{GM}{r}\left(\frac{10773}{40}v^2 -\frac{99277}{360}\dot r^2\right)+\frac{737}{36}\frac{G^2M^2}{r^2}\right]\,.\nonumber\\
\eea
The integral along a hyperboliclike orbit of $\Delta {\mathcal F}_{P }{}_i{}^{{\rm inst}\, I,J}$
can be explicitly evaluated. After projection on the $x$ and $y$ axes defined in Eq. \eqref{exydef},
one finds 
\begin{widetext}
\bea
\label{25instP}
\Delta P^{\rm rad}_x{}^{\,{\rm inst}\,I,J}&=&(Mc)\frac{m_2-m_1}{M}\nu^3\eta^5\frac{e_r}{[\bar a_r(e_r^2-1)]^{13/2}} \times \nonumber\\
&&
\left[\arccos\left(-\frac1{e_r}\right)\left(\frac{491447}{33600} e_r^8+\frac{123798}{175} e_r^6+\frac{30800977}{9450} e_r^4+\frac{13714844}{4725} e_r^2+\frac{2125082}{4725}\right)\right.\nonumber\\
&&\left.
+\frac{(e_r^2-1)^{1/2}}{e_r^2}\left(\frac{797859313}{3528000} e_r^8+\frac{7556008631}{3175200} e_r^6+\frac{4935155857}{1323000} e_r^4+\frac{652923197}{661500} e_r^2+\frac{5266216}{496125}\right)
\right]
\nonumber\\
&=& (Mc)\frac{m_2-m_1}{M}\nu^3\eta^5\left[
\frac{491447}{67200}\frac{\pi p_\infty^9}{j^4}
+\frac{13272832}{55125}\frac{p_\infty^8}{j^5}
+\frac{494871}{1280}\frac{\pi p_\infty^7}{j^6}
+\frac{1954525568}{496125}\frac{p_\infty^6}{j^7}
+O\left(\frac{1}{j^8}\right)
\right]
\,,\nonumber\\
\Delta P^{\rm rad}_y{}^{\,{\rm inst}\,I,J}&=& 0\,.
\eea
\end{widetext}
In the last line of the first equation, we have also given the first few terms of its large-$j$ expansion. Let us note that this contribution is
(contrary to the other 2.5PN contribution discussed in the previous Section) 
purely oriented along the $x$ axis, i.e. along the vectorial distance of closest approach.

\section{New contributions to the radiated energy}
\label{addition1}

Let us repeat for the radiated energy the above treatment for the radiated linear momentum, namely
\bea
E^{\rm rad}&=&E^{{\rm rad \, inst}\, I,J\, \leq{\rm 3PN, cons}}
+\delta^{\rm rr}E^{{\rm rad \, inst}\, I,J}\nonumber\\
&+&\Delta E^{{\rm rad \, inst}\, I,J}+E^{\rm rad\, tail}+E^{\rm rad\, higher-order \, tails}
\,.\nonumber\\
\eea
Here, we have indicated the (fractional) 3PN level of accuracy for the instantaneous term $E^{{\rm rad \, inst}\, I,J\, {\rm 3PN,cons}}$.
The 2PN-accurate instantaneous energy loss $E^{{\rm rad \, inst}\, I,J\, {\rm 2PN,cons}}$ was first obtained in \cite{Bini:2021gat} (see Eqs. (C7)-(C13)); its extension at the 3PN level was obtained in \cite{Cho:2021onr}. We have redone an independent 3PN-accurate computation
of the energy loss and found agreement with the final results of Ref. \cite{Cho:2021onr}  (after correcting several typos in the 
3PN quasi-Keplerian expressions of Ref. \cite{Cho:2018upo}, see Appendix \ref{3PN_orb_par}).
The leading-PN-order contribution to the linear-tail  $E^{\rm rad\, tail}$ has been obtained  in \cite{Bini:2021gat} (see Eq. (D26)), while its 1PN correction  is given in Eq. (5.20) of Ref. \cite{Bini:2022xpp}; see also Ref. \cite{Bini:2021jmj}
for a Fourier space analysis.  The higher-order tail contribution $E^{\rm rad\, higher-order \, tails}$ 
has been derived in Refs. \cite{Bini:2021qvf,Cho:2022pqy}.
As discussed in the text below Eq. (3.1) of Ref. \cite{Bini:2022xpp}, the last contribution $\Delta E^{{\rm rad \, inst}\, I,J}$
vanishes (because of the time-odd character of its integrand):
\beq
\Delta E^{{\rm rad \, inst}\, I,J}=0\,.
\eeq
Ref. \cite{Cho:2021onr} claimed  (see below Eq. (42) there) that, because of the time-odd character of radiation reaction, 
the term $\delta^{\rm rr}E^{{\rm rad \, inst}\, I,J}$ was similarly vanishing. We found that this was not correct because of
the time-asymmetric character of the motion perturbation $ \delta^{\rm rr} x(t)$, $ \delta^{\rm rr} v(t)$. We got a
non-zero result for $\delta^{\rm rr}E^{{\rm rad \, inst}\, I,J}$.
We further found that this non-vanishing contribution plays a crucial role in obtaining a correct mass-polynomiality behavior for
 the radiated (four) momentum. 
 
 The exact expression of $\delta^{\rm rr}E^{{\rm rad \, inst}\, I,J}$ in terms of $\bar a_r$ and $e_r$ reads
\begin{widetext}
\bea
\label{E_rad_inst_25PNexact}
\delta^{\rm rr}E^{{\rm rad \, inst}\, I,J}&=& (Mc^2)\nu^3\eta^5\frac{1}{[\bar a_r(e_r^2-1)]^6}\left[
{\rm arccos}^2\left(-\frac{1}{e_r}\right)\left(
\frac{7696 }{225}e_r^6+\frac{53936 }{75}e_r^4+\frac{150272 }{225}e_r^2+\frac{14336}{25}
\right)\right.\nonumber\\
&+&
(e_r^2-1)^{1/2} {\rm arccos}\left(-\frac{1}{e_r}\right)\left(
\frac{592 }{25}e_r^6+\frac{465952 }{675}e_r^4+\frac{44176 }{27}e_r^2+\frac{1106464}{675}
\right)\nonumber\\
&+&\left.
\frac{e_r^2-1}{e_r^2}\left(
\frac{44848 }{225}e_r^6+\frac{11056 }{25}e_r^4+\frac{313024 }{225}e_r^2-\frac{8576}{225}
\right)
\right]\,.
\eea
The beginning of its expansion in $\frac1j$ (i.e., of its PM expansion in powers of $G$) reads
\beq
\label{E_rad_inst_25PN}
\delta^{\rm rr}E^{{\rm rad \, inst}\, I,J}=(Mc^2)\nu^3\eta^5\left[\frac{296}{25}\pi\frac{p_\infty^7}{j^5}
+\left(\frac{50176}{225}+\frac{1924}{225}\pi^2\right)\frac{p_\infty^6}{j^6}
+\frac{56008}{135}  \pi \frac{p_\infty^5}{j^7}
+O\left(\frac{p_\infty^4}{j^8}\right)\right]\,.
\eeq
Adding this term to the 1PN corrections to the LO tails \cite{Bini:2021qvf,Cho:2022pqy,Bini:2022xpp} then gives the following complete expression for the 2.5PN radiated energy
\bea
\label{Erad2p5pn}
E^{\rm rad}_{\rm 2.5PN}&=&(Mc^2)\nu^2\eta^5\left[
\left(\frac{1216}{105}-\frac{2848 \nu }{15}\right) \frac{p_\infty^8}{j^4}\right.\nonumber\\
&+&
\left(\left(\frac{296}{25}-\frac{15291 \pi ^2}{280}\right) \nu
   -\frac{24993 \pi ^2}{1120}+\frac{9216}{35}\right) \pi\frac{p_\infty^7}{j^5}\nonumber\\
&+&
\left(\left(-\frac{71488}{75}-\frac{2974508 \pi
   ^2}{4725}\right) \nu +\frac{2898 \zeta (3)}{5}+\frac{1024 \pi
   ^2}{135}+\frac{56708}{105}\right) \frac{p_\infty^6}{j^6}\nonumber\\
&+&\left.
\left(\left(\frac{56008}{135}-\frac{23514 \pi
   ^2}{7}+\frac{30285 \pi ^4}{112}\right) \nu +\frac{689985 \pi
   ^4}{3584}-\frac{13138915 \pi
   ^2}{7392}+\frac{210176}{225}\right) \pi\frac{p_\infty^5}{j^7}
+O\left(\frac{p_\infty^4}{j^8}\right)
\right]
\,.
\eea
Let us also exhibit the $\frac1j$ expansion of the full 3PN-level  contribution to the energy loss, which combines terms
from several sources: the (exact) instantaneous contribution linked to 3PN-level multipole moments \cite{Cho:2021onr},
and the higher-order tails (tails-of-tails and tail squared) \cite{Bini:2021qvf,Cho:2022pqy,Bini:2022xpp}
\bea
\label{Erad3pn}
E^{\rm rad}_{\rm 3PN}&=&(Mc^2)\nu^2\eta^6\left[
\left(-\frac{148 \nu ^3}{15}+\frac{321 \nu ^2}{280}-\frac{2699\nu }{504}-\frac{676273}{354816}\right)\pi\frac{p_\infty^{10}}{j^3}\right.\nonumber\\
&+&
\left(-\frac{2366 \nu ^3}{9}+\frac{164 \nu ^2}{3}-\frac{1223594
   \nu }{33075}-\frac{151854}{13475}\right) \frac{p_\infty^9}{j^4}\nonumber\\
&+&
\left(-\frac{1823 \nu ^3}{5}+\frac{12269 \nu
   ^2}{80}+\left(\frac{76897}{480}-\frac{4059 \pi ^2}{640}\right)
   \nu -\frac{10593}{350} \ln \left(\frac{p_\infty}{2}\right)+\frac{99
   \pi ^2}{10}+\frac{29573617463}{310464000}\right) \pi\frac{p_\infty^8}{j^5}\nonumber\\
&+&
\left(-\frac{150892 \nu ^3}{45}+\frac{4201976 \nu
   ^2}{1575}+\left(\frac{875976284}{297675}-\frac{212216 \pi
   ^2}{1575}\right) \nu -\frac{18955264 }{23625}\ln (2
   p_\infty)\right.\nonumber\\
&+&\left.
\frac{177152 \pi
   ^2}{675}+\frac{36589282372}{11694375}\right)\frac{p_\infty^7}{j^6}\nonumber\\
&+&
\left(-\frac{13955 \nu ^3}{6}+\frac{1419153 \nu
   ^2}{448}+\left(\frac{68898691}{36288}-\frac{51947 \pi
   ^2}{384}\right) \nu -\frac{337906}{315} \ln
   \left(\frac{p_\infty}{2}\right)\right.\nonumber\\
&-&\left.\left.
\frac{58957 \zeta (3)}{32}+\frac{3158
   \pi ^2}{9}+\frac{37546579757}{8467200}\right)\pi\frac{p_\infty^6}{j^7}
+O\left(\frac{p_\infty^5}{j^8}\right)
\right]\,.
\eea
An equivalent expression   (and extended up to $1/j^{15}$), can be found 
in Ref. \cite{Cho:2022pqy}. [Note that Eq. (B3) of the published version (and of the arxiv version 1) uses a different parametrization, $p\not=p_\infty$, while Eq. (C3) of the arxiv version 2 has been  updated with the notation $p=p_\infty$.]

\section{New contributions to the radiated angular momentum}

Similarly, for the angular momentum, we have
\bea
J^{\rm rad}&=&J^{{\rm rad \, inst}\, I,J\, \leq{\rm 3PN, cons}}
+\delta^{\rm rr}J^{{\rm rad \, inst}\, I,J}+J^{{\rm rad \, mem}\, I,J}\nonumber\\
&+&\Delta J^{{\rm rad \, inst}\, I,J}+J^{\rm rad\, tail}+J^{\rm rad\, higher-order \, tails}
\,.
\eea
The 2.5PN instantaneous term is also vanishing in this case \cite{Bini:2022xpp}.
Therefore, the only contributions at that order come from the 1PN corrections to the LO tails, a memory term \cite{Bini:2021qvf,Bini:2022xpp}, and the radiation-reaction correction to hyperbolic motion.
The latter turns out to be
\bea
\label{J_rad_inst_25PNexact}
\delta^{\rm rr}J^{{\rm rad \, inst}\, I,J}&=& \frac{GM^2}{c}\nu^3\eta^5\frac{1}{[\bar a_r(e_r^2-1)]^{9/2}}\left[
{\rm arccos}^2\left(-\frac{1}{e_r}\right)\left(
\frac{5264 }{75}e_r^4+\frac{1792 }{15}e_r^2+\frac{8192}{25}
\right)\right.\nonumber\\
&+&
(e_r^2-1)^{1/2} {\rm arccos}\left(-\frac{1}{e_r}\right)\left(
\frac{752 }{25}e_r^4+\frac{59792 }{225}e_r^2+\frac{33248}{45}
\right)\nonumber\\
&+&\left.
\frac{e_r^2-1}{e_r^2}\left(
\frac{128 }{25}e_r^6+\frac{1328 }{75}e_r^4+\frac{23968 }{45}e_r^2-\frac{8576}{225}
\right)
\right]
\,,
\eea
as an exact expression in terms of $\bar a_r$ and $e_r$.
The beginning of its $\frac1j$ expansion is:
\beq
\label{J_rad_inst_25PN}
\delta^{\rm rr}J^{{\rm rad \, inst}\, I,J}= \frac{GM^2}{c}\nu^3\eta^5\left[
\frac{128}{25}\frac{p_\infty^6}{j^3}+\frac{376}{25}\pi \frac{p_\infty^5}{j^4}
+\left(\frac{4352}{75}+\frac{1316}{75}\pi^2\right)\frac{p_\infty^4}{j^5}
+\frac{52456}{225}\pi \frac{p_\infty^3}{j^6}
+O\left(\frac{p_\infty^2}{j^7}\right)
\right]\,.
\eeq
Adding all terms leads to the final result 
\bea
\label{Jrad2p5pn}
J^{\rm rad}_{\rm 2.5PN}&=&\frac{GM^2}{c}\nu^2\eta^5\left[
\left(\frac{1184}{21}-\frac{431936 \nu }{1575}\right) \frac{p_\infty^6}{j^3}\right.\nonumber\\
&+&
\left(\left(\frac{7816}{525}-\frac{2232 \pi ^2}{35}\right) \nu
   -\frac{1305 \pi ^2}{112}+\frac{7488}{25}\right) \pi\frac{p_\infty^5}{j^4}\nonumber\\
&+&
\left(\left(-\frac{225536}{525}-\frac{201724 \pi
   ^2}{33075}\right) \nu +\frac{4116 \zeta (3)}{5}-\frac{130688
   \pi ^2}{6615}+\frac{147064}{315}\right) \frac{p_\infty^4}{j^5}\nonumber\\
&+&\left.
\left(\left(\frac{365392}{1575}-\frac{57037 \pi
   ^2}{21}+\frac{102619 \pi ^4}{448}\right) \nu +\frac{163083 \pi
   ^4}{1792}-\frac{18227 \pi ^2}{28}+\frac{32}{15}\right) \pi\frac{p_\infty^3}{j^6}
+O\left(\frac{p_\infty^2}{j^7}\right)
\right]
\,.
\eea
New with this work is also the computation of the full 3PN-level contribution to the angular momentum loss. It is obtained
by combining the (exact) instantaneous contribution of Ref. \cite{Cho:2021onr}  (which we independently recomputed), and higher-order tails \cite{Bini:2021qvf}. 
We got
\bea
\label{Jrad3pn}
J^{\rm rad}_{\rm 3PN}&=&\frac{GM^2}{c}\nu^2\eta^6\left[
\left(-\frac{16 \nu ^3}{5}+\frac{24 \nu ^2}{7}+\frac{878 \nu
   }{315}+\frac{3712}{3465}\right)\frac{p_\infty^9}{j}\right.\nonumber\\
&+&
\left(-\frac{553 \nu ^3}{24}+\frac{9235 \nu ^2}{672}+\frac{1469
   \nu }{504}+\frac{115769}{126720}\right) \pi\frac{p_\infty^8}{j^2}\nonumber\\
&+&
\left(-\frac{6224 \nu ^3}{15}+\frac{67432 \nu
   ^2}{315}+\frac{1459694 \nu
   }{11025}-\frac{4955072}{121275}\right) \frac{p_\infty^7}{j^3}\nonumber\\
&+&
\left(-\frac{861 \nu ^3}{2}+\frac{74693 \nu
   ^2}{280}+\left(\frac{2048629}{7560}-\frac{123 \pi
   ^2}{32}\right) \nu -\frac{4922}{175} \ln
   \left(\frac{p_\infty}{2}\right)+\frac{46 \pi
   ^2}{5}-\frac{561803611}{10584000}\right) \pi\frac{p_\infty^6}{j^4}\nonumber\\
&+&
\left(-\frac{136976 \nu ^3}{45}+\frac{13320808 \nu
   ^2}{4725}+\left(\frac{85939786}{42525}-\frac{3362 \pi
   ^2}{75}\right) \nu -\frac{931328 }{1575}\ln (2 p_\infty)+\frac{8704
   \pi ^2}{45}+\frac{7781823776}{16372125}\right) \frac{p_\infty^5}{j^5}\nonumber\\
&+&
\left(-\frac{6517 \nu ^3}{4}+\frac{794749 \nu
   ^2}{336}+\left(\frac{46277}{432}-\frac{861 \pi ^2}{64}\right)
   \nu -\frac{21614}{35} \ln
   \left(\frac{p_\infty}{2}\right)-\frac{45261 \zeta (3)}{40}+202 \pi
   ^2+\frac{5288341351}{4233600}\right) \pi\frac{p_\infty^4}{j^6}\nonumber\\
&+&\left.
O\left(\frac{p_\infty^3}{j^7}\right)
\right]
\,.
\eea
\end{widetext}


\section{1PN-accurate tail contribution to the radiated linear momentum}

Let us now tackle  the technically challenging (fractionally 1PN) tail contribution to the radiated linear momentum,
namely the term $P_i^{\rm rad\, tail}$ in Eq. \eqref{decompPrad}. It is the time integral of the
following linear-momentum-flux integrand  (see Eq. (2.5) of Ref. \cite{Mishra:2011qz})
\bea 
\label{1PNtailflux}
{\mathcal F}_{P }^i{}^{\rm tail}&=& \frac{G^2{\mathcal M}}{c^{10}}\left\{
\frac{4}{63}\left(F^i_{I_3^{(4)}I_2^{(5)}}+F^i_{I_2^{(3)}I_3^{(6)}}\right)\right.\nonumber\\
&+&
\frac{32}{45}\left({}^*F^i_{I_2^{(3)}J_2^{(5)}}+{}^*F^i_{J_2^{(3)}I_2^{(5)}}\right)\nonumber\\
&+& 
\frac{1}{c^2}\left[
\frac{1}{567}\left(F^i_{I_4^{(5)}I_3^{(6)}}+F^i_{I_3^{(4)}I_4^{(7)}} \right)\right.\nonumber\\
&+&
\frac{1}{63}\left({}^*F^i_{I_3^{(4)}J_3^{(6)}}+{}^*F^i_{J_3^{(4)}I_3^{(6)}}\right)\nonumber\\
&+&\left.\left.
\frac{8}{63}\left(F^i_{J_3^{(4)}J_2^{(5)}}+F^i_{J_2^{(3)}J_3^{(6)}} \right) 
\right]
\right\}\,,
\eea
where $\mathcal{M}=M (1+\nu\bar E)$ is the total ADM mass of the system, and the definitions of the quantities $F^i_{X_L^{(n)}Y_M^{(m)}}$ in terms of the source multipole moments are given in Table \ref{tab:variousF}.


\begin{table}  
\caption{\label{tab:variousF}  Definition of the various terms $F^i_{X_L^{(n)}Y_M^{(m)}}$ entering the tail part of the linear momentum flux.
Here we have introduced the following set of multipolar tail time scales 
$C_{I_2}= 2\tau_0 e^{-11/12}$, $C_{I_3}= 2\tau_0 e^{-97/60}$, $C_{J_2}= 2\tau_0 e^{-7/6}$, $C_{J_3}= 2\tau_0 e^{-5/3}$, and $C_{I_4}= 2\tau_0 e^{-59/30}$.
}
\begin{ruledtabular}
\begin{tabular}{ll}
$F^i_{I_3^{(4)}I_2^{(5)}}$&$I_{ijk}^{(4)}(t)\int_0^\infty d\tau  I_{jk}^{(5)}(t-\tau)\ln \left( \frac{\tau}{C_{I_2}} \right)$  \\
$F^i_{I_2^{(3)}I_3^{(6)}}$&$I_{jk}^{(3)}(t)\int_0^\infty d\tau  I_{ijk}^{(6)}(t-\tau)\ln \left( \frac{\tau}{C_{I_3}} \right) $\\
${}^*F^i_{I_2^{(3)}J_2^{(5)}}$&$\epsilon_{ijk} I_{ja}^{(3)}(t)\int_0^\infty d\tau  J_{ka}^{(5)}(t-\tau)\ln \left( \frac{\tau}{C_{J_2}} \right)  $\\
${}^*F^i_{J_2^{(3)}I_2^{(5)}}$&$\epsilon_{ijk} J_{ka}^{(3)}(t)\int_0^\infty d\tau  I_{ja}^{(5)}(t-\tau)\ln \left( \frac{\tau}{C_{I_2}} \right)  $\\
$F^i_{I_4^{(5)}I_3^{(6)}}$&$I_{ijkl}^{(5)}(t)\int_0^\infty d\tau  I_{jkl}^{(6)}(t-\tau)\ln \left( \frac{\tau}{C_{I_3}} \right)  $\\
$F^i_{I_3^{(4)}I_4^{(7)}}$&$I_{jkl}^{(4)}(t)\int_0^\infty d\tau  I_{ijkl}^{(7)}(t-\tau)\ln \left( \frac{\tau}{C_{I_4}} \right)  $\\ 
${}^*F^i_{I_3^{(4)}J_3^{(6)}}$&$\epsilon_{ijk} I_{jab}^{(4)}(t)\int_0^\infty d\tau  J_{kab}^{(6)}(t-\tau)\ln \left( \frac{\tau}{C_{J_3}} \right)   $\\
${}^*F^i_{J_3^{(4)}I_3^{(6)}}$&$\epsilon_{ijk} J_{kab}^{(4)}(t)\int_0^\infty d\tau  I_{jab}^{(6)}(t-\tau)\ln \left( \frac{\tau}{C_{I_3}} \right)  $\\
$F^i_{J_3^{(4)}J_2^{(5)}}$&$J_{ijk}^{(4)}(t)\int_0^\infty d\tau  J_{jk}^{(5)}(t-\tau)\ln \left( \frac{\tau}{C_{J_2}} \right)   $\\
$F^i_{J_2^{(3)}J_3^{(6)}}$&$J_{jk}^{(3)}(t)\int_0^\infty d\tau  J_{ijk}^{(6)}(t-\tau)\ln \left( \frac{\tau}{C_{J_3}} \right)$\\  
\end{tabular}
\end{ruledtabular}
\end{table}

Introducing the shorthand notation
\beq
\label{aver}
\langle f \rangle = \int_{-\infty}^{\infty}dt\, f(t)\,,
\eeq
for the total time-integral of an arbitrary function $f(t)$ over the full scattering process,
we need to evaluate 
\beq
\label{Pitail}
P_i^{\rm rad\,tail}\equiv\langle {\mathcal F}_{P }^i{}^{\rm tail} \rangle\,.
\eeq
We found useful to evaluate this integral
in the frequency domain by using a quasi-Keplerian parametrization of the motion in harmonic coordinates.
We refer to previous works for a review of all necessary tools (see, e.g., Ref. \cite{Bini:2021jmj}).

Expanding the various multipole moments as Fourier integrals
\bea
X_L(t)& \equiv &\int_{-\infty}^\infty \frac{d\omega}{2\pi} e^{-i\omega t}\hat X_L(\omega)\,,\nonumber\\
{}\hat X_L(\omega)&\equiv&\int_{-\infty}^\infty dt e^{i\omega t}X_L(t)\,,
\eea
leads to (denoting $\int_\omega \equiv\int_0^\infty \frac{d\omega}{2\pi}$),
\bea
\label{PitailNpz1}
\langle F^i_{I_3^{(4)}I_2^{(5)}}+F^i_{I_2^{(3)}I_3^{(6)}} \rangle &=& 
\int_\omega
\omega^8 \left(i \pi {\mathcal S}_i^--\frac{7}{10}{\mathcal S}_i^+   \right)
\,,\qquad\nonumber\\
\langle {}^*F^i_{I_2^{(3)}J_2^{(5)}}+{}^*F^i_{J_2^{(3)}I_2^{(5)}}\rangle &=& 
\int_\omega
\omega^7 \left(\pi {\mathcal R}_i^+-\frac{i}{4}{\mathcal R}_i^- \right)
\,,\qquad\nonumber\\
\langle F^i_{I_4^{(5)}I_3^{(6)}}+F^i_{I_3^{(4)}I_4^{(7)}}\rangle &=& 
\int_\omega
\omega^{10} \left(i\pi {\mathcal U}_i^-- \frac{7}{20}{\mathcal U}_i^+ \right)
\,,\qquad\nonumber\\ 
\langle {}^*F^i_{I_3^{(4)}J_3^{(6)}}+{}^*F^i_{J_3^{(4)}I_3^{(6)}}\rangle &=& 
\int_\omega
\omega^{9} \left(\pi {\mathcal V}_i^+-\frac{i}{20}{\mathcal V}_i^-\right)
\,,\qquad\nonumber\\
\langle F^i_{J_3^{(4)}J_2^{(5)}}+ F^i_{J_2^{(3)}J_3^{(6)}}\rangle &=& 
\int_\omega
\omega^8 \left(i\pi {\mathcal Z}_i^--\frac{1}{2}{\mathcal Z}_i^+ \right)
\,,\qquad\nonumber\\
\eea
where
\bea
{\mathcal S}_i^\pm(\omega)&=&\hat I_{ijk}(-\omega)\hat I_{jk}(\omega)\pm \hat I_{ijk}(\omega)\hat I_{jk}(-\omega)
\,,\nonumber\\
{\mathcal R}_i^\pm(\omega)&=& \epsilon_{ijk}\left[\hat I_{ja}(\omega)\hat J_{ka}(-\omega)\pm \hat I_{ja}(-\omega)\hat J_{ka}(\omega)\right]
\,,\nonumber\\
{\mathcal U}_i^\pm(\omega)&=&  \hat I_{ijkl}(-\omega)\hat I_{jkl}(\omega)\pm \hat I_{ijkl}(\omega)\hat I_{jkl}(-\omega) 
\,,\nonumber\\
{\mathcal V}_i^\pm(\omega)&=&\epsilon_{ijk}\left[\hat I_{jab}(\omega)\hat J_{kab}(-\omega)\pm \hat I_{jab}(-\omega)\hat J_{kab}(\omega)\right]
\,,\nonumber\\
{\mathcal Z}_i^\pm(\omega)&=&\hat J_{ijk}(-\omega)\hat J_{jk}(\omega)\pm \hat J_{ijk}(\omega)\hat J_{jk}(-\omega)
\,.
\eea

The leading PN order tail contribution \eqref{Pitail} (i.e., the first two lines in Eqs. \eqref{PitailNpz1}) has been already computed in Ref. \cite{Bini:2021gat} (see also Ref. \cite{Bini:2022yrk}). 
We focus here on the next-to-leading order (fractionally 1PN) tail contribution.
We need to take into account the fractional 1PN corrections to the first two lines in Eqs. \eqref{PitailNpz1}, whereas the leading PN order is enough for the remaining three lines in Eqs. \eqref{PitailNpz1}.
The final results for the  large-$j$ expansions of the (nonvanishing) components $P_x^{\rm rad\,tail}$ and $P_y^{\rm rad\,tail}$ are
\begin{widetext}
\bea
\label{DeltaPxypasttails}
P_x^{\rm rad\,tail}&=& -(Mc)\frac{m_2-m_1}{M}\nu^2\eta^3\left[
 \pi\frac{\frac{1491}{400}p_\infty^7+\eta^2(-\frac{9529}{67200}\nu - \frac{26757}{5600} ) p_\infty^9}{j^4}  
+ \frac{\frac{20608}{225} p_\infty^6+\eta^2(\frac{72512}{7875}\nu - \frac{1143232}{7875}) p_\infty^8}{j^5} \right.\nonumber\\
&&
+ \pi\frac{\frac{267583}{2400} p_\infty^5+\eta^2(\frac{1711123}{57600}\nu - \frac{12566143}{67200}) p_\infty^7}{j^6} 
+ \frac{\frac{64576}{75} p_\infty^4+\eta^2( \frac{29244128}{70875} \nu - \frac{9920672}{7875}) p_\infty^6}{j^7}\nonumber\\
&&\left.
+O\left(\frac{1}{j^8}\right)
\right]
\,,\nonumber\\
P_y^{\rm rad\,tail}&=& -(Mc)\frac{m_2-m_1}{M}\nu^2\eta^3\left[
\frac{-\frac{128}{3} p_\infty^7
+\eta^2\left(\frac{320}{3} \nu - \frac{192}{175}\right) p_\infty^9}{j^4} 
+\pi\frac{-\frac{1509\pi^2}{140}p_\infty^6
+\eta^2\left(\frac{2721}{80}\pi^2\nu - \frac{2432}{15}  + \frac{75661}{4480}\pi^2\right) p_\infty^8}{j^5} \right.\nonumber\\
&&
+ \frac{\left(-\frac{8768}{45} - \frac{521216\pi^2}{4725}\right) p_\infty^5
+\eta^2\left[\left(\frac{43457024}{99225}\pi^2  + \frac{37792}{45}\right)\nu - \frac{77389}{1050} + \frac{42827264}{1091475}\pi^2  - \frac{9489}{20}\zeta(3)\right] p_\infty^7}{j^6} \nonumber\\
&& 
+ \pi\frac{\left(\frac{36885}{896}\pi^4 - \frac{142391}{280}\pi^2\right) p_\infty^4
+\eta^2\left[\left(-\frac{208525}{1024}\pi^4 + \frac{8537719}{3360}\pi^2\right)\nu - \frac{44900896}{55125} - \frac{328765}{1792}\pi^4   + \frac{989879573}{549120}\pi^2\right] p_\infty^6}{j^7}\nonumber\\
&&\left.
+O\left(\frac{1}{j^8}\right)
\right]
\,.
\eea
\end{widetext}
These tail contributions take into account the physical retarded-tail  interaction between the bodies, so that they are asymmetric under time-reversal (they were called \lq\lq past tails'' in Refs. \cite{Bini:2021gat,Bini:2021qvf}). 
Let us note in passing that  replacing the retarded-kernel in the time-domain tail integral by its
{\it time-symmetric} projection, would lead to the following integral:
\bea
\label{final_Pi_int_ts}
P_i^{\rm rad\,sym\, tail}&=&\frac{G^2{\mathcal M}}{c^{10}}\left\{
\frac{32}{45}\pi \int_{0}^\infty \frac{d\omega}{2\pi} \omega ^{7}{\mathcal R}_i^+(\omega)\right.\nonumber\\
&+&
\frac{4}{63}i\pi \int_{0}^\infty \frac{d\omega}{2\pi} \omega ^{8}{\mathcal S}_i^-(\omega)\nonumber\\
&+&
\frac1{c^2}\left[
\frac{1}{567}i\pi \int_{0}^\infty \frac{d\omega}{2\pi} \omega ^{10}{\mathcal U}_i^-(\omega)\right.\nonumber\\
&+&
\frac{1}{63}\pi \int_{0}^\infty \frac{d\omega}{2\pi} \omega ^{9}{\mathcal V}_i^+(\omega)\nonumber\\
&+&\left.\left.
\frac{8}{63}i\pi \int_{0}^\infty \frac{d\omega}{2\pi} \omega ^{8}{\mathcal Z}_i^-(\omega)
\right]
\right\}
\,,
\eea
implying
\beq
P_x^{\rm rad\,sym\, tail}=0\,,\qquad
P_y^{\rm rad\,sym\, tail}=P_y^{\rm rad\,tail}\,.
\eeq

The complete 2.5PN radiated linear momentum is then obtained by summing up all contributions, Eqs. \eqref{delta_rr_P_xy_rad_inst_IJ}, \eqref{25instP}, \eqref{DeltaPxypasttails}.
The final result is listed in Tables \ref{tab:EnJnPyn}--\ref{tab:Pxn} as a double PM-PN expansion (see Eq. \eqref{various_expansions} below).

\section{3PN-level contribution to the radiated linear momentum}

The radiated instantaneous linear momentum at the fractional 3PN accuracy can be obtained by integrating 
the  3PN instantaneous linear momentum flux,
\bea
{\mathcal F}_{P }{}_i{}^{{\rm inst}\,I,J \, \leq {\rm 3PN}}
&=& \frac{G}{c^7}\left(
f^0_i +\frac{1}{c^2}f^1_i +\frac{1}{c^4}f^2_i\right.\nonumber\\
&&\left.+\frac{1}{c^5}f^{2.5}_i+\frac{1}{c^6}f^3_i
\right)\,,
\eea
where
\bea
f^0_i&=&\frac{2}{63}I^{(4)}_{ijk}I^{(3)}_{jk} +\frac{16}{45}\epsilon_{ijk}I^{(3)}_{jc}J^{(3)}_{kc}\,,\nonumber\\
f^1_i&=&\frac{4}{63}J^{(4)}_{ijk}J^{(3)}_{jk}+\frac{1}{1134}I^{(5)}_{ijkl}I^{(4)}_{jkl}+\frac{1}{126}\epsilon_{ijk} I^{(4)}_{jab}J^{(4)}_{kab}\,,\nonumber\\
f^2_i&=&\frac{2}{945}J^{(5)}_{ijkl}J^{(4)}_{jkl}+\frac{1}{59400}I^{(6)}_{ijklm}I^{(5)}_{jklm}\nonumber\\
&+&\frac{2}{14175}\epsilon_{ijk}I^{(5)}_{jabc}J^{(5)}_{kabc}\,, \nonumber\\
f^3_i&=&\frac{1}{22275}J^{(6)}_{ijklm}I^{(5)}_{jklm}+\frac{1}{4343625}I^{(7)}_{ijklmn}I^{(6)}_{jklmn}\nonumber\\
&+&\frac{1}{534600}\epsilon_{ijk}I^{(6)}_{jabcd}J^{(6)}_{kabcd} \,,
\eea
with $f^0_i$ (namely $I_{ij}$, $I_{ijk}$ and $J_{ij}$) to be evaluated at the 3PN level of accuracy, $f^1_i$ at 2PN, etc. The 2.5PN contribution, $f^{2.5}_i$  has already been discussed in the previous sections.
 
Moreover, all multipoles are needed in modified harmonic coordinates and several of them already exist in the literature (mainly from Ref. \cite{Mishra:2015bqa}), while for the others only the expression in harmonic coordinates is known, and one has to transform their expression to modified harmonic coordinates, following Ref. \cite{Arun:2007sg}, Section IV.B. More precisely, 
\begin{enumerate}
\item $I_{ij}$, needed at 3PN, see Eqs. (3.1)-(3.2c) of Ref. \cite{Arun:2007sg}; see also  Eqs. (3.19)-(3.20) of  Ref. \cite{Mishra:2015bqa};
\item $I_{ijk}$, needed at 3PN, see Eqs. (4.9)-(4.10) of  Ref. \cite{Faye:2014fra} for the expression in standard harmonic coordinates; 
\item $I_{ijkl}$, needed at  2PN, see  Eq. (3.23a) of  Ref. \cite{Mishra:2015bqa};   
\item $I_{ijklm}$, needed at  1PN, see Eq. (3.23b) of  Ref. \cite{Mishra:2015bqa}; 
\item $I_{ijklmn}$, needed at  N, see Eq. (3.23c) of  Ref. \cite{Mishra:2015bqa};  
\item $J_{ij}$, needed at  3PN, see Eqs. (3.6)-(3.7) of  Ref. \cite{Henry:2021cek} for the expression in standard harmonic coordinates;
\item $J_{ijk}$, needed at  2PN, see Eq. (3.26a) of  Ref. \cite{Mishra:2015bqa};
\item $J_{ijkl}$, needed at  1PN, see Eq. (3.26b) of  Ref. \cite{Mishra:2015bqa};
\item $J_{ijklm}$, needed at  N, see Eq. (3.26c) of  Ref. \cite{Mishra:2015bqa}.
\end{enumerate}
The final 3PN instantaneous term for a generic orbit reads
\bea
{\mathcal F}_{P }{}_i{}^{{\rm inst}\,I,J \, {\rm 3PN}}&=&\frac{G^3 M^3 \nu^2}{r^4c^7} (m_2-m_1)\eta^6 (A^{\rm 3PN} \dot r n^i\nonumber\\
&& +B^{\rm 3PN} v^i)\,,
\eea
with
\begin{widetext}
\bea
A^{\rm 3PN}&=& \left(\frac{50647}{4095}-\frac{6891347}{45045} \nu+\frac{378098}{715} \nu^2-\frac{17700712}{45045} \nu^3\right) v^8\nonumber\\
&+&\left[\left(-\frac{1486192}{15015}+\frac{38072087}{45045} \nu-\frac{124611538}{45045} \nu^2+\frac{7420632}{5005} \nu^3\right)  \dot r^2\right.\nonumber\\
&+&\left. \left(-\frac{5742794}{35035}+\frac{2875777}{3465} \nu-\frac{10668793}{9009} \nu^2+\frac{3851017}{6435} \nu^3\right)\frac{GM}r\right]  v^6 \nonumber\\
&+&\left[\left(\frac{2039066}{5005}-\frac{95290066}{45045} \nu+\frac{18801898}{4095} \nu^2-\frac{9617408}{5005} \nu^3\right) \dot r^4\right.\nonumber\\
&+&\left(\frac{925151368}{945945}-\frac{682213787}{135135} \nu+\frac{865924949}{135135} \nu^2-\frac{95805097}{45045} \nu^3\right) \frac{GM\dot r^2}{r}\nonumber\\
&+&\left.
\left(\frac{173961024956}{165540375}-\frac{486464}{3675} \ln\left(\frac{r}{r_0}\right)+\left(-\frac{1845}{28} \pi^2-\frac{312503921}{675675}\right) \nu+\frac{33465314}{27027} \nu^2-\frac{16820996}{45045} \nu^3
\right)\frac{G^2M^2}{r^2}\right] v^4\nonumber\\
&+&\left[ \left(-\frac{683368}{1287}+\frac{20252888}{9009} \nu-\frac{27439754}{9009} \nu^2+\frac{9326368}{9009} \nu^3\right) \dot r^6\right.\nonumber\\
&+& \left(-\frac{836106314}{525525}+\frac{88323799}{10010} \nu-\frac{21042751}{2310} \nu^2+\frac{103758968}{45045} \nu^3\right)\frac{GM\dot r^4}{r}   \nonumber\\
&+&\left(-\frac{742542259516}{165540375}+\frac{1647104}{3675} \ln\left(\frac{r}{r_0}\right)+\left(\frac{66462078}{25025}+\frac{3075}{14} \pi^2\right) \nu-\frac{50439274}{10395} \nu^2+\frac{13610134}{12285} \nu^3\right) \frac{G^2M^2 \dot r^2}{r^2}
\nonumber\\
&+&\left. \left(-\frac{69490090246}{55180125}+\frac{60992}{3675} \ln\left(\frac{r}{r_0}\right)+\left(\frac{56867}{840} \pi^2-\frac{1698895}{2457}\right) \nu-\frac{993904}{2457} \nu^2+\frac{891622}{7371} \nu^3\right)\frac{G^3M^3}{r^3}\right] v^2
\nonumber\\
&+&\left(\frac{58349}{273}-\frac{2486416}{3003} \nu+\frac{2067616}{3003} \nu^2-\frac{195680}{1001} \nu^3\right) \dot r^8\nonumber\\
&+&\left(\frac{1228477436}{1576575}-\frac{408167713}{90090} \nu+\frac{69532495}{18018} \nu^2-\frac{1022414}{1287} \nu^3\right) \frac{GM\dot r^6}{r}\nonumber\\
&+&\left(\frac{111525644752}{33108075}-\frac{42496}{147} \ln\left(\frac{r}{r_0}\right)+\left(-\frac{615}{4} \pi^2-\frac{67591807}{27027}\right) \nu+\frac{153157904}{45045} \nu^2-\frac{4165558}{6435} \nu^3\right) \frac{G^2M^2\dot r^4}{r^2}\nonumber\\
&+&\left(\frac{4786348054}{5016375}+\frac{174592}{3675} \ln\left(\frac{r}{r_0}\right)+\left(-\frac{20869}{280} \pi^2+\frac{1498465697}{4729725}\right) \nu+\frac{15571532}{27027} \nu^2-\frac{9925382}{135135} \nu^3\right)\frac{G^3M^3\dot r^2}{r^3}\nonumber\\
&+&\left(-\frac{21844644124}{496621125}+\frac{11904}{1225} \ln\left(\frac{r}{r_0}\right)+\left(-\frac{41}{70} \pi^2-\frac{2315974202}{4729725}\right) \nu+\frac{179768}{3003} \nu^2-\frac{40396}{6237} \nu^3\right)\frac{G^4M^4}{r^4}
\eea
and
\bea
B^{\rm 3PN}&=& \left(-\frac{438226}{15015}+\frac{40657}{231} \nu-\frac{1010414}{3003} \nu^2+\frac{3600536}{15015} \nu^3\right) v^8\nonumber\\
&+&\left[\left(\frac{9523744}{45045}-\frac{395467}{315} \nu+\frac{7045306}{3465} \nu^2-\frac{3058024}{3465} \nu^3\right) \dot r^2\right.\nonumber\\
&+&\left. \left(\frac{143914678}{1576575}-\frac{3249853}{6435} \nu+\frac{1744231}{3003} \nu^2-\frac{2536901}{9009} \nu^3\right)\frac{GM}{r}\right]v^6\nonumber\\
&+&\left[\left(-\frac{1756652}{3003}+\frac{5370304}{5005} \nu^3-\frac{159792518}{45045} \nu^2+\frac{19699546}{6435} \nu\right) \dot r^4\right. \nonumber\\
&+&\left(-\frac{230962616}{315315}+\frac{2612939}{715} \nu-\frac{24270509}{6435} \nu^2+\frac{44195497}{45045} \nu^3\right) \frac{GM\dot r^2}{r}\nonumber\\
&+&\left. \left(-\frac{429821166328}{496621125}+\frac{337504}{3675} \ln\left(\frac{r}{r_0}\right)+\left(\frac{6519}{280} \pi^2+\frac{80267816}{405405}\right) \nu-\frac{1917296}{4095} \nu^2+\frac{43333016}{405405} \nu^3\right)\frac{G^2M^2}{r^2}\right] v^4\nonumber\\
&+&\left[\left(\frac{3974752}{6435}-\frac{27078616}{9009} \nu+\frac{1667210}{693} \nu^2-\frac{2625664}{5005} \nu^3\right) \dot r^6\right. \nonumber\\
&+& \left(\frac{46881654}{35035}-\frac{124322753}{18018} \nu+\frac{115944133}{20790} \nu^2-\frac{6226289}{6435} \nu^3\right)\frac{GM \dot r^4}{r}  \nonumber\\
&+&\left(\frac{205470694976}{55180125}-\frac{1286752}{3675} \ln\left(\frac{r}{r_0}\right)+\left(-\frac{47299991}{45045}-\frac{3813}{35} \pi^2\right) \nu+\frac{8684332}{3861} \nu^2-\frac{7223842}{19305} \nu^3\right)\frac{G^2M^2\dot r^2}{r^2}\nonumber\\
&+&\left. \left(\frac{25042228006}{70945875}+\frac{704}{175} \ln\left(\frac{r}{r_0}\right)+\left(-\frac{21607}{840} \pi^2+\frac{783374999}{1289925}\right) \nu+\frac{411716}{10395} \nu^2-\frac{8327414}{405405} \nu^3\right)\frac{G^3M^3}{r^3}
\right] v^2 \nonumber\\ 
&+&\left(-\frac{1974958}{9009}+\frac{440666}{429} \nu-\frac{4954028}{9009} \nu^2+\frac{767248}{9009} \nu^3\right) \dot r^8\nonumber\\
&+&\left(-\frac{219642736}{315315}+\frac{47166701}{12870} \nu-\frac{3852731}{1638} \nu^2+\frac{629897}{2145} \nu^3\right) \frac{GM\dot r^6}{r}\nonumber\\ 
&+&\left(-\frac{875401568}{315315}+\frac{1600}{7} \ln\left(\frac{r}{r_0}\right)+\left(\frac{24149}{280} \pi^2+\frac{58379561}{45045}\right) \nu-\frac{31792976}{19305} \nu^2+\frac{23299642}{135135} \nu^3\right)\frac{G^2M^2\dot r^4}{r^2}\nonumber\\
&+&\left(-\frac{1458606746}{165540375}-\frac{716608}{11025} \ln\left(\frac{r}{r_0}\right)+\left(\frac{9389}{280} \pi^2-\frac{5177354543}{14189175}\right) \nu-\frac{34654208}{135135} \nu^2+\frac{596782}{81081} \nu^3\right)\frac{G^3M^3\dot r^2}{r^3}\nonumber\\
&+&\left(\frac{5316518908}{45147375}-\frac{44032}{11025} \ln\left(\frac{r}{r_0}\right)+\left(-\frac{41}{14} \pi^2+\frac{5234646748}{14189175}\right)\nu-\frac{140216}{12285} \nu^2+\frac{1524088}{405405} \nu^3\right)\frac{G^4M^4}{r^4}\,.
\eea

The integration along hyperboliclike orbits (see Appendix \ref{3PN_orb_par}) can be carried on exactly and the sought for 3PN contribution reads
\bea
\label{PxPy3PNinst}
P_x^{{\rm rad\,inst},\, I,J\,{\rm 3PN}}&=&0
\,,\nonumber\\
P_y^{{\rm rad\,inst},\, I,J\,{\rm 3PN}}&=&(Mc)\frac{m_2-m_1}{M}\nu^2\eta^6\frac{1}{e_r[\bar a_r(e_r^2-1)]^7}\left(
Q_y^{A^0} + Q_y^{A^1}A+Q_y^{A^2}A^2+Q_y^{A^3}A^3\right)\,,
\eea
with $A={\rm arccos}(-1/e_r)$ and
\bea
Q_x^{A^3}&=&
e_r^8 \left(-\frac{37 \nu }{20}-\frac{375}{112}\right)+e_r^6 \left(\frac{72427}{420}-\frac{7 \nu }{2}\right)+e_r^4 \left(\frac{1104 \nu}{5}+482\right)+e_r^2 \left(\frac{804 \nu }{5}+\frac{8798}{105}\right)+\frac{32}{5}
\,,\nonumber\\
Q_y^{A^2}&=& 
\sqrt{e_r^2-1} \left[e_r^6 \left(\frac{311517}{2800}-\frac{849 \nu }{20}\right)+e_r^4 \left(\frac{1831 \nu }{5}+\frac{6287443}{4200}\right)+e_r^2\left(\frac{3861 \nu }{5}+\frac{61543}{105}\right)+\frac{164 \nu }{5}+\frac{15473}{525}\right]
\,,\nonumber\\
Q_y^{A^1}&=& 
e_r^2 \left(-\frac{41053 e_r^8}{1225}-\frac{3180022 e_r^6}{2205}-\frac{4558096 e_r^4}{735}-\frac{19581152 e_r^2}{3675}-\frac{9002752}{11025}\right)\ln\left(\frac{2\bar a_r(e_r^2-1)}{e_r r_0}\right)\nonumber\\
&&
+e_r^{12} \left(-\frac{37 \nu ^3}{48}+\frac{2447 \nu ^2}{448}-\frac{349061 \nu }{20160}+\frac{26726213}{591360}\right)\nonumber\\
&&
+e_r^{10} \left(-\frac{351 \nu^3}{40}-\frac{4891 \nu ^2}{630}+\left(\frac{5171197}{22400}-\frac{9143 \pi ^2}{1920}\right) \nu -\frac{5601182987}{32928000}\right)\nonumber\\
&&
+e_r^8\left(\frac{46 \nu ^3}{15}-\frac{230705 \nu ^2}{252}+\left(\frac{691093489}{453600}-\frac{3284551 \pi ^2}{15360}\right) \nu
   +\frac{253778681389}{17385984}\right)\nonumber\\
&&
+e_r^6 \left(\frac{47 \nu ^3}{5}-\frac{407258 \nu ^2}{105}+\left(-\frac{7006220329}{226800}-\frac{644561 \pi^2}{960}\right) \nu +\frac{15212690520617}{244490400}\right)\nonumber\\
&&
+e_r^4 \left(\frac{8 \nu ^3}{5}-\frac{1305494 \nu ^2}{315}+\left(-\frac{772592833}{14175}-\frac{54899 \pi ^2}{192}\right) \nu +\frac{326345642761}{7546000}\right)\nonumber\\
&&
+e_r^2 \left(-\frac{7040 \nu^2}{9}+\left(\frac{2419 \pi ^2}{60}-\frac{127239209}{9450}\right) \nu +\frac{253937658533}{50935500}\right)\nonumber\\
&&
-\frac{208 \nu ^2}{15}+\left(-\frac{9398}{175}-\frac{82 \pi ^2}{15}\right) \nu -\frac{183451039}{1984500}
\,,\nonumber\\
Q_y^{A^0}&=& 
e_r^2 \left(-\frac{41053 e_r^8}{1225}-\frac{3180022 e_r^6}{2205}-\frac{4558096 e_r^4}{735}-\frac{19581152 e_r^2}{3675}-\frac{9002752}{11025}\right){\rm Cl}_2(2A)\nonumber\\
&&
+\sqrt{e_r^2-1}\left[
\left(-\frac{79892213 e_r^8}{165375}-\frac{153055244 e_r^6}{33075}-\frac{42259956 e_r^4}{6125}-\frac{296405824 e_r^2}{165375}-\frac{3215456}{165375}\right)\ln\left(\frac{\bar a_r e_r}{2r_0}\right)\right.\nonumber\\
&&
+e_r^{10} \left(-\frac{283 \nu ^3}{48}+\frac{292361 \nu ^2}{11200}-\frac{78936947 \nu }{2116800}+\frac{679015961}{186278400}\right)\nonumber\\
&&+e_r^8 \left(-\frac{214 \nu ^3}{45}-\frac{2544931 \nu ^2}{7200}+\left(\frac{364880588983}{190512000}-\frac{1027583 \pi ^2}{13440}\right) \nu +\frac{4811241578461}{1629936000}\right)\nonumber\\
&&
+e^6 \left(\frac{2143 \nu ^3}{180}-\frac{103970549 \nu^2}{37800}+\left(-\frac{2926723837}{235200}-\frac{935846033 \pi ^2}{1612800}\right) \nu +\frac{189910151942603}{4656960000}\right)\nonumber\\
&&
+e^4\left(\frac{143 \nu ^3}{45}-\frac{18824609 \nu ^2}{3780}+\left(-\frac{478614974947}{7938000}-\frac{417611363 \pi ^2}{806400}\right) \nu +\frac{2221845272842369}{48898080000}\right)\nonumber\\
&&
+e^2 \left(\frac{4 \nu ^3}{45}-\frac{7722154 \nu ^2}{4725}+\left(\frac{2537449 \pi^2}{67200}-\frac{154305135683}{5953500}\right) \nu +\frac{6582692584319}{814968000}\right)\nonumber\\
&&\left.
-\frac{70376 \nu ^2}{1575}-\frac{395368697 \nu}{992250}-\frac{143459 \pi ^2 \nu }{33600}-\frac{369018184091}{6112260000}
\right]
\,,
\eea
where
\beq
{\rm Cl}_2(x)=\frac{i}{2}\left[{\rm Li}_2(e^{-ix})-{\rm Li}_2(e^{ix})\right]\,,
\eeq
is the Clausen function of order 2.

As expected, these terms involve the arbitrary length scale $r_0$ (entering the retarded time as well as the relation connecting harmonic to modified harmonic coordinates), which disappears in the complete expression when all 3PN hereditary terms are included, i.e.,
\bea
P_i^{\rm rad\, 3PN}&=&P_i^{{\rm rad \, inst}\, I,J\, {\rm 3PN}}
+P_i^{\rm rad\, higher-order \, tails}
\,.
\eea
Indeed, this is exactly the case when using the results of Ref. \cite{Bini:2022yrk} for the higher-order tail contributions.
We list below the final  large-$j$ expansion (including terms from $1/j^3$ up to terms $1/j^7$) of both $P_x^{\rm rad}$ and $P_y^{\rm rad}$

\bea
\label{PxPy3PNall}
P_x^{\rm rad\, 3PN}&=& -(Mc)\frac{m_2-m_1}{M}\nu^2\eta^6 \left[  \frac{196096}{945}\frac{p_\infty^9}{j^5} 
+ \frac{20719\pi^3}{320}\frac{p_\infty^8}{j^6}
+\left(\frac{ 42739712\pi^2 }{55125} + \frac{9226496}{4725}\right)\frac{p_\infty^7}{j^7}
+O\left(\frac{p_\infty^6}{j^8}\right)\right]
\,,\nonumber\\
P_y^{\rm rad\, 3PN}&=&(Mc)\frac{m_2-m_1}{M}\nu^2\eta^6 \left\{
\left(-\frac{1531643}{1182720}  - \frac{27581}{10080}\nu  - \frac{197}{560}\nu^2- \frac{74}{15}\nu^3\right) \pi\frac{p_\infty^{11}}{j^3}\right.\nonumber\\
&+& 
\left(- \frac{1218176}{72765}-\frac{118676}{6615}\nu  - \frac{76}{15}\nu^2 - 140\nu^3 \right)\frac{p_\infty^{10}}{j^4} \nonumber\\
&+& 
\left[  \frac{37806320227}{790272000}+\frac{503\pi^2}{70} - \frac{41053}{2450}\ln\left(\frac{p_\infty}{2}\right) 
 +\left(\frac{945563}{8064}-\frac{4059 \pi^2}{1280}\right)\nu   + \frac{17617\nu^2}{840} - \frac{6199\nu^3}{30}\right]\pi \frac{p_\infty^9}{j^5} \nonumber\\
&+& 
\left[ \frac{393851925056}{191008125}+ \frac{1042432\pi^2}{4725} - \frac{85434368}{165375}\ln(2 p_\infty)
+\left(\frac{815056834}{297675} - \frac{8528\pi^2 }{105}\right)\nu\right.\nonumber\\ 
&+&\left. 
 \frac{174074}{225}\nu^2  - \frac{30422\nu^3}{15} 
 \right]\frac{p_\infty^8}{j^6} \nonumber\\
&+& 
\left[\frac{1006741665001549}{312947712000} + \frac{907691\pi^2}{2688} - \frac{303491\zeta(3)}{224}
-\frac{35125513}{44100}\ln\left(\frac{p_\infty}{2}\right) 
+\left( \frac{2124695071}{725760}- \frac{3017083\pi^2}{30720} \right)\nu \right.\nonumber\\
&+& \left.\left.
\frac{1209467\nu^2}{960} - \frac{30181\nu^3}{20}\right]\pi\frac{p_\infty^7}{j^7}
+O\left(\frac{p_\infty^6}{j^8}\right)\right\}\,.
\eea 

\end{widetext}

\section{Summary of results for the energy, angular momentum and linear momentum losses in the c.m. frame}

For the convenience of the reader, let us summarize here the new results derived in this work concerning
the losses of energy, angular momentum, and linear momentum (radiated as gravitational waves), 
as recorded in the (initial) c.m. frame. In this section we use the notation of our previous work \cite{Bini:2021gat}
for parametrizing the PM expansions of the radiative losses by the coefficients of their
 power expansion in $\frac1j$, namely
\bea
\label{various_expansions}
\frac{E^{\rm rad}}{M} &=&  + \nu^2   \sum_{n=3}^\infty \frac{ E_{n}}{j^n} \,, \nonumber\\
\frac{ J^{\rm rad}}{J_{\rm c.m.}} &=& + \nu^1  \sum_{n=2}^\infty \frac{{ J}_{n}}{j^n} \,, \nonumber\\
\frac{P_x^{\rm rad}}{M}&=& + \frac{m_2-m_1}{M} \nu^2 \sum_{n=4}^\infty \frac{P_{x n}}{j^n}\,, \nonumber\\
\frac{P_y^{\rm rad}}{M}&=& + \frac{m_2-m_1}{M} \nu^2 \sum_{n=3}^\infty \frac{P_{y n}}{j^n}\,.
\eea
Here the left-hand sides have been adimensionalized, and we pulled out some powers of $\nu$ on the 
right-hand sides, to ensure that the expansion coefficients $E_n$, $J_{n}$,  $P_{x n}$,   $P_{y n}$ are
dimensionless, and that their LO PN contribution is $\nu$-independent. [We recall that $J_{\rm c.m.} = b \mu \pinf/h=GM^2\nu j$.] Note that in Ref. \cite{Bini:2021gat} we focussed on the PM expansion of $P_y^{\rm rad}$, because $P_x^{\rm rad}$
was subdominant, and linked to time-asymmetric hereditary tail effects. See Eq. (H3) there, giving the LO contribution
to $P_x^{\rm rad}$.

\subsection{Energy loss in the c.m. frame}

The radiated c.m. energy $E^{\rm rad}$ has been evaluated at the 2PN fractional accuracy in our previous work 
Ref. \cite{Bini:2021gat}. The corresponding $\frac1j$-expansion PM coefficients were given (up to $\frac1{j^7}$)
 in the first five lines of Table IX there. In the present work, we have computed
 the heretofore unevaluated fractional 2.5PN instantaneous contribution due the radiation-reaction correction to hyperbolic motion (incorrectly argued to vanish in \cite{Cho:2021onr}),
 and we have used the results of  \cite{Bini:2021qvf,Cho:2021onr,Cho:2022pqy} when computing the fractional 3PN contribution
 in the form of a $\frac1j$-expansion (see Eqs. \eqref{Erad2p5pn} and \eqref{Erad3pn}). In order to confirm the value
 of the fractional 3PN contribution to the radiated energy, we have done an independent computation of the instantaneous,
 3PN-level contribution. The technically most challenging part of the latter computation comes
  from inserting the 3PN-accurate hyperbolic motion in the 3PN-accurate quadrupole moment. Following Ref. \cite{Cho:2018upo},
  the computation uses a 3PN-level, hyperbolic version of the quasi-Keplerian representation of binary motion. 
  In redoing the  computation of the latter hyperbolic quasi-Keplerian representation, we found that there were several
  typos in the results displayed in Ref. \cite{Cho:2018upo}. For the convenience of the reader, we give the corresponding
  corrected results in Appendix \ref{3PN_orb_par}.
  
 Our results are displayed in Table \ref{tab:EnJnPyn}.
 Many of the $\nu$-dependent terms can be directly
 checked by using the polynomiality rule satisfied by the coefficients $E_n$, namely
 \beq \label{ruleEn}
 h^{n+1}E_n=P^{\gamma}_{[(n-2)/2]}(\nu)\,,
 \eeq 
 where $P^{\gamma}_{N}(\nu)$ denotes a polynomial of order $N$ in $\nu$, having $\g$-dependent coefficients.
 This rule was pointed out in  Ref. \cite{Bini:2020hmy} (see also Eq. (7.7) in Ref. \cite{Bini:2021gat}). We shall give
 below another simple proof of this polynomiality rule.
 Our results on the coefficients $E_{n}$ satisfy this polynomiality rule after adding all separate contributions. 
 For instance, at the 4PM order ($n=4$), if one would consider separately the 3PN contribution ($\frac1{j^4}$ term on
 the second line of Eq. \eqref{Erad3pn}) it would violate the polynomiality rule \eqref{ruleEn} because of the terms
 $(-\frac{2366 \nu ^3}{9}+\frac{164 \nu ^2}{3})$. In fact, these terms precisely cancel the rule-violating terms in
 $h^5 E_4$ coming from lower PN contributions in $E_4$.

While writing up our results,  a PN-exact computation 
of  the $G^4$ energy coefficient $E_4$  was made public \cite{Dlapa:2022lmu}. 
Our (fractionally 3PN accurate) PN-expanded result listed in Eq. (D27) of Ref. \cite{Bini:2021gat},
and Table \ref{tab:EnJnPyn} here agrees (when expressed in terms of $\tilde E_4 \equiv h^5 E_4$)
with the 3PN expansion of the curly bracket on the right-hand side of Eq. (8) in Ref. \cite{Dlapa:2022lmu}.

 Let us also note that we have included in Table II the PN-acquired knowledge of the 3PM-level contribution $E_3$, though
 $E_3$ has been determined as an exact function of $\pinf$ \cite{Bern:2021dqo,Herrmann:2021tct}. It agrees with the
 corresponding term in Refs. \cite{Bern:2021dqo,Herrmann:2021tct}, and thereby provides an additional check of our PN calculations.


\begin{table*}[h]  
\caption{\label{tab:EnJnPyn} 
New terms at the 2.5PN and 3PN level of fractional accuracy improving the PN expansion given in Table IX of Ref. \cite{Bini:2021gat} of the coefficients $E_n$, $J_n$, and $P_{yn}$, entering the PM expansion \eqref{various_expansions} of the radiated energy, angular momentum, and $y$-component of the linear momentum, respectively.
}
\begin{ruledtabular}
\begin{tabular}{ll}
$E_3^{>\rm 2PN}$ &  $ \pi\left[(-\frac{676273}{354816} -\frac{2699}{504}\nu+\frac{321}{280}\nu^2-\frac{148}{15}\nu^3) p_\infty^{10}
+O(p_\infty^{11})\right]$\\
$E_4^{>\rm 2PN}$ &$ (\frac{1216}{105}-\frac{2848}{15} \nu) p_\infty^8+(-\frac{151854}{13475}-\frac{1223594}{33075}\nu+\frac{164}{3}\nu^2-\frac{2366}{9}\nu^3) p_\infty^9 
+O(p_\infty^{10})$\\
$E_5^{>\rm 2PN}$ &$\pi\left\{\left[(\frac{296}{25}-\frac{15291 \pi ^2}{280}) \nu -\frac{24993 \pi ^2}{1120}+\frac{9216}{35}\right] p_\infty^7\right.$\\
&$\left.+\left[ \frac{29573617463}{310464000}+\frac{99}{10}\pi^2-\frac{10593}{350}\ln(\frac{p_\infty}{2}) +(\frac{76897}{480}-\frac{4059}{640}\pi^2)\nu  +\frac{12269}{80}  \nu^2-\frac{1823}{5}  \nu^3\right] p_\infty^8
+O(p_\infty^9)\right\} $\\
$E_6^{>\rm 2PN}$ &$\left[\left(-\frac{2974508}{4725} \pi^2-\frac{71488}{75}\right) \nu+\frac{56708}{105}+\frac{1024}{135} \pi^2+\frac{2898}{5}\zeta(3)\right] p_\infty^6$\\
&$ +\left[
-\frac{18955264}{23625}\ln(2p_\infty)+\frac{36589282372}{11694375}+\frac{177152}{675}\pi^2+ (-\frac{212216}{1575}\pi^2+\frac{875976284}{297675})\nu+\frac{4201976}{1575}\nu^2 -\frac{150892}{45} \nu^3\right] p_\infty^7 
+O(p_\infty^8)$\\
$E_7^{>\rm 2PN}$ &$\pi\left\{\left[(\frac{56008}{135}-\frac{23514 \pi^2}{7}+\frac{30285 \pi ^4}{112}) \nu +\frac{689985 \pi
   ^4}{3584}-\frac{13138915 \pi^2}{7392}+\frac{210176}{225}  \right] p_\infty^5 \right.$\\
&$\left.+\left[  \frac{3158}{9}\pi^2-\frac{337906}{315}\ln(\frac{p_\infty}{2})+\frac{37546579757}{8467200}-\frac{58957}{32}\zeta(3) + (-\frac{51947}{384}\pi^2+\frac{68898691}{36288})\nu+\frac{1419153}{448}\nu^2 -\frac{13955}{6}  \nu^3\right]  p_\infty^6
+O(p_\infty^7)\right\} $\\
\hline
$J_{2}^{>\rm 2PN}$ &$ \left(\frac{3712}{3465}+\frac{878}{315} \nu+\frac{24 }{7}\nu^2-\frac{16 }{5}\nu ^3 \right) p_\infty^9
+O(p_\infty^{10})$\\
$J_{3}^{>\rm 2PN}$& $ \pi \left[\left(\frac{115769}{126720}-\frac{553}{24} \nu ^3+\frac{9235 }{672}\nu
   ^2+\frac{1469}{504} \nu \right) p_\infty^8
+O(p_\infty^9)\right]$\\
$J_{4}^{>\rm 2PN}$ &$\left(-\frac{431936}{1575}\nu+\frac{1184}{21} \right)p_\infty^6
+\left(
 -\frac{4955072}{121275}+\frac{1459694}{11025}\nu 
+\frac{67432}{315} \nu ^2
 -\frac{6224 }{15}\nu
   ^3\right) p_\infty^7
+O(p_\infty^8)$\\
$J_{5}^{>\rm 2PN}$ &$\pi  \left\{\left[(\frac{7816}{525}-\frac{2232 \pi
   ^2 }{35})\nu-\frac{1305 \pi ^2
   }{112}+\frac{7488 }{25}\right]p_\infty^5\right. $\\
&$\left. +\left[ 
-\frac{4922}{175}\ln(\frac{p_\infty}{2})+\frac{46}{5}\pi^2-\frac{561803611}{10584000}
+(-\frac{123}{32}\pi^2+\frac{2048629}{7560})\nu 
+\frac{74693 }{280}\nu ^2
 -\frac{861 }{2}\nu
   ^3\right]p_\infty^6
+O(p_\infty^7)\right\}$\\
$J_{6}^{>\rm 2PN}$ &$\left[\left(-\frac{201724}{33075}\pi^2-\frac{225536}{525}\right)\nu+\frac{4116
   \zeta (3)}{5}+\frac{147064}{315}-\frac{130688 \pi
   ^2}{6615}\right]p_\infty^4$\\
&$+\left[ \frac{8704}{45}\pi^2+\frac{7781823776}{16372125}-\frac{931328}{1575}\ln(2p_\infty)
+(-\frac{3362}{75}\pi^2+\frac{85939786}{42525})\nu+\frac{13320808}{4725}\nu^2 -\frac{136976}{45} \nu ^3\right]p_\infty^5
+O(p_\infty^6)$\\
$J_{7}^{>\rm 2PN}$ &$\pi  \left\{
   \left[  \left(\frac{365392}{1575}-\frac{57037 \pi ^2
   }{21}+\frac{102619 \pi
   ^4 }{448}\right)\nu+\frac{163083 \pi ^4
  }{1792} -\frac{18227 \pi ^2
   }{28}+\frac{32 }{15}\right]p_\infty^3\right. $\\
&$\left.+\left[ -\frac{21614}{35}\ln(\frac{p_\infty}{2})-\frac{45261}{40}\zeta(3)+\frac{5288341351}{4233600}+202\pi^2 + (\frac{46277}{432}-\frac{861}{64}\pi^2)\nu+\frac{794749}{336}\nu^2-\frac{6517 }{4}\nu
   ^3\right] p_\infty^4
+O(p_\infty^5)\right\}$\\
\hline
$P_{y3}^{>\rm 2PN}$ & $\pi \left(-\frac{1531643}{1182720}-\frac{27581}{10080}\nu-\frac{197}{560}\nu^2-\frac{74}{15}\nu^3\right)p_\infty^{11}+O(p_\infty^{12})$\\
$P_{y4}^{>\rm 2PN}$ &$\left(\frac{192}{175}-\frac{320}{3}\nu\right)p_\infty^9+\left(-140\nu^3-\frac{118676}{6615}\nu-\frac{76}{15}\nu^2-\frac{1218176}{72765} 
\right) p_\infty^{10}
+O(p_\infty^{11})$\\
$P_{y5}^{>\rm 2PN}$ &$\pi\left\{\left[\frac{2432}{15}-\frac{75661}{4480}\pi^2+(-\frac{2721}{80}\pi^2+\frac{148}{25})\nu\right] p_\infty^8\right.$\\
&$+\left[
-\frac{41053}{2450}\ln\left(\frac{p_\infty}{2}\right)+\frac{503}{70}\pi^2+\frac{37806320227}{790272000}
+(\frac{945563}{8064}-\frac{4059}{1280}\pi^2)\nu+\frac{17617}{840}\nu^2-\frac{6199}{30}\nu^3\right] p_\infty^9
+O(p_\infty^{10})\left.\right\}$\\
$P_{y6}^{>\rm 2PN}$ &$\left[\frac{9489}{20}\zeta(3)+\frac{77389}{1050}-\frac{42827264}{1091475}\pi^2+(-\frac{172734857}{396900}\pi^2-\frac{31648}{45})\nu\right] p_\infty^7  $\\
&$+\left(
-\frac{85434368}{165375}\ln(2p_\infty)+\frac{1042432}{4725}\pi^2+\frac{393851925056}{191008125}+\left( -\frac{8528}{105}\pi^2+\frac{815056834}{297675}\right)\nu
+\frac{174074}{225}\nu^2-\frac{30422}{15}\nu^3\right) p_\infty^8
+O(p_\infty^9)$\\
$P_{y7}^{>\rm 2PN}$ & $\pi\left\{\left[(\frac{208525}{1024}\pi^4-\frac{8537719}{3360}\pi^2+\frac{160406}{675})\nu+\frac{328765}{1792}\pi^4-\frac{989879573}{549120}\pi^2
+\frac{44900896}{55125}\right] p_\infty^6\right.$\\
&$
+\left(-\frac{35125513}{44100}\ln\left(\frac{p_\infty}{2}\right)-\frac{303491}{224}\zeta(3)+\frac{907691}{2688}\pi^2+\frac{1006741665001549}{312947712000}+\left( -\frac{3017083}{30720}\pi^2+\frac{2124695071}{725760}\right)\nu
+\frac{1209467}{960}\nu^2-\frac{30181}{20}\nu^3\right)  p_\infty^7$\\
&$\left.+O(p_\infty^8)\right\}$\\
\end{tabular}
\end{ruledtabular}
\end{table*}

\subsection{Angular momentum loss in the c.m. frame}

The fractionally 2PN-accurate expansion of the PM coefficients $J_n$ of the radiated c.m. angular momentum $J^{\rm rad}$ can also be found in Table IX of Ref. \cite{Bini:2021gat}, up to $n=7$.
In the present work we have raised their accuracy to the 3PN order, by computing the missing term in the instantaneous part of the radiated angular momentum at the 2.5PN level due the radiation-reaction correction to hyperbolic motion, thereby completing partial results available in the literature for the various contributions through the 3PN order \cite{Bini:2021qvf,Cho:2021onr,Cho:2022pqy,Bini:2022xpp}.
The final result is given by Eqs. \eqref{Jrad2p5pn} and \eqref{Jrad3pn} as an expansion in inverse angular momentum.  
The post-2PN coefficients are listed in Table \ref{tab:EnJnPyn}.
The 2PM and 3PM coefficients $J_2$ and $J_3$ are known exactly (see Refs. \cite{Damour:2020tta} and \cite{Manohar:2022dea}, respectively), but are also shown in their PN expanded form for completeness.

Concerning the $\nu$-structure of the coefficients $J_n$, they satisfy the polynomiality rule \cite{Bini:2021gat}
\beq \label{ruleJn} 
h^nJ_n+h^{n-1}\nu E_n=P^{\gamma}_{[(n-2)/2]}(\nu)\,, 
\eeq 
with $n\geq3$, whereas $h^2J_2$ is independent of $\nu$.

\subsection{Linear momentum loss in the c.m. frame}

Table IX of Ref. \cite{Bini:2021gat}  listed the PN expansion of the coefficients $P_{yn}$ of the PM expansion of the $y $-component of the radiated linear momentum $J^{\rm rad}$ in the c.m. frame, accurate to the 2PN fractional order.
The corresponding post-2PN contributions up to the 3PN order are listed in Table \ref{tab:EnJnPyn}.

As pointed out in \cite{Bini:2021gat} (and as is further discussed below)  the coefficients $P_{yn}$
must satisfy the polynomiality property 
\beq \label{rulePyn} 
h^{n+1}P_{y n}=P^{\gamma}_{[(n-3)/2]}(\nu)\,.
\eeq
Our results on the coefficients $P_{yn}$ satisfy this polynomiality rule after adding all separate contributions.
For instance, at order $n=4$ the term proportional to $+ \eta^2 \frac{320}{3}\frac{\pinf^9}{j^4}$ in the fractionally 1PN
tail term \eqref{DeltaPxypasttails} would separately violate the rule \eqref{rulePyn}, but is needed to cancell corresponding
rule-violating terms in $h^5 P_{y4}$.

We recall that $P_{y 3}$ is exactly known in PM sense, being related to $E_3$ by
\beq 
P_{y 3}=\sqrt{\frac{\gamma-1}{\gamma+1}}E_3\,.
\eeq

The PN expansion of the coefficients $P_{xn}$ are instead listed in Table \ref{tab:Pxn}. 
These expansions include the leading-order (past-tail) contribution computed in
\cite{Bini:2021gat}, and complete them by two further terms in the PN expansion
(fractionally 2.5PN and 3PN).
 
The coefficients $P_{xn}$ satisfy (see below) the polynomiality property 
\beq 
\label{rulePxn} 
h^nP_{x n}=P^{\gamma}_{[(n-4)/2]}(\nu)\,.
\eeq
Our results on the coefficients $P_{xn}$ were found to satisfy this polynomiality rule after adding all separate contributions, 
and notably the one linked  to radiation-reaction modifications of the orbital motion.
E.g., at order $G^4$ the term proportional to $- \eta^2 \frac{9529}{67200}\frac{\pinf^9}{j^4}$ in the fractionally 1PN
tail term \eqref{DeltaPxypasttails} would separately violate the rule \eqref{rulePxn}, but is needed to cancell corresponding
rule-violating terms in $h^4 P_{x4}$, while, at order $G^5$, the term $-(Mc)\frac{m_2-m_1}{M}\nu^3\eta^5
\frac{15872}{6125}\frac{p_\infty^8}{j^5} $ in $\delta^{\rm rr} P_x^{{\rm rad\,inst},\, I,J}$, Eq. \eqref{delta_rr_P_xy_rad_inst_IJ},
is non-polynomial by itself, but corrects the non-polynomiality of other contributions.


\begin{table*}  
\caption{\label{tab:Pxn}  PN-expansion of the coefficients $P_{xn}$ of the $x$-component of the radiated linear momentum through the 3PN fractional accuracy. 
}
\begin{ruledtabular}
\begin{tabular}{ll}
$P_{x4}$ &$\pi\left[-\frac{1491}{400}  p_\infty^7+(\frac{1491}{200} \nu+\frac{26757}{5600})  p_\infty^9+O(p_\infty^{11})\right]$\\
$P_{x5}$ & $-\frac{20608}{225} p_\infty^6+(\frac{10304}{45}\nu+\frac{1143232}{7875}) p_\infty^8 -\frac{196096}{945}  p_\infty^9  
+O(p_\infty^{10})$\\
$P_{x6}$ & $\pi\left[-\frac{267583}{2400} p_\infty^5+(\frac{2509097}{7200}\nu+\frac{12566143}{67200})p_\infty^7-\frac{20719}{320}\pi^2 p_\infty^8 
+O(p_\infty^9)\right]$\\
$P_{x7}$ & $-\frac{64576}{75} p_\infty^4+(\frac{3802976}{1125}\nu+\frac{9920672}{7875})p_\infty^6+\left(-\frac{42739712}{55125}\pi^2-\frac{9226496}{4725}\right)p_\infty^7
+O(p_\infty^8)$\\
\end{tabular}
\end{ruledtabular}
\end{table*}

\section{Lorentz-invariant form factors for $P^\mu_{\rm rad}$, and mass-polynomiality rules} 
\label{formfactors}

In the sections above, we have discussed the values of the losses of energy, angular momentum and linear momentum
{\it in the c.m. frame}. This was motivated by the fact that the multipolar-post-Minkowskian approach \cite{Blanchet:1985sp,Blanchet:1989ki,Damour:1990ji} to gravitational radiation is conveniently applied within the c.m. frame of
the binary system. Let us now re-express these c.m.-based, and PN-expanded, results in a Lorentz-invariant way.

As was pointed out in previous works (e.g. \cite{Damour:2019lcq,Bini:2021gat}), if one expresses the individual
momentum changes (or impulses), $\Delta p^\mu_1, \Delta p^\mu_2$, during gravitational scattering, and therefore
also the radiated 4-momentum  $P^\mu_{\rm rad}= -(\Delta p^\mu_1+ \Delta p^\mu_2)$, 
 in terms of the incoming 4-velocities,  $u^\mu_{1-}, u^\mu_{2-}$ and of
the vectorial impact parameter $b_{12}^\mu \equiv b^\mu_1- b^\mu_2$, their expansion coefficients in powers of $G$ must
be  polynomials in the two masses $m_1, m_2$. Let us show here what information we can thereby get from
such mass-polynomiality.

We can decompose $P^\mu_{\rm rad}$ as follows
\bea \label{decompPmu}
P^\mu_{\rm rad}&=& P^{\rm rad}_{1+2}(m_1,m_2,\g, b) (u^\mu_{1-}+ u^\mu_{2-}) \nonumber \\  
&+&  P^{\rm rad}_{1-2}(m_1,m_2,\g,b) (u^\mu_{1-} -  u^\mu_{2-})\nonumber\\
&+&  P^{\rm rad}_{b_{12}}(m_1,m_2,\g,b) {\hat b}_{12}^\mu  \,.
\eea
The basis $u^\mu_{1-}+ u^\mu_{2-}$, $u^\mu_{1-}- u^\mu_{2-}$, $ {\hat b}_{12}^\mu $ is
{\it orthogonal}, though not orthonormal. While $(\hat b_{12})^2=+1$  we have
\beq
(u^\mu_{1-}+ u^\mu_{2-})^2=-2(\gamma+1)\,,\quad (u^\mu_{1-}- u^\mu_{2-})^2=+2(\gamma-1)\,.
\eeq

Taking into account the symmetry of $P^\mu_{\rm rad}$ under the $1 \leftrightarrow 2$ exchange,
and the (anti-)symmetry of $u^\mu_{1-}+ u^\mu_{2-}$ ($u^\mu_{1-}- u^\mu_{2-},  {\hat b}_{12}^\mu $),
we see that the first form factor $ P^{\rm rad}_{1+2}(m_1,m_2,\g, b)$ must be  $1 \leftrightarrow 2$-symmetric, while
 $ P^{\rm rad}_{1-2}(m_1,m_2,\g, b)$ and  $ P^{\rm rad}_{b_{12}}(m_1,m_2,\g, b)$ must be 
  $1 \leftrightarrow 2$-antisymmetric. We can then use the further facts that: (i) radiative losses of energy and linear momentum
  being quadratic in the retarded-time derivative of the waveform  must contain a factor $(m_1 m_2)^2$; and (ii) $ P^{\rm rad}_{1+2}(m_1,m_2,\g, b)$ starts at order $G^3$, while $ P^{\rm rad}_{1-2}(m_1,m_2,\g, b)$ and  $ P^{\rm rad}_{b_{12}}(m_1,m_2,\g, b)$  start at order  $G^4$. The mass-polynomiality of the PM expansion coefficients of $P^\mu_{\rm rad}$
  then allows us to  write
  \bea
\label{gen_rels}
  P^{\rm rad}_{1+2}(m_1,m_2,\g, b) &=& \frac{G^3}{b^3} m_1^2 m_2^2  {\hat P}^{\rm rad}_{1+2} \,,  \nonumber\\
  P^{\rm rad}_{1-2}(m_1,m_2,\g, b) &=&  \frac{G^4}{b^4} m_1^2 m_2^2 (m_2-m_1){\hat P}^{\rm rad}_{1-2}\,,  \nonumber\\
   P^{\rm rad}_{b_{12}}(m_1,m_2,\g, b) &=& \frac{G^4}{b^4} m_1^2 m_2^2  (m_2-m_1){\hat P}^{\rm rad}_{b_{12}}\,,\qquad
  \eea
  where the dimensionless factors $ {\hat P}^{\rm rad}_{1+2}$, $ {\hat P}^{\rm rad}_{1-2}$, $ {\hat P}^{\rm rad}_{b_{12}}$
  have PM expansions of the form
  \bea
   {\hat P}^{\rm rad}_{1+2} &=& \sum_{n\geq3} \frac{G^{n-3}}{b^{n-3}} SP_{n-3}^{1+2}(m_1,m_2)\nonumber\\
&=&  \sum_{n\geq3} \frac{G^{n-3} M^{n-3}}{b^{n-3}} p^{1+2, G^n}_{[\frac{n-3}{2}]}(\g, \nu) 
\,, \nonumber\\
   {\hat P}^{\rm rad}_{1-2} &=& \sum_{n\geq4} \frac{G^{n-4}}{b^{n-4}} SP_{n-4}^{1-2}(m_1,m_2)\nonumber\\
&=&  \sum_{n\geq4} \frac{G^{n-4} M^{n-4}}{b^{n-4}} p^{1-2, G^n}_{[\frac{n-4}{2}]}(\g, \nu)
\,,\nonumber \\ 
    {\hat P}^{\rm rad}_{b_{12}} &=& \sum_{n\geq4} \frac{G^{n-4}}{b^{n-4}} SP_{n-4}^{b_{12}}(m_1,m_2)\nonumber\\
&=&  \sum_{n\geq4} \frac{G^{n-3} M^{n-3}}{b^{n-3}} p^{b_{12}, G^n}_{[\frac{n-4}{2}]}(\g, \nu)  \,.\qquad
  \eea
 Here, $  SP_{N}^X(m_1,m_2)$ denotes a {\it symmetric} polynomial of order $N$  in the two masses. By scaling out
 the total mass $M = m_1+m_2$, each such polynomial can be rewritten as  
 \beq
\label{form_fact_def}
 SP_{N}^X(m_1,m_2) = M^N p^{X, G^n}_{[\frac{N}{2}]}(\g,\nu)\,, 
 \eeq
 where $p^{X, G^n}_{[\frac{N}{2}]}(\g,\nu)$ is
  a polynomial in $\nu$ of order  $[\frac{N}{2}]$ (the integer part of $\frac{N}{2}$), with $\g$-dependent coefficients. 
  In order to keep track of the PM order $n$, we add a label $G^n$, and we also sometimes keep the notation $[\frac{N}{2}]$,
  with $N=n-3$ or $N=n-4$ (e.g. we write $[\frac{1}{2}]$
  instead replacing it by its numerical value 0). 
 
 We thereby see that, while at order $G^3$ (3PM order), $P^\mu_{\rm rad}$ was described by only one function of $\g$,
 namely (see Eq. \eqref{PradG3})
 \beq
 SP_{0}^{1+2}(m_1,m_2)=  p^{1+2, G^3}_{[\frac{0}{2}]}(\g)=  \frac{{\mathcal E (\g)}}{\g+1}\,,
 \eeq
 it will involve three functions of $\g$ at order $G^4$, namely
 \bea
 P^{\rm rad, G^4}_{1+2} &=&  \frac{G^4}{b^4} m_1^2 m_2^2 SP_{1}^{1+2}(m_1,m_2)\nonumber\\
&=& \frac{G^4}{b^4} m_1^2 m_2^2 (m_1+m_2)  p^{1+2, G^4}_{[\frac{1}{2}]}(\g)\,, \nonumber\\ 
  P^{\rm rad, G^4}_{1-2} &=&  \frac{G^4}{b^4} m_1^2 m_2^2 (m_2-m_1) SP_{0}^{1-2}(m_1,m_2)\nonumber\\
&=& \frac{G^4}{b^4} m_1^2 m_2^2 (m_2-m_1)  p^{1-2, G^4}_{[\frac{0}{2}]}(\g)\,,  \nonumber\\ 
 P^{\rm rad, G^4}_{b_{12}} &=&  \frac{G^4}{b^4} m_1^2 m_2^2 (m_2-m_1) SP_{0}^{b_{12}}(m_1,m_2)\nonumber\\
&=& \frac{G^4}{b^4} m_1^2 m_2^2 (m_2-m_1)  p^{b_{12}, G^4}_{[\frac{0}{2}]}(\g)\,.    
 \eea
 At order $G^5$, we have four functions of $\g$:
  \bea
\label{G5_rel}
 P^{\rm rad, G^5}_{1+2} &=&  \frac{G^5}{b^5} m_1^2 m_2^2 SP_{2}^{1+2}(m_1,m_2)\nonumber\\
&=& \frac{G^5}{b^5} m_1^2 m_2^2 (m_1+m_2)^2  p^{1+2, G^5}_{[\frac{2}{2}]}(\g, \nu)\,, \nonumber\\ 
  P^{\rm rad, G^5}_{1-2} &=&  \frac{G^5}{b^5} m_1^2 m_2^2 (m_2-m_1) SP_{1}^{1-2}(m_1,m_2)\nonumber\\
&=& \frac{G^5}{b^5} m_1^2 m_2^2 (m_2-m_1) (m_1+m_2) p^{1-2, G^5}_{[\frac{1}{2}]}(\g)\,,  \nonumber\\ 
 P^{\rm rad, G^5}_{b_{12}} &=&  \frac{G^5}{b^5} m_1^2 m_2^2 (m_2-m_1) SP_{1}^{b_{12}}(m_1,m_2)\nonumber\\
&=& \frac{G^5}{b^5} m_1^2 m_2^2 (m_2-m_1) (m_1+m_2) p^{b_{12}, G^5}_{[\frac{1}{2}]}(\g)\,, \nonumber\\   
 \eea
 where  $p^{1+2, G^5}_{[\frac{2}{2}]}(\g, \nu)$, being linear in $\nu$, involves two independent functions of $\g$.
 At order $G^n$, $P^\mu_{\rm rad}$ generally involves
 \beq
 N^{G^n} _{P_{\rm rad}}= \left[\frac{n-1}{2}\right] + 2 \times \left[\frac{n-2}{2}\right]
 \eeq
 functions of $\g$.
 
 Let us now discuss how to relate 
 the Lorentz-invariant building blocks $ p^{1+2, G^n}_{[\frac{n-3}{2}]}(\g, \nu)$,  $p^{1-2, G^n}_{[\frac{n-4}{2}]}(\g, \nu)$,
  $p^{b_{12}, G^n}_{[\frac{n-4}{2}]}(\g, \nu)$ parametrizing the PM expansion of $P^\mu_{\rm rad}$ to our
  previous   c.m.-frame, PN-expanded, results on $E^{\rm rad}, P_x^{\rm rad}, P_y^{\rm rad}$.
  
  A first step in this direction consists in computing the projections of $P^\mu_{\rm rad}$ on the three unit vectors
  $U^\mu$, $n_{-}^\mu$ and ${\hat b}_{12}$, where $U^\mu$ is the c.m. time axis, such that
  \beq
  M h U^\mu= m_1 u^\mu_{1-}+ m_2 u^\mu_{2-}\,,
  \eeq
  and where $n_{-}^\mu$ is the unit vector in the c.m.-frame direction of  $u^\mu_{1-}$, such that
  \beq
   M h \pinf n_{-}^\mu= (m_2 + \g m_1)  u^\mu_{1-} -  (m_1 + \g m_2)  u^\mu_{2-} \,.
  \eeq
  The definition of $E^{\rm rad}$, namely $E^{\rm rad}= - U^\mu P_\mu^{\rm rad}$ then yields
  \beq
  M h E^{\rm rad}= ( m_1 u^\mu_{1-}+ m_2 u^\mu_{2-})  P_\mu^{\rm rad}\,.
  \eeq
  From the definition Eq. \eqref{exydef} of ${\mathbf e}_x$ and ${\mathbf e}_y$, we deduce that
  \beq \label{defPn}
  P_n^{\rm rad} \equiv  n_{-}^\mu  P_\mu^{\rm rad}= \sin \frac{\chi_{\rm cons}}{2}  P_x^{\rm rad} +  \cos \frac{\chi_{\rm cons}}{2}  P_y^{\rm rad}\,,
  \eeq
  while
  \beq \label{defPb}
    P_b^{\rm rad} \equiv  {\hat b}_{12}^\mu  P_\mu^{\rm rad}= \cos \frac{\chi_{\rm cons}}{2}  P_x^{\rm rad} -  \sin \frac{\chi_{\rm cons}}{2}  P_y^{\rm rad}\,.
  \eeq
 Inserting the parametrization  \eqref{decompPmu} into these results then yields the following links between
  $E^{\rm rad}, P_x^{\rm rad}, P_y^{\rm rad}$ (remembering the definitions \eqref{defPn}, \eqref{defPb})
  and the form factors of  $P_\mu^{\rm rad}$:
  \bea
  M h E^{\rm rad} &=& M (\g+1)  P^{\rm rad}_{1+2} + (m_2-m_1) (\g -1)  P^{\rm rad}_{1-2}\,, \nonumber\\ 
   M h P_n^{\rm rad} &=& (m_2-m_1) \pinf  P^{\rm rad}_{1+2} +  M \pinf  P^{\rm rad}_{1-2}\,, \nonumber\\ 
    P_b^{\rm rad} &=&  P^{\rm rad}_{b_{12}}\,.
 \eea
 These simple links can be easily inverted to express $ P^{\rm rad}_{1+2} $ and  $ P^{\rm rad}_{1-2} $
 as linear combinations of $ h E^{\rm rad}$ and $h P_n^{\rm rad}$, and we have used them to extract the values
 of  $ P^{\rm rad}_{1+2} $ and  $ P^{\rm rad}_{1-2} $. Before exhibiting our results, several remarks are in order.
 
 Let us first note that while the mass-polynomiality of the form factor $ P^{\rm rad}_{b_{12}}$ immediately implies
 the mass-polynomiality of $ P_b^{\rm rad} \equiv  {\hat b}_{12}^\mu  P_\mu^{\rm rad}$, the mass-polynomiality 
 of the two other form factors,  $ P^{\rm rad}_{1+2} $ and  $ P^{\rm rad}_{1-2} $, implies the mass-polynomiality
 of the combinations $M h E^{\rm rad}$ and $ M h P_n^{\rm rad}$. In these combinations it is crucial to include
 the factor  $M h= M \sqrt{1+ 2\nu(\g-1)}=E^{\rm c.m.}_-$ (including the extra mass factor $M$, which cannot be,
generally,  factored out on the right-hand sides). 

In more detail, we have 
\bea
  M h E^{\rm rad} &=& \frac{G^3}{b^3} m_1^2 m_2^2 M (\g+1)  {\hat P}^{\rm rad}_{1+2} \nonumber\\
&+&  \frac{G^4}{b^4} m_1^2 m_2^2 (m_2-m_1)^2 (\g -1)  {\hat P}^{\rm rad}_{1-2}\,, \nonumber\\
   M h P_n^{\rm rad} &=& \frac{G^3}{b^3} (m_2-m_1) \pinf {\hat P}^{\rm rad}_{1+2} \nonumber\\
&+&  \frac{G^4}{b^4} m_1^2 m_2^2 (m_2-m_1) M \pinf  {\hat P}^{\rm rad}_{1-2}\,, 
 \eea
 where we recall that the various dimensionless factors $ {\hat P}^{\rm rad}_{X} $ have the more explicit structure
 \beq
{\hat P}^{\rm rad}_{X} =\sum_{N\geq 0}\frac{G^N}{b^N} SP_N^X(m_1,m_2)= \sum_{N\geq 0}\left(\frac{GM}{b}\right)^N P^X_{[\frac{N}{2}]}(\nu)\,.
\eeq
These expressions give a direct proof of the $\nu$-structures pointed out in our previous works, notably\footnote{Here
we use the expansion in  powers of $\frac{G}{b}$. When using the expansion in $\frac1j= \frac{GMh}{b \pinf}$ one
must add an extra factor $h^n$ at order $\frac1{j^n}$, as used in Eq. \eqref{ruleEn}.}, 
 \beq
\label{Erad_formfact}
\left(\frac{ h  E^{\rm rad}}{M}\right)^{\frac{G^n}{b^n}}= \left(\frac{GM}{b}\right)^n  \nu^2 P^{\gamma}_{[(n-2)/2]}(\nu)\,,
 \eeq 
 and also 
\beq
\label{Pnrad_formfact}
\left(\frac{ h  P_n^{\rm rad}}{M}\right)^{\frac{G^n}{b^n}}= \left(\frac{GM}{b}\right)^n  \nu^2 \frac{m_2-m_1}{M} P^{\gamma}_{[(n-3)/2]}(\nu)\,.
 \eeq 
 Note also that, while in $ M h E^{\rm rad} $ the dimensionless form factor $ {\hat P}^{\rm rad}_{1-2}$ is multiplied
by the small PN factor $\g-1 = O(\pinf^2)$, in   $M h P_n^{\rm rad}$ the two form factors 
$ {\hat P}^{\rm rad}_{1+2}$ and $ {\hat P}^{\rm rad}_{1-2}$ contribute with the same PN weight (at any given order in $G$).
 
Inserting the mass-polynomiality structures of $ P_b^{\rm rad}$ and $ M h P_n^{\rm rad}$ in the expressions
of $P_x^{\rm rad}$ and  $P_y^{\rm rad}$ in terms of  $ P_b^{\rm rad}$ and $ P_n^{\rm rad}$, and using the
mass-polynomiality of the magnitude of the conservative momentum transfer
\bea
\frac{\sf Q}{2}&=& P_{\rm c.m.}  \sin \frac{\chi_{\rm cons}}{2}= \frac{G m_1 m_2}{b} \left[ \frac{2 \g^2-1}{\gamma^2-1}\right.\nonumber\\
&+&\left. \frac{G}{b} SP_1(m_1,m_2) +  \frac{G^2}{b^2} SP_2(m_1,m_2) + \cdots \right]\,,\nonumber\\
\eea
which yields
\bea
 \sin \frac{\chi_{\rm cons}}{2}&=& \frac{G M h}{b} \left[ \frac{2 \g^2-1}{\gamma^2-1}\right.\nonumber\\
&+& \left. \frac{G}{b} SP_1(m_1,m_2) +  \frac{G^2}{b^2} SP_2(m_1,m_2) + \cdots \right]\,,\nonumber\\
\eea
one can easily derive the following mass-polynomiality structures
\bea \label{PxG4}
P_x^{\rm rad}&=&\frac{G^4}{b^4} m_1^2 m_2^2 (m_2-m_1)  \left[SP_0(m_1,m_2)\right.\nonumber\\
&+&\left. \frac{G}{b} SP_1(m_1,m_2) +  \frac{G^2}{b^2} SP_2(m_1,m_2) + \cdots \right]\,,\nonumber\\
\eea
and
\bea
M h P_y^{\rm rad}&=&\frac{G^3}{b^3} m_1^2 m_2^2 (m_2-m_1)  \left[SP_0(m_1,m_2)\right.\nonumber\\
&+&\left. \frac{G}{b} SP_1(m_1,m_2) +  \frac{G^2}{b^2} SP_2(m_1,m_2) + \cdots \right]\,.\nonumber\\
\eea
As above, each such mass-polynomiality structure leads, after scaling out the appropriate power of $\frac{GM}{b}$, a polynomial
structure in the symmetric mass ratio $\nu$ (with $\g$-dependent coefficients), namely
\beq
\frac{G^N}{b^N} SP_N^X(m_1,m_2)= \left(\frac{GM}{b}\right)^N P^X_{[\frac{N}{2}]}(\nu)\,.
\eeq
One then easily checks that relations such as Eq. (7.27) in Ref. \cite{Bini:2021gat} and its $G^5$-generalization indicated in the caption of Table II there,  
follow from Eqs. \eqref{Erad_formfact} and \eqref{Pnrad_formfact} above.

We have already mentioned above that our c.m.-based, and PN-based, results on $E^{\rm rad}$,  $P_x^{\rm rad}$ and  $P_y^{\rm rad}$ were all in agreement (after adding all separate contributions, and notably the one linked to radiation-reaction
modifications of the orbital motion) with the $\nu$-polynomiality rules rederived here. We can therefore encapsulate
the full, current PN-expanded information on  $P_{\rm rad}^\mu$ in the values of the $\g$-dependent 
$\nu$-polynomials $p^X_{[\frac{N}{2}]}(\g,\nu)$ parametrizing the form factors, see Eqs. \eqref{gen_rels}--\eqref{G5_rel}. 

At order $G^3$ our results  yield
\bea
p^{1+2, G^3}_{[\frac{0}{2}]}(\gamma)&=& \pi  \left(\frac{37}{30}p_\infty+\frac{839}{1680}p_\infty^3+\frac{2699}{2016}p_\infty^5\right. \nonumber\\
&-& \left.\frac{1531643}{1182720}p_\infty^7+O(p_\infty^9)\right)\,,
\eea
which agrees with the fractionally 3PN-level expansion of the exact result
\beq
p^{1+2, G^3}_{[\frac{0}{2}]}(\gamma)= \pi \frac{\hat {\mathcal E}(\g)}{\gamma+1}\,.
\eeq
At order $G^4$ we find 
\bea
\label{form_factors_G4}
p^{1+2, G^4}_{[\frac{1}{2}]}(\gamma)&=& \frac{784}{45 p_\infty}+\frac{2168}{175}p_\infty+\frac{1568}{45} p_\infty^2+\frac{98666}{11025}p_\infty^3\nonumber\\
&-&\frac{512}{105}p_\infty^4-\frac{2702747}{363825}p_\infty^5
+O(p_\infty^6)
\,,\nonumber\\
p^{1-2, G^4}_{[\frac{0}{2}]}(\gamma)&=& \frac{176}{45 p_\infty}-\frac{72}{25} p_\infty+\frac{352}{45}p_\infty^2-\frac{9746}{4725} p_\infty^3\nonumber\\
&+&\frac{448}{75}p_\infty^4-\frac{484019}{51975}p_\infty^5
+O(p_\infty^6)
\,,\nonumber\\
p^{b_{12}, G^4}_{[\frac{0}{2}]}&=& -\pi \left[ \frac{37}{30}+\frac{1661}{560} p_\infty^2+\frac{1491}{400} p_\infty^3+\frac{23563}{10080} p_\infty^4\right.\nonumber\\
&-&\left.  \frac{26757}{5600} p_\infty^5+
\frac{700793}{506880} p_\infty^6
+O(p_\infty^7)\right] .
\eea
While writing up our results, a PN-exact computation of  the 4PM contribution to  $P^{\rm rad}_\mu$, and notably, its
${\hat b}_{12}^\mu$ projection, appeared on arXiv \cite{Dlapa:2022lmu}.
Our (fractionally 3PN accurate) results, Eqs. \eqref{form_factors_G4},  are compatible with those given in Ref. \cite{Dlapa:2022lmu}.

Similarly at $O(G^5)$ we have   
\begin{widetext}
\bea
p^{1+2,G^5}_{[\frac{2}{2}]}(\gamma,\nu)&=&\pi \left[ \frac{61}{5p_\infty^3} +\frac{34073}{1680 p_\infty}+\frac{297}{40}\pi^2-\frac{23923}{2880} p_\infty
+\left(-\frac{31029}{2240}\pi^2+\frac{1484997}{11200} \right)p_\infty^2\right.\nonumber\\
&&\left.
+\left(\frac{99}{20}\pi^2+\frac{34695068413}{620928000} -\frac{10593}{700}\ln\left(\frac{p_\infty}{2}\right) \right) p_\infty^3\right]\nonumber\\
&& 
+\nu \pi \left[-\frac{55}{12p_\infty} +\frac{6427}{10080} p_\infty+\left(\frac{877}{400 }-\frac{939}{560}\pi^2\right) p_\infty^2\left(\frac{255491}{10080}-\frac{4059}{1280}\pi^2\right) p_\infty^3
+O(p_\infty^4)\right] 
\,, \nonumber\\
p^{1-2,G^5}_{[\frac{1}{2}]}(\gamma)&=&\pi\left[ \frac{82}{15p_\infty^3}-\frac{5207}{630 p_\infty}-\frac{1491}{400}+\frac{939}{280}\pi^2-\frac{963239}{40320} p_\infty+\left(\frac{902743}{33600}-\frac{13603}{4480}\pi^2\right)p_\infty^2\right.\nonumber\\
&&\left.+\left(-\frac{4809573323}{434649600}   -\frac{1591}{980}\ln\left(\frac{p_\infty}{2}\right)+\frac{313}{140}\pi^2 \right)p_\infty^3
+O(p_\infty^4)\right]
\,, \nonumber\\
p^{b_{12},G^5}_{[\frac{1}{2}]}(\gamma)&=& -\frac{64}{3p_\infty^2}-\frac{37}{20}\pi^2-\frac{27392}{525}-\frac{30208}{225}p_\infty
+\left(-\frac{856768}{33075}-\frac{3429}{1120}\pi^2\right) p_\infty^2+\frac{462592}{7875} p_\infty^3\nonumber\\
&&+\left(-\frac{74417152}{363825}- \frac{7915}{2688}\pi^2\right)p_\infty^4
+O(p_\infty^5)
 \,.
\eea
\end{widetext}

\section{Information on the individual impulses $\Delta p^\mu_a$ derivable from $P_{\rm rad}^\mu$
}
\label{individual_impulses}

Let us now discuss what information on the individual momentum changes (or impulses), $\Delta p^\mu_1, \Delta p^\mu_2$,
can be extracted from our results on $P_{\rm rad}^\mu$ by combining {\it six} different facts:

First,  the coefficients of the PM expansion of $\Delta p^\mu_1, \Delta p^\mu_2$ in terms of the incoming 4-velocities,  $u^\mu_{1-}, u^\mu_{2-}$ and of
the vectorial impact parameter $b^\mu \equiv b^\mu_1- b^\mu_2$, must be
 polynomials in the two masses $m_1, m_2$. 
More precisely, one has (for the first particle)
\beq \label{deltapmu}
\Delta p_{1 \mu}= - 2G m_1 m_2 \frac{2 \g^2-1}{\sqrt{\g^2-1}} \frac{b_{12\mu}}{b^2} +  \sum_{n\geq 2} \Delta p_{1 \mu}^{n \rm PM}\,,
\eeq 
 where each term $ \Delta p_{1 \mu}^{n \rm PM}$
is a combination of the three vectors $b_{12}^\mu/b$, $u_{1 -}^{\mu}$ and 
$u_{2 -}^{\mu}$, with coefficients that are, at each order in $G$,  {\it homogeneous polynomials} in $m_1$ and $m_2$, containing the product $m_1 m_2$ as an overall factor. Symbolically
\bea \label{deltapmugennPM}
 \Delta p_{1 \mu}^{n \rm PM} &\sim& \frac{G m_1 m_2}{b^n} \left[  (Gm_1)^{n-1} \right.\nonumber\\
&+&\left.  (Gm_1)^{n-2} G m_2+ \cdots + (Gm_2)^{n-1}\right]\,,\qquad\quad
\eea  
 where each term is a combination of the three vectors $b^\mu/b$, $u_{1 -}^{\mu}$ and 
$u_{2 -}^{\mu}$, with coefficients that are functions of $\g$. [Note that contrary to the case of  $P_{\rm rad}^\mu$,
 $\Delta p_{1 \mu}^{n \rm PM}$  is not symmetric under particle exchange.]

Second, linear momentum conservation implies that the radiated momentum is equal to
\beq \label{pconservation}
\Delta  p^\mu_1 + \Delta p^\mu_2 = -  P^\mu_{\rm rad} .
\eeq
Third, we have the decomposition
\beq \label{Dpa}
\Delta p_{ a \mu}= \Delta p_{ a \mu}^{\rm cons} + \Delta p_{ a \mu}^{\rm rr \, lin } + \Delta p_{ a \mu}^{\rm rr \, nonlin}\,.
\eeq
Here: (i) the conservative part  $\Delta p_{ a \mu}^{\rm cons}$ is known  up to the sixth PN order (modulo 6 still unknown parameters, \cite{Bini:2020wpo,Bini:2020nsb,Bini:2020hmy,Bini:2020rzn}), while its $G$ expansion is known exactly up to order $G^4$
 included \cite{Bern:2021yeh,Dlapa:2022lmu}; (ii) the linear-response contribution $\Delta p_{ a \mu}^{\rm rr \, lin }$ is known 
 (modulo some linear, time-even radiation-reaction effects discussed below) from our previous work \cite{Bini:2021gat}; while (iii)
the remainder term $\Delta p_{ a \mu}^{\rm rr \, nonlin}$ can  be described as containing the contributions that are
higher-order in radiation-reaction (starting with the quadratic order $O({\cal F^\mu_{\rm rr}}^2)$).

Fourth, as we are going to show, the linear-response contribution happens to satisfy, by itself, the momentum conservation law
\eqref{pconservation}, namely
\beq \label{plinconservation}
\Delta  p_{1 \mu}^{\rm rr \, lin } + \Delta p_{2 \mu}^{\rm rr \, lin }= -  P^\mu_{\rm rad} .
\eeq

Fifth, the linear response contribution satisfies a linearized version of the mass-shell condition that must hold for the outgoing momenta, namely
\beq
\label{mass_shell_smt}
p_{ a \mu}^{+\rm cons} \Delta p^\mu_{ a }{}^{\rm rr \, lin }=0\,.
\eeq

Sixth, the nonlinear contribution $\Delta p_{ a \mu}^{\rm rr \, nonlin}$ to the impulse of the $a$th particle
  (as well as the additional contribution 
$\Delta  p_{a \mu}^{\rm rr \, \Delta c_\phi }$ to $\Delta  p_{a \mu}^{\rm rr \, lin }$ linked to
the time-even part of ${\cal F^\mu_{\rm rr}}$ discussed below) must involve a factor $m_a^3$.

In the following, we explain the origin of these facts, and then show how they determine the 
 conservativelike radiative contributions 
 $\Delta  p_{a \mu}^{\rm rr \, \Delta c_\phi } + \Delta  p_{a \mu}^{\rm rr \, nonlin }$
 at the fourth post-Minkowskian order ($O(G^4)$), and
 strongly constrain them at the fifth  post-Minkowskian order ($O(G^5)$).

 \subsection{Proof of the identity \eqref{plinconservation} and antisymmetry property of $\Delta  p_{a \mu}^{\rm rr \, nonlin }$}

The linear-response contribution $\Delta  p_{a \mu}^{\rm rr \, lin }$ was obtained in \cite{Bini:2021gat} as the sum
of two terms: a {\it relative motion} term  $\Delta  p_{a \mu}^{\rm rr \, rel }$, and a {\it recoil} term $\Delta  p_{a \mu}^{\rm rr \, rec }$:
\beq
\Delta  p_{a \mu}^{\rm rr \, lin }= \Delta  p_{a \mu}^{\rm rr \, rel }+ \Delta  p_{a \mu}^{\rm rr \, rec }\,.
\eeq
From Eqs. (3.32) and (3.33) in \cite{Bini:2021gat}, we have
\bea
\Delta  p_{a \mu}^{\rm rr \, rel }&=& \chi^{\rm  rr \, rel } \frac{d}{d\chi_{\rm cons}}\Delta p_{a \mu}^{\rm cons} + 
\frac{\Delta P_{\rm c.m.}}{ P_{\rm c.m.}} p_{a \mu}^+ \nonumber\\
& -& \frac{m_a^2}{E_a} \frac{\Delta P_{\rm c.m.}}{ P_{\rm c.m.}} U_\mu\,,
\eea
and
\beq
\Delta  p_{a \mu}^{\rm rr \, rec }= -\frac{E_a}{E_{\rm c.m.}}  P_\mu^{\rm rad} - \frac{(p_{a \nu}^+   P^\nu_{\rm rad})}{E_{\rm c.m.}} U_\mu\,.
\eeq
Here, 
\beq
\Delta P_{\rm c.m.} =  - \frac{E_1 E_2}{E_{\rm c.m.} P_{\rm c.m.}} E_{\rm rad}\,,
\eeq
and the quantities $E_a$, $E_{\rm c.m.}= E_1+E_2= M h$, $ P_{\rm c.m.} = \frac{m_1 m_2 \pinf}{E_{\rm c.m.}}$, $p_{a \nu}^+ $ (outgoing momenta), $U_\mu\equiv (p_{1 \nu}^- + p_{2 \nu}^-)/E_{\rm c.m.}$
 are all taken along the unperturbed, conservative motion.

When summing over the particle label $a$, taking into account the fact that $\sum_a \Delta p_{a \mu}^{\rm cons}=0$
and $\sum_a p_{a \nu}^+ = \sum_a p_{a \nu}^-= E_{\rm c.m.} U_\mu$, one easily finds that Eq. \eqref{plinconservation}
is (exactly) satisfied. This identity (together with the fact that  $\sum_a \Delta p_{a \mu}^{\rm cons}=0$) implies the
somewhat remarkable identity that the remainder (nonlinear) term in the linear-response formula \eqref{Dpa} must 
separately satisfy the identity
\beq \label{premainconservation}
\Delta  p_{1 \mu}^{\rm rr \, nonlin } + \Delta p_{2 \mu}^{\rm rr \, nonlin }= 0\,.
\eeq
In other words, the nonlinear contribution $\Delta  p_{a \mu}^{\rm rr \, nonlin }$ must be {\it antisymmetric} under particle exchange.

Another constraint on $\Delta p_{a \mu}^{\rm rr \, nonlin }$ is the mass-shell condition
\beq
p_{a\mu}^{+\rm tot}p_a^{\mu}{}^{+\rm tot}=-m_a^2\,,
\eeq
where the total outgoing momentum is 
\beq
p_{a\mu}^{+\rm tot}=p_{a\mu}^{+\rm cons}+\Delta p_{a \mu}^{\rm rr \, lin }+\Delta p_{a \mu}^{\rm rr \, nonlin }\,.
\eeq
Using the fact that $\Delta p_{a \mu}^{\rm rr \, lin }$ satisfies (independently of the value of $\chi^{\rm rel}$) 
Eq. \eqref{mass_shell_smt} we get the following additional constraint on $\Delta p_{a \mu}^{\rm rr \, nonlin }$
\beq
\label{mass_shell_smt2}
2 p_{a\mu}^{+\rm cons} \Delta p_{a}^{\mu}{}^{\rm rr \, nonlin }+(\Delta p_{a \mu}^{\rm rr \, lin }+\Delta p_{a \mu}^{\rm rr \, nonlin })^2=0\,.
\eeq

 \subsection{Completing the linear-response formula when ${\mathcal F}^{\rm rr}_a$ is time-asymmetric, 
 without  being time-antisymmetric.}
 
At this point we need to complete one result derived in Ref. \cite{Bini:2021gat}, namely Eq. (3.25) there, giving the value of
the radiation-reaction contribution,  $\chi^{\rm  rr \, rel }$, to the relative scattering angle. Note first that the actual value of
 $\chi^{\rm  rr \, rel }$ did not matter in the proof  of the validity of Eq. \eqref{plinconservation} we have just given.
 Indeed, after summing over $a$, the coefficient of $\chi^{\rm  rr \, rel }$ is
 \beq
  \frac{d}{d\chi_{\rm cons}}\left[ \sum_a\Delta p_{a \mu}^{\rm cons} \right]\,,
\eeq
which vanishes because  $\sum_a \Delta p_{a \mu}^{\rm cons}$ vanishes, independently of the value of $\chi_{\rm cons}$.

The only place were the assumption of time-antisymmetry of the radiation reaction force was crucial in
the derivation of the linear-response formula in Ref. \cite{Bini:2021gat} was in the derivation of the value of 
$\chi^{\rm  rr \, rel }$ (leading to Eq. (3.25) there). Going back to the previous derivation of 
$\chi^{\rm  rr \, rel }$ in Ref. \cite{Bini:2012ji}, it was explained, around Eq. (5.98) there,  that one could (when using
Lagrange's method of variation of constants) directly relate $\chi^{\rm  rr \, rel }$ to the radiative losses of (c.m.) energy 
and angular momentum if the time-derivatives of $ \frac{ d c_l(t)}{dt}$ and $ \frac{ d c_\phi(t)}{dt}$ 
were odd functions of time (around the moment of
closest approach in the conservative motion). As $ \frac{ d c_l(t)}{dt}$ and $ \frac{ d c_\phi(t)}{dt}$ 
are linear expressions in the radiation-reaction force,
their time-odd character  is directly linked to the time-odd character of ${\mathcal F}_{\rm rr}$
(as was discussed at the end of section \ref{2.5PNmotion} above, when working with the LO, 2.5PN radiation-reaction force).
As we were aware of this limitation in  Ref. \cite{Bini:2021gat}, we limited our study of radiation-reaction effects to the
4.5PN level, because we had shown there (see Eq. (H3) there), that, at the 5PN level there arose a non-zero value
of $P_x^{\rm rad}$ (while a time-odd  ${\mathcal F}_{\rm rr}$ implies a vanishing value for  $P_x^{\rm rad}$).

When staying at the level of linear effects in ${\mathcal F}_{\rm rr}$, a re-examination of the proof of the linear-response
formula in Ref. \cite{Bini:2021gat} shows that the only  $O({\mathcal F}_{\rm rr})$ modification to take into
account is the presence of an extra contribution in Eq. (3.25) there.  
One gets an explicit expression for the latter extra contribution by using the varying-constant version of the
quasi-Keplerian representation, Eqs. \eqref{sys0}.  From the equation parametrizing $\phi(t)$,
and the link $\chi = [\phi]_{-\infty}^{+\infty} -\pi$ between the total scattering angle, $\chi$,  and the variation of $\phi$,
we get (using $c_1 = E$ and $c_2=J$) 
\beq
\chi= [{\bar W}(l, E(t), J(t))] - \pi+ [c_\phi(t)]\,.
\eeq
The first term yields (when separating out the conservative contribution and linearly expanding in the
radiative losses of energy and angular momentum)
our usual linear-response formula for the radiative contribution to the  c.m. relative scattering angle.
The second contribution is new (and exists only when ${\mathcal F}_{\rm rr}(t)$ is time-asymmetric,
rather than time-odd). This yields the result 
\bea \label{chirrrel}
\chi^{\rm rr \, rel}&=&- \left(\frac12 \frac{\partial \chi^{\rm cons}}{\partial  E  } E_{\rm rad} +\frac12 \frac{\partial \chi^{\rm cons}}{\partial  J } J_{\rm rad} \right)+  \Delta c_\phi \nonumber\\
&\equiv & \bar \chi^{\rm rr \, rel} + \Delta c_\phi\,, 
\eea
where the first contribution, $\bar \chi^{\rm rr \, rel}$, has been evaluated at the $O({\mathcal F}_{\rm rr})$ accuracy,
and where a formal, but explicit, expression for the additional contribution $ \Delta c_\phi= [c_\phi]=\int_{- \infty}^{+ \infty} dt  \frac{ d c_\phi(t)}{dt}$ is obtained from the last equation in Eqs.  \eqref{LagrangeHyp}, and reads
\bea
\label{Deltacphi_eq}
\Delta c_\phi&=&  \int_{- \infty}^{+ \infty} dt \left[  \frac{\partial \bar W}{\partial l} \left(\frac{\partial S}{\partial l}\right)^{-1}\left[\frac{\partial S}{\partial E}\frac{d E}{dt}+\frac{\partial S}{\partial  J}\frac{d J}{dt}  \right]\right.\nonumber\\
&-&\left.\frac{\partial \bar W}{\partial E}\frac{d  E}{dt}-\frac{\partial \bar W}{\partial  J}\frac{d  J}{dt} \right]\,.
\eea
Here $\frac{d  E}{dt}$ and $\frac{d  J}{dt}$ are linear expressions in  ${\mathcal F}_{\rm rr}(t)$, defined
by the first two equations in  Eqs.  \eqref{LagrangeHyp} (or, explicitly, Eqs. \eqref{EJevolution}
in the Hamiltonian formalism).

We leave to future work the use of this result to directly estimate the additional term 
(starting at the 5PN level),  $\Delta c_\phi$,
in $\chi^{\rm rr \, rel}$, linked to time-asymmetric radiation-reaction effects.

 \subsection{Proof that time-asymmetric radiation-reaction contributions to $\Delta  p_{a \mu}$ involve $m_a^3$.}
 
 One of the aims of the present paper is to go beyond the limitations of Ref. \cite{Bini:2021gat}, and to discuss
the physical effects present in $P_{ a \mu}^{\rm rad}$ and in $\Delta p_{ a \mu}$ that are related to
time-asymmetric (rather than simply time-odd) radiative processes. Time-asymmetric effects in the equations
of motion first enter at the 4PN (and 4PM) level via  tail-transported hereditary processes \cite{Blanchet:1987wq}.
However, at the 4PN level one can still uniquely decompose these contributions to the dynamics into a
nonlocal-in-time conservative (time-symmetric) contribution, and a nonlocal-in-time dissipative (time-antisymmetric) 
one \cite{Damour:2014jta}. This postpones the presence of genuinely time-asymmetric effects 
to the 5PN level (still being at the 4PM level).

Additional information on the structure of time-asymmetric contributions to, say, the impulse of particle 1, 
is obtained by considering the small mass-ratio limit (say $m_1 \ll m_2$).  This limit is usefully
tackled by using the gravitational self-force approximation method (i.e., perturbations around the probe limit
in which a test-particle of infinitesimal mass $m_1 $ moves around a Schwarzschild black hole of mass $m_2$).
It was shown in Ref. \cite{Mino:2003yg} that,  if one works at the {\it first-order self-force} approximation,
i.e. if one keeps only terms of order $m_1$ in the acceleration of particle 1, i.e. terms of order $m_1^2$ 
in the force acting on particle 1, one can
 uniquely decompose the dynamics in a conservative (time-symmetric) contribution, and a 
nonlocal-in-time dissipative (time-antisymmetric) one.    This proves that the level where the separation time-even versus time-odd becomes
ambiguous is the second-order self-force approximation, corresponding to terms of order 
$m_1^3$ in the force acting on particle 1. The corresponding contributions to $\Delta p_{ 1 \mu}$
will therefore also involve a factor $m_1^3$.
 [When scaling out the total mass, such terms contain a factor $\nu^3$.]

 \subsection{Contribution to the impulses proportional to $P^{\rm rad}_x$ and its nonpolynomiality
 in the masses.}

As recalled above, Ref. \cite{Bini:2021gat} generalized the linear-response formula of Ref. \cite{Bini:2012ji}
by  including recoil\footnote{As $P_\mu^{\rm rad}=O(G^3)$, it is enough to work linearly in recoil to
reach the $O(G^6)$ accuracy.} effects. However, while 
the effects proportional to the ${\bf e}_y$  component,   $ P^{\rm rad}_y$,  of the recoil were kept 
(and analyzed) in all the formulas derived in  Ref. \cite{Bini:2021gat}, in some of the formulas there the contributions
proportional  to the ${\bf e}_x$  component,  $ P^{\rm rad}_x$, were set to zero.
Here we explicitly include (and analyze) the contribution to the impulses
proportional to  $ P^{\rm rad}_x$.

Accordingly, it is henceforth useful to decompose the radiation-reaction
contribution $\Delta  p_{a \mu}^{\rm rr }$ to the impulses in the following  new way:
\beq \label{newdecompDpa}
\Delta  p_{a \mu}^{\rm rr }= \Delta  p_{a \mu}^{\rm rr \, lin-odd } +  \Delta  p_{a \mu}^{\rm rr \, P^{\rm rad}_x }+ \Delta  p_{a \mu}^{\rm rr \, remain }\,.
\eeq
Here: $ \Delta  p_{a \mu}^{\rm rr \, lin-odd }$ denotes the part of our linear-response formula obtained when assuming
that ${\mathcal F}_{\rm rr}$ is time-odd (keeping the full\footnote{The adjective \lq\lq full" means here that we keep all the time-asymmetric (tail) contributions to the radiative losses.} $E^{\rm rad}$,  $J^{\rm rad}$
and $ P^{\rm rad}_y $ contributions, but setting $ \Delta c_\phi=0$, and $ P^{\rm rad}_x=0$);
\beq \label{DpaPx}
\Delta  p_{a \mu}^{\rm rr \, P^{\rm rad}_x }\equiv -\frac{E_a}{E_{\rm c.m.}}  P_x^{\rm rad} e_{x \,\mu}- \frac{(p_{a x}^+   P_x^{\rm rad})}{E_{\rm c.m.}} U_\mu\,,
\eeq
is the contribution  linked to a non-zero value of $ P^{\rm rad}_x$ contained in Eq. (3.33) of Ref. \cite{Bini:2021gat}; and, finally,
\beq
\Delta  p_{a \mu}^{\rm rr \, remain }\equiv \Delta  p_{a \mu}^{\rm rr \, \Delta c_\phi }+ \Delta  p_{a \mu}^{\rm rr \, nonlin }\,,
\eeq
where
\beq
\Delta  p_{a \mu}^{\rm rr \, \Delta c_\phi }= \Delta c_\phi  \frac{d}{d\chi_{\rm cons}} \Delta p_{a \mu}^{\rm cons} \,,
\eeq
is the additional term linked to a non-zero $ \Delta c_\phi$, and where
$\Delta  p_{a \mu}^{\rm rr \, nonlin }$ is the same remainder term as in our previous decomposition (nonlinear
in radiation-reaction and satisfying the antisymmetry constraint  Eq. \eqref{premainconservation}).

An important fact for the following reasonings is that, as $ \Delta c_\phi$ is  symmetric under
particle exchange, while $ \sum_a\frac{d}{d\chi_{\rm cons}} \Delta p_{a \mu}^{\rm cons}=0$, the
contribution $\Delta  p_{a \mu}^{\rm rr \, \Delta c_\phi }$ is antisymmetric under particle exchange.
As the same was proven to be true for $\Delta  p_{a \mu}^{\rm rr \, nonlin }$ 
(see Eq. \eqref{premainconservation}), we conclude that $\Delta  p_{a \mu}^{\rm rr \, remain }$
also satisfies the antisymmetry constraint
\beq \label{premainconservation2}
\Delta  p_{1 \mu}^{\rm rr \, remain } + \Delta p_{2 \mu}^{\rm rr \, remain }= 0 .
\eeq
From our previous work, and from the considerations above, we know that both $\Delta c_\phi$ and  $ P^{\rm rad}_x$
start at order $\frac{G^4}{c^{10}}$, i.e. at 4PM and 5PN. Therefore $\Delta  p_{a \mu}^{\rm rr \, remain }$
starts also at order $\frac{G^4}{c^{10}}$.

One useful source of information on the various contributions
to $\Delta  p_{a \mu}^{\rm rr }$ in the decomposition \eqref{newdecompDpa}
 is that they should combine to ensure the mass-polynomiality of $\Delta  p_{a \mu}^{\rm rr }$.
[We assume here, consistently with previous works, that $\Delta  p_{a \mu}^{\rm cons }$ has been defined so as
to be mass-polynomial.]

 It was shown in Ref. \cite{Bini:2021gat}, that $\Delta  p_{a \mu}^{\rm rr \, lin-odd } $ (in the precise sense defined
above) is polynomial in the masses under some constraints on the mass structure of $E_{\rm rad}$, $J^{\rm rad}$
and $ P^{\rm rad}_y $. It is easily checked that the constraints discussed in Ref. \cite{Bini:2021gat} are all
implied by the more general constraints on the mass structure of $E_{\rm rad}$, $J^{\rm rad}$
and $ P^{\rm rad}_y $ which have been deduced above from the mass polynomiality of $ P^{\rm rad}_\mu $,
considered as a function of $b$ (see the Section \ref{formfactors} above). Therefore, the contribution 
$\Delta  p_{a \mu}^{\rm rr \, lin-odd } $ to $\Delta  p_{a \mu}^{\rm rr }$ in the decomposition \eqref{newdecompDpa}
is separately polynomial in masses.

By contrast, we see that the presence of denominators $E_{\rm c.m.}$ in 
$\Delta  p_{a \mu}^{\rm rr \, P^{\rm rad}_x }$, Eq. \eqref{DpaPx},
implies that the $P^{\rm rad}_x $ contribution to $\Delta  p_{a \mu}^{\rm rr }$ is {\it non-polynomial} in the masses.
We are going to see that the need to cancell the nonpolynomiality of $P^{\rm rad}_x $ by the remaining contribution
$\Delta  p_{a \mu}^{\rm rr \, remain}$, together with the antisymmetric character, Eq. \eqref{premainconservation2},
and the second-self-force character ($\propto m_a^3$),
of the remaining contribution, uniquely determines $\Delta  p_{a \mu}^{\rm rr \, remain}$ (and therefore 
$\Delta  p_{a \mu}^{\rm rr }$) at order $G^4$, and determines it nearly completely at order $G^5$.

\subsection{Uniqueness of $\Delta  p_{a \mu}^{\rm rr \, remain}$ and $\Delta  p_{a \mu}^{\rm rr }$ at 4PM,
and strong constraints on them at 5PM.}

To discuss the uniqueness of  $\Delta  p_{a \mu}^{\rm rr \, remain}$, it is useful to consider its form factors on the
same basis as the one used in Section \ref{formfactors}, namely $u_{1-}^\mu+ u_{2-}^\mu$, $u_{1-}^\mu- u_{2-}^\mu$, and ${\hat b}_{12}^\mu$.  Namely, for $a=1$, and for  any label $X= {\rm  rr \, remain}$, ${\rm rr P_x^{rad}}$, ${\rm rr lin-odd}$, etc., we write
\bea \label{decompPmu}
\Delta  p_{1 }^{\mu, X}&=& c^{ 1  X}_{1+2}(m_1,m_2,\g, b) (u^\mu_{1-}+ u^\mu_{2-}) \nonumber \\  
&+&   c^{ 1  X}_{1-2}(m_1,m_2,\g,b) (u^\mu_{1-} -  u^\mu_{2-})\nonumber\\
&+&  c^{ 1  X}_{b}(m_1,m_2,\g,b) {\hat b}_{12}^\mu  \,.
\eea
For $a=2$, one should exchange $1\leftrightarrow 2$, including in the basis vectors.

Among the basis vectors, the first one is symmetric under particle exchange, while the other two are
antisymmetric. The exchange antisymmetry of $\Delta  p_{1 \mu}^{\rm rr \, remain}$ then implies that its
component $ c^{1 \, \rm remain}_{1+2}$ along $u_{1-}^\mu+ u_{2-}^\mu$ will be antisymmetric, while its components, $ c^{1 \, \rm remain}_{1-2},  c^{1\, \rm remain}_{b}$ along  
$u_{1-}^\mu- u_{2-}^\mu$, and ${\hat b}_{12}^\mu$ will be symmetric. Let us assume that we can construct 
(as we will do next) one particular
$\Delta  p_{a \mu}^{\rm rr \, remain}$ that satisfies the needed conditions of cancelling the nonpolynomiality of
$\Delta  p_{a \mu}^{\rm rr \, P^{\rm rad}_x }$ (so as to lead to a mass-polynomial $\Delta  p_{a \mu}^{\rm rr }$),
and of being $\propto m_a^3$.
The most general $\Delta  p_{a \mu}^{\rm rr \, remain}$ satisfying the latter condition will then be obtained by
adding to this particular solution a general additional term,  say $ \Delta  p_{a \mu}^{\rm rr \, remain \, add}$ 
that must satisfy several conditions. Namely: (i) it must be antisymmetric; (ii) it must be mass-polynomial;
and (iii) it must contain a factor $m_a^3$ (in addition to containing the factor $m_1^2 m_2^2$
which is a common factor of all contributions to $ \Delta  p_{a \mu}^{\rm rr }$). 

Let us prove that there cannot exist such a $ \Delta  p_{a \mu}^{\rm rr \, remain \, add}$ at order $G^4$. 
Indeed, at order $G^4$, mass-polynomiality of an impulse means that it must be quintic in masses. 
After factoring the universal factor $m_1^2 m_2^2$, we find that the mass dependence of the
(antisymmetric) component of $ \Delta  p_{a \mu}^{\rm rr \, remain \, add}$ along $u_{1-}^\mu+ u_{1-}^\mu$
must be proportional to  $m_1^2 m_2^2(m_1-m_2)$, while the (symmetric) components of $ \Delta  p_{a \mu}^{\rm rr \, remain \, add}$ along $u_{a-}^\mu- u_{a'-}^\mu$ and ${\hat b}_{aa'}^\mu$ (with $a'\not= a$) 
must be proportional to  $m_1^2 m_2^2(m_1+m_2)$. Neither of these types of components can also
satisfy the last condition of containing a factor $m_a^3$. 

When going at order $G^5$, we must discuss antisymmetric, or symmetric, {\it sextic} polynomials in masses.
In the antisymmetric case ($u_{1-}^\mu+ u_{2-}^\mu$-component) such polynomials must 
be proportional to  $m_1^2 m_2^2(m_1-m_2)(m_1+m_2)$. And the $m_a^3$ condition does not allow such terms.
By contrast, in the symmetric case ($u_{1-}^\mu- u_{2-}^\mu$ and $b_{12}^\mu$ components) such polynomials
must be proportional to a combination  $m_1^2 m_2^2( c_{M^2} (m_1+m_2)^2 +  c_{m_1 m_2} m_1 m_2)$.
The first combination (with coefficient $ c_{M^2}$) is forbidden by the $m_a^3$ condition. However, the
second combination, namely $c_{m_1 m_2} m_1^3 m_2^3$ is compatible with the $m_a^3$ condition.
The conclusion is that at order $G^5$ there are two different types of contributions that can be added
to any specific solution of all the conditions, namely
\bea \label{Dpadd}
\Delta { p_{1}^{ \mu}}^{\rm rr \, remain \, add} &=&\frac{G^5m_1^3 m_2^3 }{b^5 } \left( f_{1-2}^{ G^5}(\g) (u_1^\mu- u_2^\mu)\right.\nonumber\\
& +& \left.  f_{b}^{ G^5}(\g) {\hat b}_{12}^\mu \right)\,,
\eea
involving two, a priori unconstrained, functions of $\g$: $ f_{1-2}^{ G^5}(\g) $ and $ f_{b}^{ G^5}(\g) $.

We show below how to construct a particular solution of all the constraints. The general solution
at order $G^5$ is then obtained by adding the  specific ($\propto m_1^3 m_2^3$) additional terms
displayed in Eq. \eqref{Dpadd}.

\subsection{Determining the unique transverse components $\Delta  p_{a b}^{\rm rr \, remain}$ and $\Delta  p_{a  b}^{\rm rr }$ at 4PM.}

For definiteness, we henceforth consider the impulse of the first particle, $a=1$.
It is easily seen from its definition in Eq. \eqref{DpaPx} that, at order $G^4$, the only non-zero component of
$\Delta  p_{1 \mu}^{\rm rr \, P^{\rm rad}_x }$ is the one along  ${\hat b}_{12}^\mu$, say
\beq
\Delta  p_{1 \, b}^{\rm rr \, P^{\rm rad}_x }\equiv \Delta  p_{1 \mu}^{\rm rr \, P^{\rm rad}_x }{\hat b}_{12}^\mu\,,
\eeq
 which is equal to
\beq
\Delta  p_{1\, b}^{\rm rr \, P^{\rm rad}_x } = -\frac{E_1}{E_{\rm c.m.}}  P_{x  \, G^4}^{{\rm rad}}\,.
\eeq
 
 The problem to be solved is the following: given the non-polynomial term in the ${\hat b}_{12}^\mu$ component
 of $\Delta  p_{1 \mu}^{\rm rr \, P^{\rm rad}_x }$
\beq
\Delta  p_{1\, b}^{\rm rr \, P^{\rm rad}_x } = -\frac{E_1}{E_{\rm c.m.}}  P_{x  \, G^4}^{{\rm rad}}= - \frac{m_1 (m_1+\g m_2)}{M^2 h^2}  P_{x  \, G^4}^{{\rm rad}}\,,
\eeq
where $ P_{x  \, G^4}^{{\rm rad}}$is  mass-polynomial and of the type (see Eq. \eqref{PxG4})
\beq 
\label{PxG4n}
P_{x  \, G^4}^{{\rm rad}}=\frac{G^4}{b^4} m_1^2 m_2^2 (m_2-m_1) p^{G^4}_x(\g)\,,
\eeq
 what type of extra contribution $\Delta  p_{1 \, b}^{\rm rr \, remain}\equiv \Delta  p_{1 \mu}^{\rm rr \, remain }{\hat b}_{12}^\mu $ (satisfying the constraints discussed above) 
can be added to it to guarantee that the sum becomes polynomial in the masses. 

It is easily seen that
\bea
\label{G4_b_remain}
\Delta  p_{1  b  \, G^4}^{\rm rr \, remain} &=& \frac{G^4}{b^4} m_1^2 m_2^2 \frac{m_1 E_2+m_2 E_1}{E} p^{G^4}_x(\g)\nonumber\\
&=&  \frac{G^4}{b^4} m_1^2 m_2^2 \frac{m_1 m_2 (\g+1)}{M h^2} p^{G^4}_x(\g)
\,,\nonumber\\
\eea
satisfies the needed constraints (symmetry, $\propto m_1^3$) and solves the problem at hand. Indeed,
\beq
 -\frac{E_1}{E_{\rm c.m.}}  P_{x  \, G^4}^{{\rm rad}}+ \Delta  p_{1  b  \, G^4}^{\rm rr \, remain} = 
 +  \frac{G^4}{b^4} m_1^3 m_2^2  p^{G^4}_x(\g)\,.
\eeq
As proven above this solution is unique.

Therefore, we have proven that the full radiation-reaction contribution to the impulse (including the
time-even contribution  $\Delta  p_{1  \mu  \, G^4}^{\rm rr \, \Delta c_\phi }$ and
the nonlinear one $ \Delta  p_{1  \mu \, G^4}^{\rm rr \, nonlin }$) is given by
\beq \label{Dp1rr1}
\Delta  p_{1 \mu \, G^4}^{{\rm rr} }=  \Delta  p_{1 \mu \, G^4}^{\rm rr \, lin-odd } +  \frac{G^4}{b^4} m_1^3 m_2^2  p^{G^4}_x(\g) {\hat b}_{12}^\mu\,,
\eeq
or, equivalently (using the definition Eq. \eqref{PxG4} of $p^{G^4}_x(\g)$)
\beq \label{Dp1rr2}
\Delta  p_{1 \mu \, G^4}^{{\rm rr} }=\Delta  p_{1 \mu \, G^4}^{\rm rr \, lin-odd } +   \frac{m_1}{m_2-m_1} P_{x  \, G^4}^{{\rm rad}} {\hat b}_{12}^\mu \,.
\eeq
In other words, the full, 4PM-level, transverse impulse of the first particle reads
\bea \label{Dp1tot1}
 \Delta  p_{1 b \, G^4}&=&   \Delta  p_{1 b \, G^4}^{\rm cons }+\Delta  p_{1 b \, G^4}^{\rm rr \, lin-odd } +  \frac{G^4}{b^4} m_1^3 m_2^2  p^{G^4}_x(\g) \nonumber \\
 &=&  \Delta  p_{1 b \, G^4}^{\rm cons }+\Delta  p_{1 b \, G^4}^{\rm rr \, lin-odd } +  \frac{m_1}{m_2-m_1} P_{x  \, G^4}^{{\rm rad}}\,.\nonumber\\
\eea
The latter equation corresponds to Eq. (18) in Ref. \cite{Manohar:2022dea}, with the value $\frac{G^4}{b^4} m_1^3 m_2^2  p^{G^4}_x(\g)$ for the (undefined) term denoted $\frac{G^4}{b^4} \nu M^5 c_{b, 4}^{\rm rr,even}$ there.
Note that our reasoning has given a direct relation between this term and the value of
$P_{x  \, G^4}^{{\rm rad}}$, namely
\beq
\frac{G^4}{b^4} m_1^3 m_2^2  p^{G^4}_x(\g)= \frac{m_1}{m_2-m_1} P_{x  \, G^4}^{{\rm rad}}\,.
\eeq
 
Our results above yield only the beginning of the PN expansion of the function $ p^{G^4}_x(\g)$, namely
\bea
\label{pxvsP4x}
 p^{G^4}_x(\g)&=&\frac{h^4P_{x4}}{(\gamma^2-1)^2}\nonumber\\
&=&\pi\left(-\frac{1491}{400}p_\infty^3+\frac{26757}{5600}p_\infty^5
+O(p_\infty^7)\right)\,.\nonumber\\
\eea
Concerning the first term,  $\Delta  p_{1 b \, G^4}^{\rm rr \, lin-odd }$, its general expression
 as a function of  $ E_{ G^3}^{{\rm rad}}$, $ J_{ G^2}^{{\rm rad}}$ and $ J_{ G^3}^{{\rm rad}}$ was derived 
in Eq. (7.16) of  Ref. \cite{Bini:2021gat}. At the time, only $ E_{ G^3}^{{\rm rad}}$ \cite{Bern:2021dqo,Herrmann:2021tct} and  $ J_{ G^2}^{{\rm rad}}$ \cite{Damour:2020tta} were known (in a PN-exact sense). Since then, the exact value of $ J_{ G^3}^{{\rm rad}}$ has
been obtained in Ref. \cite{Manohar:2022dea}. This leads to the following exact value of $\Delta  p_{1 b \, G^4}^{\rm rr \, lin-odd }$:
\beq \label{Dp1G4linodd}
\Delta  p_{1 b \, G^4}^{\rm rr \, lin-odd }= \frac{G^4}{b^4}m_1^2m_2^2\left[ C_{b M}^{\rm 4PM}(\gamma) M + C_{b m_1}^{\rm 4PM}(\gamma)m_1 \right] \,,
\eeq
with coefficients (see Eq. (7.31) of Ref. \cite{Bini:2021gat},  and Eq. (19) of Ref. \cite{Manohar:2022dea}) 
\bea
 C_{b M}^{\rm 4PM}(\gamma) &=&   \pi \widehat{\mathcal E} \frac{\gamma(6\gamma^2-5)}{(\gamma^2-1)^{3/2}} 
-\pi \frac34 \widehat J_2  \frac{(5\gamma^2-1)}{(\gamma^2-1)^{3/2}}\nonumber\\
&&
-\widehat J_3 \frac{(2\gamma^2-1)}{(\gamma^2-1)^2}
\,, \nonumber\\
C_{b m_1}^{\rm 4PM}(\gamma)&=& -\pi \widehat{\mathcal E}\frac{2\gamma^2-1}{(\gamma+1)\sqrt{\gamma^2-1}}\,.
\eea
Here, $\widehat{\mathcal E}={\mathcal E}/\pi $, $\widehat J_2 = 2(2\gamma^2-1)(\gamma^2-1)^{1/2}{\mathcal I}$
(with ${\mathcal I}$ defined in \cite{Damour:2020tta}), and $\widehat J_3 = (\gamma^2-1)({\mathcal C}+2{\mathcal D})$ 
(with ${\mathcal C}$ and ${\mathcal D}$ defined in \cite{Manohar:2022dea}).

When separating out the 4PM conservative
contribution  $\Delta  p_{1 b \, G^4}^{\rm cons }$ \cite{Bern:2021yeh,Dlapa:2022lmu} from
the ${\hat b}_{12}^\mu$-projected impulse in our Eq. \eqref{Dp1tot1}, the term
$\Delta  p_{1 b \, G^4}^{\rm rr \, lin-odd }$
coincides with the term $c^{(4) \rm diss}_{\rm 1b, 1rad}$ in Eq. (15) of \cite{Dlapa:2022lmu},
while the remaining term $\frac{G^4}{b^4} m_1^3 m_2^2  p^{G^4}_x(\g)$ has the same mass
structure as the term $c^{(4) \rm diss}_{\rm 1b, 2rad}$ in Eq. (16) of \cite{Dlapa:2022lmu}.
Moreover, not only the first two terms in the PN expansion of $c^{(4) \rm diss}_{\rm 1b, 2rad}$ 
given in Eq. (16) of \cite{Dlapa:2022lmu} agree with those given by inserting our PN-derived result
Eq. \eqref{pxvsP4x} in the last term in Eq. \eqref{Dp1tot1}, but the PN-exact value of 
$c^{(4) \rm diss}_{\rm 1b, 2rad}$  \cite{Dlapa:2022lmu} satisfies  
the exact relation $c^{(4) \rm diss}_{\rm 1b, 2rad}
=  \frac{m_1}{m_2-m_1} P_{x  \, G^4}^{{\rm rad}}$ derived here between this remaining term
and the $x$ component of the radiated momentum.

\subsection{High-energy behavior of  $ \Delta  p^{G^4}_{1 b}$}

Let us  remark in passing that, if one considers the result Eq. \eqref{Dp1tot1}, the mass-scaling of the term
$ \frac{G^4}{b^4} m_1^3 m_2^2  p^{G^4}_x(\g)$ makes it impossible to tame the high-energy behavior
of $ \Delta  p^{G^4}_{1 b}$. 

When considering the high-energy (HE) limit $\g \to \infty$ for a fixed
 value of the scattering angle $\chi_1 \sim \frac{G E_{\rm c.m.}}{b}$, with $E_{\rm c.m.} = M h \propto \g^{\frac12}$, 
  one would expect, in this limit, (suitably scaled\footnote{E.g., one should consider the ratio $\Delta  p_{1 b }/P_{\rm c.m.}$.}) 
  scattering observables to admit a finite limit.  
 If the formal $ G \to 0$ limit commuted with the HE limit, this would imply, in particular, that each term
 in the PM expansion of the impulse would admit a finite HE limit (at fixed $\chi_1 \sim \frac{G E_{\rm c.m.}}{b}$).
 This is the case at orders $G^1$ and $G^2$. At the $G^3$ level, the conservative 
  contribution  $ \Delta  p^{\rm cons, G^3}_{1 b}/P_{\rm c.m.}$ \cite{Bern:2019nnu} is logarithmically larger
  than its expected contribution $\sim \chi_1^3$. However, it was  found \cite{Damour:2020tta,DiVecchia:2021ndb} 
  that this logarithmic divergence is tamed when completing the conservative impulse by the radiative correction
  $ \Delta  p^{\rm rr, G^3}_{1 b}$. This raises the hope that a similar taming might occur at order $G^4$.
  
  At order $G^4$ the ratio $\Delta  p^{\rm cons, G^4}_{1 b }/(P_{\rm c.m.} \chi_1^4)$ is power-law divergent,
  being proportional to $\g^{\frac12}$. In terms of the un-rescaled impulse this divergence is
   $\Delta  p^{\rm cons, G^4}_{1 b } \propto \g^3$. Parametrizing the various contributions to the HE limit
   of the impulse according to
   \beq
  \Delta  p^{X, G^4}_{1 b } \approx \frac{G^4 m_1^2 m_2^2}{b^4}\pi  C^{X, G^4}  \g^3 \,,
  \eeq
  the coefficient entering the conservative contribution $\Delta  p^{\rm cons, G^4}_{1 b }$ is
  \beq
  C^{\rm cons, G^4} = -\frac{105}{8}  (4\ln(2)-1+4\ln(2)^2) (m_1+m_2) \,.
  \eeq
  As pointed out in \cite{Manohar:2022dea}, the linear-response radiative contribution
$\Delta  p_{1 b \, G^4}^{\rm rr \, lin-odd }$  is similarly $\propto \g^3$. However, the corresponding coefficient is
\beq
 C^{\rm rr, lin, G^4} = \frac{35}{4 }[m_1 (1+8\ln(2))+2 m_2 (1+5\ln(2))]\,,
\eeq
which has the correct sign, but not the correct value to cancell the ``bad"  high-energy behavior of 
the conservative contribution. If we assume that the function $p_x^{G^4}(\gamma)$ entering our 
additional contribution has a HE behavior of the type $p_x^{G^4}(\gamma) \approx\pi c_{x}\gamma^3 $,
it will contribute another term of order $\g^3$, with a coefficient  
\beq
 C^{\rm rr, p_x, G^4}= c_x m_1\,.
\eeq
It is, however, easy to see that, whatever the value of $c_x$, such an additional term (proportional only to $m_1$)
cannot tame the contribution proportional to $m_2$, i.e. cannot yield a vanishing total coefficient $ C^{\rm tot, G^4}= C^{\rm cons, G^4}+C^{\rm rr, lin, G^4}+C^{\rm rr, p_x, G^4}$.
Indeed, the latter turns out to be\footnote{The recent result of Ref. \cite{Dlapa:2022lmu} happens to lead
to a coefficient $c_x$ which precisely annull the coefficient of $m_1$ in $C^{\rm tot, G^4}$. }
\bea
 C^{\rm tot, G^4}&=&
\left(\frac{35}{2}\ln(2)+\frac{175}{8}-\frac{105}{2}\ln(2)^2+c_x\right)m_1\nonumber\\
&+&
\left(35\ln(2)+\frac{245}{8}-\frac{105}{2}\ln(2)^2\right)m_2 
\,.
\eea
In order to tame the HE behavior of $ \Delta  p^{\rm tot, G^4}_{1 b } $, i.e. to reduce it from $\g^3$ to, say, 
$\g^2$, or $\g^2 \ln \g$ \footnote{Such a reduction, would ensure the HE vanishing of the ratio  
$\Delta  p^{\rm tot, G^4}_{1 b }/(P_{\rm c.m.} \chi_1^4)$, as expected from the structure of the 
massless scattering discussed in Ref. \cite{Amati:1990xe}.}  one would need to add a suitable extra
contribution of the (disallowed) symmetric type $\frac{G^4}{b^4}m_1^2 m_2^2(m_1+m_2) f_{\rm sym}(\g)$.

We do not view the inability of the additional term to tame the HE behavior of the {\it $G$-expanded} impulse
as a blemish. It seems indeed probable  that the $ G \to 0$ limit does not commute with the HE limit $\g \to \infty$.
This is notably indicated by the studies of the HE limit of the total gravitational-wave energy emitted during
the collision of massless particles \cite{Gruzinov:2014moa,Ciafaloni:2015xsr,DiVecchia:2022nna}. 
While the HE limit of the $O(G^3)$ leading-order
radiative energy loss exceeds the energy $E_{\rm c.m.}$ available in the system by a factor $\propto \g^{\frac12}$, the works
\cite{Gruzinov:2014moa,Ciafaloni:2015xsr,DiVecchia:2022nna} suggest that (due to coherence effects in the beamed radiation) the HE
limit of radiative losses is finite, and of order $\chi_1^2 \ln \frac1{\chi_1}$.

\subsection{Longitudinal components of  $\Delta  p_{1 \mu}^{\rm rr }$ at 4PM}

To end our discussion of the radiative contributions to the  impulse of the first particle $\Delta  p_{1 \mu}^{\rm rr }$,
let us also consider its longitudinal components, i.e. the components along $u_{1-}$ and $u_{2-}$. We have shown above
that the only source of nonpolynomiality 
(namely the $P_x$-related contribution $\Delta  p_{1 \mu}^{\rm rr \, P^{\rm rad}_x }$)
does not contribute to the longitudinal components. In addition, we have shown that there was, at the 4PM level,
 a unique value of  $\Delta  p_{1 \mu}^{\rm rr }$ satisfying all the needed constraints. 
Namely, the one given by Eq. \eqref{Dp1rr1} or \eqref{Dp1rr2}.

 In view of Eq. \eqref{Dp1rr1}, at order $G^4$, the longitudinal components of $\Delta  p_{1 \mu}^{\rm rr }$ are
 fully described by the time-odd-linear-response formula of Ref. \cite{Bini:2021gat}, i.e.  the term denoted
 $\Delta p^{\mu \rm rr \, lin-odd}_{1 }$ above. Using the notation of Ref. \cite{Bini:2021gat}, its
 longitudinal components are defined as follows:
 \bea
 \Delta p^{\mu \rm rr \,  longit}_{1 } &=& \Delta p^{\mu \rm rr \, lin-odd \, longit}_{1 }\nonumber\\
 &=& c^{1 \rm rr}_{u_1} u_{1 -}^\mu +  c^{1 \rm rr}_{u_2} u_{2 -}^\mu \nonumber\\
  &=& c^{1 \rm rr,  lin-odd }_{u_1} u_{1 -}^\mu +  c^{1 \rm rr,  lin-odd }_{u_2} u_{2 -}^\mu \,.\nonumber\\
 \eea
 Using the expressions given in Table II of Ref. \cite{Bini:2021gat}\footnote{We also use  Eq. (7.26) there to replace the original expression in terms of
the 4PM component $P_4$ of the $y$ component, $P^{\rm rad}_y$, of the recoil in terms of the
rescaled 4PM component $\tilde E_4$ of the energy loss.}, 
 we find that the coefficients $c_{u_1}^{1\,\rm  rr}$ and $c_{u_2}^{1\,\rm  rr}$ are given by
 \begin{widetext}
 \bea
c_{u_1}^{1\,\rm  rr,  \, 4PM}&=&-\frac{G^4 m_1^2m_2^2}{b^4(\gamma^2-1)^3}\left[\gamma M \tilde E_{4}^0+\frac12 \gamma m_1 \tilde E_{4}^1+2(2\gamma^2-1)^2(m_1\gamma+m_2)\hat J_2\right]
\,,\nonumber\\
c_{u_2}^{1\,\rm  rr, \, 4PM}&=&\frac{G^4 m_1^2m_2^2}{b^4(\gamma^2-1)^3}\left[ M \tilde E_{4}^0+\frac12 m_1 \tilde E_{4}^1+2(2\gamma^2-1)^2(m_2\gamma+m_1)\hat J_2\right]\,,
\eea
\end{widetext}
where $\tilde E_{4}^0$, and $\tilde E_{4}^1$ (defined by $h^5E_4=\tilde E_{4}^0+\nu \tilde E_{4}^1$), 
as well as $\hat J_2 \equiv h^2J_2$ are all functions only of $\gamma$. [See Eq. (8) of \cite{Dlapa:2022lmu}
for the exact value of $h^5E_4$.]

The combination 
\beq
b^4\left(c_{u_1}^{1\,\rm  rr, 4PM}+\gamma c_{u_2}^{1\,\rm  rr, 4PM}\right)= 
m_1^2m_2^3 \frac{2 \hat J_2 (2\gamma^2-1)^2}{(\gamma^2-1)^2} 
\,,
\eeq
coincides with the impulse coefficient $c^{(4) \rm diss}_{ 1\check{u}_1, \rm 1rad}$ given in Eq. (15) of  Ref. \cite{Dlapa:2022lmu}.
The other combination
\bea \label{cchecku2}
&& b^4\left(c_{u_2}^{1\,\rm  rr, 4PM}+\gamma c_{u_1}^{1\,\rm  rr, 4PM}\right)=-\frac{m_1^2m_2^2}{(\gamma^2-1)^2}  \left[m_1\left(\tilde E_{4}^0\right.\right.\nonumber\\ 
&&\left.\left.+\frac12\tilde E_{4}^1+2(2\gamma^2-1)^2 \hat J_2\right) +m_2\tilde E_{4}^0\right]\,,
\eea 
coincides with the sum $c^{(4) \rm diss}_{ 1\check{u}_2, \rm 1rad}+c^{(4) \rm diss}_{ 1\check{u}_2, \rm 2rad}$  
of the two $\check{u}_2$-type impulse coefficients given in Eqs.  (15)--(16) in Ref. \cite{Dlapa:2022lmu}.
More precisely, the part called  $c^{(4) \rm diss}_{ 1\check{u}_2, \rm 1rad}$ corresponds to the part of the
right-hand side of Eq. \eqref{cchecku2} featuring {\it odd} powers of $p_\infty$ in its PN expansion, while
the part called  $c^{(4) \rm diss}_{ 1\check{u}_2, \rm 2rad}$ 
corresponds to the part of the
right-hand side of Eq. \eqref{cchecku2} featuring {\it even} powers of $p_\infty$ in its PN expansion
(the latter part is the one generated by the tail contribution to the radiated energy).

\subsection{Radiative contribution to the impulse coefficients at 5PM: transverse component}

As in the above discussion of the impulse at 4PM, it is convenient to project the various radiative
contributions (labelled by $X={\rm rr \, lin-odd }, {\rm rr \, P^{\rm rad}_x }, {\rm rr \, remain }$) to the impulse,
\beq \label{newdecompDpa2}
\Delta  p_{a \mu}^{\rm rr }= \Delta  p_{a \mu}^{\rm rr \, lin-odd } +  \Delta  p_{a \mu}^{\rm rr \, P^{\rm rad}_x }+ \Delta  p_{a \mu}^{\rm rr \, remain }\,,
\eeq
on the basis given in  Eq. \eqref{decompPmu}.
For instance, for $a=1$, the transverse ($ {\hat b}_{12}^\mu$) component is the sum of the following contributions
\beq
c_b^{1, \rm rr }= c_b^{1, \rm rr \, lin-odd  }+ c_b^{1, \rm rr \, P^{\rm rad}_x }+ c_b^{1, \rm rr \, remain }\,.
\eeq
Similarly to what happened at 4PM, the nonpolynomial contribution generated by $c_b^{1, \rm rr \, P^{\rm rad}_x }$
reads, at the 5PM level
\beq
c_{b}^{1, \rm rr \, P^{\rm rad}_x \, G^5} = -\frac{E_1}{E_{\rm c.m.}}  P_{x  \, G^5}^{{\rm rad}}\,.
\eeq
Again, the simplest solution (satisfying all the needed constraints) for the remaining
contribution  $c_b^{1, \rm rr \, remain \, G^5 }$ to cancell the nonpolynomiality of $c_{b}^{1, \rm rr \, P^{\rm rad}_x \, G^5}$ is  
\beq
c_{b}^{1, \rm rr \, remain \, simplest, G^5} = +\frac{m_1 E_2+m_2 E_1}{(m_2 - m_1)E_{\rm c.m.}}  P_{x  \, G^5}^{{\rm rad}}\,.
\eeq
Indeed, we have
\beq
c_{b}^{1, \rm rr \, P^{\rm rad}_x \, G^5}+ c_{b}^{1, \rm rr \, remain \, simplest, G^5} = +\frac{m_1 }{(m_2 - m_1)}  P_{x  \, G^5}^{{\rm rad}}\,,
\eeq
which is polynomial in masses because $ P_{x  \, G^5}^{{\rm rad}}$ contains a factor $(m_2 - m_1)$.

As was discussed above, the most general solution for $c_b^{1 , \rm rr \, remain \, G^5 }$ is
\beq
c_{b}^{1, \rm rr \, remain , G^5} = +\frac{m_1 E_2+m_2 E_1}{(m_2 - m_1)E_{\rm c.m.}}  P_{x  \, G^5}^{{\rm rad}}
+   \frac{G^5}{b^5} m_1^3 m_2^3   f_{b}^{ G^5}(\g)\,.
\eeq
Writing  $ P_{x  \, G^5}^{{\rm rad}}$ as
\beq
 P_{x  \, G^5}^{{\rm rad}}= \frac{G^5}{b^5} m_1^2 m_2^2 (m_2-m_1) (m_1+m_2)  p_{x  \, G^5}(\g)\,,
\eeq
we finally get
\bea
&& c_{b}^{1, \rm rr \, P^{\rm rad}_x \, G^5}+ c_{b}^{1, \rm rr \, remain, G^5} = \nonumber\\
&&
+\frac{G^5}{b^5} m_1^3 m_2^2  (m_1+m_2)  p_{x  \, G^5}(\g) 
+\frac{G^5}{b^5} m_1^3 m_2^3   f_{b}^{ G^5}(\g)\,.\nonumber\\
\eea
In other words, the most general 5PM transverse radiative impulse reads
\bea
\label{cb1_G5_all}
&& c_{b}^{1, \rm rr \, G^5}= c_{b}^{1, \rm rr \, lin-odd G^5}\nonumber\\
&&
+\frac{G^5}{b^5} m_1^3 m_2^2  (m_1+m_2)  p_{x  \, G^5}(\g) 
+\frac{G^5}{b^5} m_1^3 m_2^3   f_{b}^{ G^5}(\g)
\,.\nonumber\\
\eea
 Table I and Table II of Ref. \cite{Bini:2021gat} gave exact expressions for $c_{b}^{1, \rm rr \, lin-odd G^5}$    in terms of $E_n$ and $J_n$ with  $n\leq 4$. 
However, the PN-exact value of $J_4$ is  unknown so that our 5.5PN accurate determination of $J_4$ currently limits the knowledge of $c_{b}^{1, \rm rr \, lin-odd G^5}$  to the 5.5PN level.
We so find
\begin{widetext}

\bea
c_{b}^{1, \rm rr \, lin-odd G^5}&=&\frac{G^5m_1^2m_2^2}{b^5} 
\left\{ \frac{416}{45}(m_1+m_2)^2\frac{1}{p_\infty^4}\right.\nonumber\\
&+&
\left[\left(\frac{169664}{1575}-\frac{47}{5}\pi^2\right) m_1^2+\left(\frac{409888}{1575}-\frac{94}{5}\pi^2\right) m_1 m_2+\left(\frac{203264}{1575}-\frac{47}{5}\pi^2\right) m_2^2\right]\frac{1}{p_\infty^2}\nonumber\\
&-&
\frac{896}{45} (m_1+m_2)^2\frac{1}{p_\infty}\nonumber\\
&+&
\left(\frac{159232}{3675}-\frac{1243}{56}\pi^2\right) m_1^2+\left(\frac{2283544}{11025}-\frac{1489}{35}\pi^2\right)  m_1m_2+\left(\frac{116992}{1225}-\frac{5697}{280}\pi^2\right) m_2^2\nonumber\\
&+&
\left(\frac{8384}{45} m_1^2+\frac{694016}{1575} m_1 m_2+\frac{10304}{45} m_2^2\right)p_\infty\nonumber\\
&+&
\left[\left(\frac{241}{120}\pi^2-\frac{22294592}{363825}\right) m_1^2+\left(\frac{67876972}{363825}+\frac{23783}{3360}\pi^2\right)m_1m_2+\left(-\frac{9728}{275}+\frac{3407}{672}\pi^2\right) m_2^2\right] p_\infty^2
\nonumber\\
&+& 
O(p_\infty^3)\bigg\}\,.
\eea
\end{widetext}
The second contribution in  $c_{b}^{1, \rm rr \, G^5}$ is known to 6.5PN absolute accuracy, because our results above give the following 6.5PN-accurate value of $p_{x  \, G^5}(\g)$
\beq
\label{Px_G5}
 p_{x  \, G^5}(\g) = -\frac{20608}{225} p_\infty+\frac{1143232}{7875}p_\infty^3-\frac{196096}{945} p_\infty^4+O(p_\infty^5)\,.
\eeq
By contrast, the only thing we know at this stage concerning the additional contribution  $\propto f_{b}^{ G^5}(\g)$ in Eq. \eqref{cb1_G5_all} is that it could start at the 5PN level and be $f_{b}^{ G^5}(\g) = O(\pinf)$. 

The latter result limits the PN accuracy of $c_{b}^{1, \rm rr \, G^5}$. However, more is known about the sum\footnote{Note that the sum becomes a difference if one exchanges $\hat b_{12}$ into $\hat b_{21}$.} 
$c_{b_{12}}^{1, \rm rr \, G^5}+c_{b_{12}}^{2, \rm rr \, G^5}$, in which the $f_b-$ term cancels.   Indeed, the  linear-odd contribution to this only depends on $E_3$ and $E_4$ (see Table II of Ref. \cite{Bini:2021gat}), which are exactly known \cite{Bern:2021dqo,Dlapa:2022lmu}. 
The beginning of its PN expansion reads
\bea
&&c_{b}^{1, \rm rr \, lin-odd G^5}+c_{b}^{2, \rm rr \, lin-odd G^5}=\nonumber\\
&& 
\frac{G^5}{b^5} m_1^2 m_2^2  (m_1+m_2)(m_1-m_2) \times\nonumber\\
&&
\left[-\frac{64}{3 p_\infty^2}  
-\left(\frac{37}{20}\pi^2+\frac{27392}{525}\right)
-\frac{128}{3} p_\infty \right.\nonumber\\
&&\left.
-\left(\frac{856768}{33075}+\frac{3429}{1120}\pi^2\right)p_\infty^2
+O(p_\infty^3)\right]\,.\qquad
\eea
The second contribution is known to 6.5PN accuracy by using Eq. \eqref{Px_G5}, and reads
\beq
\frac{G^5}{b^5} m_1^2 m_2^2  (m_1+m_2)(m_1-m_2)  p_{x  \, G^5}(\g)\,,
\eeq
where the 6.5PN value of $p_{x  \, G^5}(\g)$ is given in Eq. \eqref{Px_G5} above.

\subsection{Radiative contribution to the impulse coefficients at 5PM: longitudinal components}

Let us finally consider the nonpolynomial contributions to the $u_{1-} \pm u_{2-}$ components of 
$\Delta  p_{1\, }^{\rm rr \, P^{\rm rad}_x } $:
\begin{widetext}
\bea
c_{1+2}^{1\,\rm  rr,  \,  P_x\, G^5}&=&\frac{G^5}{2}\frac{m_1^3m_2^2}{b^5}(m_1-m_2)\frac{(2\gamma^2-1)(\gamma-1)}{(\gamma^2-1)^{3/2}}\frac{-2m_2^2\gamma+m_1^2-2m_1m_2-m_2^2}{(m_1^2+ 2 \g m_1 m_2 +m_2^2) }  p^{G^4}_x(\g)
\,,\nonumber\\
c_{1-2}^{1\,\rm  rr,  \,  P_x\, G^5}&=&\frac{G^5}{2}\frac{m_1^3m_2^2}{b^5}(m_1-m_2)\frac{(2\gamma^2-1)(\gamma+1)}{(\gamma^2-1)^{3/2}}\frac{2m_2^2\gamma-m_2^2+2m_1m_2+m_1^2}{(m_1^2+ 2 \g m_1 m_2 +m_2^2 )}  p^{G^4}_x(\g)
\,,
\eea 
\end{widetext}
where $ p^{G^4}_x(\g)$ is the same function of $\g$ as defined above, Eq. \eqref{pxvsP4x}.

As before, we look for corresponding components of $ \Delta  p_{a \mu}^{\rm rr \, remain }$ that will
cancel the nonpolynomiality of the above longitudinal components. As discussed above, there is a unique
way to do so for the $u_{1-} + u_{2-}$ component, while the $u_{1-} - u_{2-}$ component is non unique, and
 can be augmented by a term of the form (see Eq. \eqref{Dpadd})
\beq \label{Dpadd1-2}
\Delta c_{1-2}^{1\,\rm  rr, remain}=\frac{G^5m_1^3 m_2^3 }{b^5 }  f_{1-2}^{ G^5}(\g)  \,.
\eeq

Let us start by considering the $u_{1-}+u_{2-}$ component,  $c_{1+2}^{1\,\rm  rr,  \,  P_x\, G^5}$,
and look for an additional mass-antisymmetric contribution $c_{1+2}^{1\,\rm  rr, remain}$ able
to cancel the nonpolynomiality of $c_{1+2}^{1\,\rm  rr,  \,  P_x\, G^5}$.
After scaling out
\beq
\frac{G^5m_1^2 m_2^2 }{2 b^5 }  (2\gamma^2-1)(\gamma-1)(\gamma^2-1)^{-3/2}  p^{G^4}_x(\g)\,,
\eeq
 and multiplying by $m_1^2+ 2 \g m_1 m_2 +m_2^2 $,
the problem to be solved involves quartic polynomials in the masses. Namely, we look for a rescaled
\beq
{\hat c}_{1+2}^{1\,\rm  rr, remain}= c_+ (m_1 - m_2) (m_1 + m_2) m_1  m_2\,,
\eeq
and  two coefficients $x, y$, such that $c_+, x , y$  satisfy the mass-polynomial equation
 \bea \label{eqG5as}
 && m_1 (m_1 - m_2) (  m_1^2 - 2  m_1 m_2 - (2 \g + 1)  m_2^2) \nonumber\\
&&+ 
 c_+ (m_1 - m_2) (m_1 + m_2) m_1  m_2 \nonumber\\
&&- (m_1^2  + 2 \g  m_1 m_2+ m_2^2) (x \, m_1^2 + 
    y \,  m_1  m_2)=0\,.\nonumber\\
\eea
Here, we imposed the constraint that the resulting contribution to $\Delta c_{1+2}^{1 \, \rm rr}$ be $\propto m_1^3$.

It is easily found that the mass-polynomiality Eq. \eqref{eqG5as} admits a unique solution, namely
\beq
c_+= 2(\g+1) \, ; \, x=1\,;\, y=-1\,.
\eeq
\begin{widetext}
This proves that 
\beq
c_{1+2}^{1\,\rm  rr, remain} = \frac{G^5 m_1^2 m_2^2}{b^5} \frac{m_1 m_2 (m_1-m_2)(m_1+m_2)(2\gamma^2-1)}{(\gamma^2-1)^{1/2}(m_1^2+2m_1 m_2\gamma+m_2^2)}p^{G^4}_x(\g)\,,
\eeq
and therefore that
\beq
c_{1+2}^{1\,\rm  rr \, G^5} = c_{1+2}^{1\,\rm  rr, lin-odd} +\frac{G^5 m_1^2 m_2^2}{b^5} m_1 (m_1-m_2)\frac{(\gamma-1)(2\gamma^2-1)}{2(\gamma^2-1)^{3/2}}  p^{G^4}_x(\g)\,.
\eeq
Proceeding in a similar way for the particle-symmetric  $u_{1-} - u_{2-}$ component, we find as general solution
for $c_{1-2}^{1\,\rm  rr, remain}$
\beq
\label{G5_1min2_remain}
c_{1-2}^{1\,\rm  rr, remain} = -2\frac{G^5m_1^2m_2^2}{b^5} \frac{m_1^2m_2^2(2\gamma^2-1)(\gamma+1)}{(m_1^2+2m_1m_2\gamma+m_2^2)\sqrt{\gamma^2-1}} p^{G^4}_x(\g) + \frac{G^5m_1^3 m_2^3 }{b^5 }  f_{1-2}^{ G^5}(\g)\,,
\eeq
and therefore that
\beq
c_{1-2}^{1\,\rm  rr \, G^5} = c_{1-2}^{1\,\rm  rr, lin-odd} +\frac{G^5m_1^2m_2^2}{b^5} \frac{m_1(m_1+m_2-2m_2\gamma) (2\gamma^2-1)(\gamma+1) }{2(\gamma^2-1)^{3/2}}  p^{G^4}_x(\g)+ \frac{G^5m_1^3 m_2^3 }{b^5 }  f_{1-2}^{ G^5}(\g)\,.
\eeq
\end{widetext}
At this stage, the constraints we used above leave undetermined the additional longitudinal term involving
the function $f_{1-2}^{ G^5}(\g)$ (in addition to the function   $f_{b}^{ G^5}(\g)$ entering the transverse
component).

However, we still have  one more constraint that we can use, namely the mass-shell-related constraints, Eqs. \eqref{mass_shell_smt} and \eqref{mass_shell_smt2}. 
When using our new decomposition the following analogue of Eq. \eqref{mass_shell_smt} holds (because the $\Delta c_\phi$ contribution vanishes separately):
\beq
\label{mass_shell_smt3}
p_{ a \mu}^{+\rm cons} (\Delta  p^\mu_{a }{}^{\rm rr \, lin-odd } +  \Delta  p^\mu_{a }{}^{\rm rr \, P^{\rm rad}_x })=0\,.
\eeq
The analogue of Eq. \eqref{mass_shell_smt2} then reads
\beq
p_a^{+\rm cons}\cdot \Delta p_a^{\rm rr\, remain}=-\frac12 (\Delta p_a^{\rm rr\, tot})^2\,,
\eeq
where $\Delta p_a^{\rm rr\, tot}$ is the full (nonlinear) radiative impulse, as determined above at orders $G^4$ and $G^5$.
\beq
\label{massshellconstraint}
\Delta p_a^{\rm rr\, tot}=\Delta  p_{a }^{\rm rr \, lin-odd } +  \Delta  p_{a }^{\rm rr \, P^{\rm rad}_x }+ \Delta  p_{a }^{\rm rr \, remain }\,.
\eeq 
Since $\Delta p_a^{\rm rr\, tot}$ starts at order $G^3$, the right-hand-side of Eq. \eqref{massshellconstraint}  starts at order $G^6$.
Inserting the decomposition (for $a=1$)
\bea
\Delta  p_{1}^{\rm rr \, remain }&=&c_{b1}^{\rm rr \, remain }\hat {\mathbf b}+c^1_{1+2}{}^{\rm rr \, remain }(u_{1-}+u_{2-})\nonumber\\
&+&c^1_{1-2}{}^{\rm rr \, remain }(u_{1-}-u_{2-})
\eea
in Eq. \eqref{mass_shell_smt3}, we find
\bea
&&p_1^{+\rm cons} \cdot \Delta  p_{1 }^{\rm rr \, remain }
=-c_{b1}^{\rm rr \, remain }p_- \sin (\chi_{\rm cons})\nonumber\\
&&+\frac{p_-^2}{m_1m_2} \cos  (\chi_{\rm cons})\left[  c_{1+2}^{1\,\rm  rr, remain}(m_2-m_1)\right.\nonumber\\
&&\left.+c_{1-2}^{1\,\rm  rr, remain}(m_1+m_2)\right]\,.\nonumber\\
\eea
Here we  used  
\bea
p_1^{+\rm cons}\cdot u_{1-}&=&\frac{p_-^2}{m_1}\cos (\chi_{\rm cons})\,,\nonumber\\ 
p_1^{+\rm cons}\cdot u_{2-}&=&-\frac{p_-^2}{m_2}\cos (\chi_{\rm cons})\,,\nonumber\\
p_1^{+\rm cons}\cdot \hat {\mathbf b}&=&-p_- \sin(\chi_{\rm cons})\,.
\eea
Working up to order $G^5$ we find
\bea
&&0=-c_{b1}^{\rm rr \, remain G^4 } \frac{2\chi_{1\rm cons}}{j}+\frac{p_-}{m_1m_2} \bigg[ \nonumber\\
&&\left.  c_{1+2}^{1\,{\rm  rr, remain}, G^5}(m_2-m_1)+c_{1-2}^{1\,{\rm  rr, remain}, G^5}(m_1+m_2)\right]\,,\nonumber\\
\eea
which determines the value of $f_{1-2}^{ G^5}(\g)$, namely
\beq
f_{1-2}^{ G^5}(\g)=\frac{(2\gamma^2-1)(\gamma+1)}{(\gamma-1)\sqrt{\gamma^2-1}}p_{x  \, G^4}(\g)\,.
\eeq
Consequently,
\bea
c_{1-2}^{1\,\rm  rr \, G^5}& =& c_{1-2}^{1\,\rm  rr, lin-odd}\nonumber\\ 
&+&\frac{G^5 m_1^3m_2^2(m_1+3m_2)}{b^5}\frac{(2\gamma^2-1)(\gamma+1)}{2(\gamma^2-1)^{3/2}}p_{x  \, G^4}(\g)\,.\nonumber\\
\eea

\section{Concluding remarks}

 In the present work, we improved the knowledge of radiative contributions to scattering observables in several directions.

We pushed the PN accuracy of the energy, angular momentum and linear-momentum radiated during a scattering encounter
 to higher levels, namely, the {\it fractional} 3PN accuracy: for energy we reached the absolute 5.5PN accuracy (see Eqs. \eqref{E_rad_inst_25PNexact}--\eqref{E_rad_inst_25PN}); for angular momentum we reached the absolute 5.5PN accuracy (see Eqs. \eqref{J_rad_inst_25PNexact}--\eqref{J_rad_inst_25PN}); for linear momentum we reached the absolute 6.5PN accuracy (see Eqs. \eqref{Pxy_rad_inst_25PNexact}, \eqref{25instP}, \eqref{DeltaPxypasttails}, \eqref{PxPy3PNinst} and \eqref{PxPy3PNall}).
See the summary of our results in Section X, and notably in Tables \ref{tab:EnJnPyn} and \ref{tab:Pxn}.

Our results have a limited PN accuracy, but are valid (at least) at order $G^7$.

We completed the linear-response computation of the radiative contribution  to the individual impulses \cite{Bini:2021gat} by including two additional terms (see  Sec. \ref{individual_impulses}): i) the additional contribution $\Delta c_\phi$ in the relative scattering angle linked to the time-asymmetric piece of teh radiation-reaction force (see Eq. \eqref{Deltacphi_eq}) and ii)  the additional contribution $\Delta p_a^{\rm rr\, nonlin}$ linked to nonlinear radiation-reaction effects.
We then  wrote the total radiative contribution to the impulses in the following form
\beq
\label{finale}
\Delta p_a^{\rm rr\, tot}=\Delta  p_{a }^{\rm rr \, lin-odd } +  \Delta  p_{a }^{\rm rr \, P^{\rm rad}_x }+ \Delta  p_{a }^{\rm rr \, remain }\,,
\eeq
with
\bea
\Delta  p_{a }^{\rm rr \, lin-odd }&=&\bar \chi^{\rm  rr \, rel } \frac{d}{d\chi_{\rm cons}}\Delta p_{a \mu}^{\rm cons} + 
\frac{\Delta P_{\rm c.m.}}{ P_{\rm c.m.}} p_{a \mu}^+ \nonumber\\
& -& \frac{m_a^2}{E_a} \frac{\Delta P_{\rm c.m.}}{ P_{\rm c.m.}} U_\mu\nonumber\\
&-& \frac{E_a}{E_{\rm c.m.}}  \bar P_\mu^{\rm rad} - \frac{(p_{a \nu}^+   \bar P^\nu_{\rm rad})}{E_{\rm c.m.}} U_\mu\,,
\eea
where
\beq
\Delta P_{\rm c.m.} =  - \frac{E_1 E_2}{E_{\rm c.m.} P_{\rm c.m.}} E_{\rm rad}\,.
\eeq
Here  $\bar \chi^{\rm  rr \, rel }$ is defined as
\beq
\bar \chi^{\rm  rr \, rel }\equiv - \left(\frac12 \frac{\partial \chi^{\rm cons}}{\partial  E  } E_{\rm rad} +\frac12 \frac{\partial \chi^{\rm cons}}{\partial  J } J_{\rm rad} \right) \,, 
\eeq
and  $\bar P^\mu_{\rm rad}$ denotes the part of $P^\mu_{\rm rad}$ orthogonal to the $x$ direction, namely
\beq
\bar P^\mu_{\rm rad}\equiv P^\mu_{\rm rad}-P^x_{\rm rad}{\mathbf e}_x^\mu\,.
\eeq
All the radiative losses (in $E_{\rm rad}$, $J_{\rm rad}$ and $P^\mu_{\rm rad}$) entering here include time-asymmetric (hereditary) effects.
The second term in Eq. \eqref{finale}, $\Delta  p_{a }^{\rm rr \, P^{\rm rad}_x }$, is the contribution linked to the $x$ component of $P^\mu_{\rm rad}$, namely
\beq
\Delta  p_{a \mu}^{\rm rr \, P^{\rm rad}_x }\equiv -\frac{E_a}{E_{\rm c.m.}}  P_x^{\rm rad} e_{x \,\mu}- \frac{(p_{a x}^+   P_x^{\rm rad})}{E_{\rm c.m.}} U_\mu\,.
\eeq
Finally, the remaining contribution in the decomposition  \eqref{finale}, is 
\beq
\Delta  p_{a \mu}^{\rm rr \, remain }\equiv \Delta c_\phi  \frac{d}{d\chi_{\rm cons}} \Delta p_{a \mu}^{\rm cons}+ \Delta  p_{a \mu}^{\rm rr \, nonlin }\,.
\eeq

We studied the consequences of the  mass-polynomiality of its Lorentz-invariant form factors as defined in Eqs. \eqref{gen_rels}-\eqref{G5_rel}.
The resulting structures were shown to imply the $\nu$-polynomiality rules introduced in \cite{Bini:2021gat}. The latter $\nu$-rules ensure the mass-polynomiality of the first contribution $\Delta  p_{a }^{\rm rr \, lin-odd }$ to the impulses (see Table II of   \cite{Bini:2021gat}).
Then we showed how the non-polynomiality of the $P_x$-related contribution $\Delta  p_{a }^{\rm rr \, P^{\rm rad}_x }$ could be cured by adding specific remaining contributions $\Delta  p_{a }^{\rm rr \, remain }$. At order $G^4$ the various constraints to be satisfied by $\Delta  p_{a }^{\rm rr \, remain }$ were shown to be sufficient to fully determine  $\Delta  p_{a }^{\rm rr \, remain }$ in terms of $P^{\rm rad}_x$, see Eq. \eqref{G4_b_remain}.
At order $G^5$ $\Delta  p_{a }^{\rm rr \, remain }$ was determined up to the addition of one extra term, see Eq. \eqref{G5_1min2_remain}.
 
All our 4PM level results are compatible with those of Ref. \cite{Dlapa:2022lmu} and provide an alternative way of understanding 4PM radiation reaction effects.
Our 5PM level results give benchmarks  for future 5PM computations, and hopefully will bring a new light on the current puzzles concerning the 5PN dynamics of binary-systems \cite{Almeida:2022jrv,Blumlein:2021txe}.

\section*{Acknowledgments}
We thank Rafael Porto for helpful exchange of information.
T.D. thanks Rodolfo Russo and Gabriele Veneziano for informative discussions. 
The present research was partially supported by the  ``2021 Balzan Prize for Gravitation: Physical and
Astrophysical Aspects", awarded to Thibault Damour.
D.B. thanks ICRANet for partial support, and acknowledges sponsorship of the Italian Gruppo Nazionale per la Fisica Matematica (GNFM) of the Istituto Nazionale di Alta Matematica (INDAM).
D.B. also acknowledges discussions with Pierpaolo Mastrolia, and Gihyuk Cho,  as well as the hospitality and the highly stimulating environment of the Institut des Hautes Etudes Scientifiques.

\appendix

\section{Notation and useful formulas} \label{appA}

We list below some  useful formulas which one often needs to have at hand. 
The incoming c.m. Lorentz factor  $ \gamma =-u_1^-\cdot u_2^-$ and its associated (dimensionless) momentumlike variable $p_\infty$ are related by 
\beq
p_\infty \equiv \sqrt{\gamma^2-1}\,.
\eeq
The dimensionless angular momentum $j$ is related to the original c.m. angular momentum $J$ by 
\beq
j \equiv \frac{c J}{G m_1 m_2}\,.
\eeq
The vectorial  impact parameter (orthogonal to $ u_1^-$ and  $ u_2^-$)
 ${\mathbf b}_{12}= {\mathbf b}_{1}- {\mathbf b}_{2}=b \hat {\mathbf b}_{12}$ together with the conservative scattering angle $\chi_{\rm cons}$ enters the definition of the Cartesianlike basis vectors ${\mathbf e}_x$ and ${\mathbf e}_y$ as follows (see Eq. (3.49) of Ref. \cite{Bini:2021gat}):
\bea
\label{exydef}
{\mathbf e}_x&=& \cos \frac{\chi_{\rm cons}}{2} \hat {\mathbf b} + \sin \frac{\chi_{\rm cons}}{2}  {\mathbf n}_- \,,\nonumber\\
{\mathbf e}_y&=& -\sin \frac{\chi_{\rm cons}}{2} \hat {\mathbf b} + \cos \frac{\chi_{\rm cons}}{2}  {\mathbf n}_-\,,
\eea
where ${\mathbf n}_-$ is the direction of the incoming momenta, 
\beq
{\mathbf n}_-=\frac{m_1 m_2}{P_{\rm c.m.}^- E_{\rm c.m.}^-}\left(\frac{E_2^-}{m_2}u_1^- -\frac{E_1^-}{m_1}u_2^-\right)\,,
\eeq
and (see Eqs. (A4) and (A5) of Ref. \cite{Bini:2021gat})
\bea
P_{\rm c.m.}^-&=& \frac{m_1 m_2}{E_{\rm c.m.}^-}\sqrt{\gamma^2-1} \,,\nonumber\\
E_{\rm c.m.}^-&=& Mc^2 h =Mc^2 \sqrt{1+2\nu(\gamma-1)}\,.
\eea
An equivalent expression for ${\mathbf n}_-$ is the following
\beq
{\mathbf n}_-=\frac{(u_{2-}\wedge u_{1-})\cdot U^-}{\sqrt{\gamma^2-1}}\,,
\eeq
where the wedge product of two vectors $A$ and $B$ is standardly defined as
\beq
A\wedge B= A\otimes B-B\otimes A\,, 
\eeq
so that the  contraction with a third vector $C$ is given by $(A\wedge B)\cdot C= A (B\cdot C)-B (A\cdot C)$. 

Boldface vectors denote spatial vectors in the c.m. frame with time axis $U^-$: $p_a^-=m_a u_a^-=E_a U^- +{\mathbf p}_a^-$ (where ${\mathbf p}_a^-$ is orthogonal to $U^-$,
and ${\mathbf p}_1^-=-{\mathbf p}_2^-={\mathbf p}^-$), with
\beq
U^-=\frac{p_1^-+p_2^-}{|p_1^-+p_2^-|}=\frac{1}{E_{\rm c.m.}^-}(m_1 u_1^-+m_2 u_2^-)\,,
\eeq
and $E_{\rm c.m.}^-=E_1^-+E_2^-$.

To ease the notation, we often remove the \lq\lq c.m." label from  both energy and linear momentum, e.g., $P_{\rm c.m.}^-\to p_-$. 
The label \lq\lq $-$\rq\rq (for incoming) is also frequently omitted: $E_{\rm c.m.}^-\to E$.

Let us also recall the following expressions (see Eqs. (A9) of Ref. \cite{Bini:2021gat}) for the incoming c.m. energy of each particle
\beq
E_1^-=\frac{m_1(m_2\gamma+m_1)}{E}\,,\qquad
E_2^-=\frac{m_2(m_1\gamma+m_2)}{E}\,, 
\eeq
as well as the relation between the dimensionless angular momentum and  the impact parameter
\beq
\frac1j = \frac{GM h}{b \pinf}=  \frac{G E}{b \pinf}\,.
\eeq
When describing the conservative scattering it is useful to introduce the c.m. direction of the (conservative) outgoing
momenta, ${\mathbf n}_+^{\rm cons}$, as well as its associated orthogonal direction $\hat {\mathbf B}$, namely
\bea
\hat {\mathbf B}&=& \cos(\chi_{\rm cons})\hat {\mathbf b} +\sin (\chi_{\rm cons}){\mathbf n}_- \,,\nonumber\\
{\mathbf n}_+^{\rm cons}&=&-\sin(\chi_{\rm cons})\hat {\mathbf b} +\cos (\chi_{\rm cons}){\mathbf n}_-
\,.
\eea
In the text we used the relation
\beq
\hat {\mathbf B}= -\frac{d}{d \chi_{\rm cons}}{\mathbf n}_+^{\rm cons}\,.
\eeq
The dyad $ (\hat {\mathbf B}, {\mathbf n}_+^{\rm cons},) $ differs from the incoming dyad
$(\hat {\mathbf b}, {\mathbf n}_-)$ by a rotation of angle $\chi_{\rm cons}$. The dyad $({\mathbf e}_x ,{\mathbf e}_y)$
is midway between the latter two dyads, being obtained from the incoming dyad by a rotation of angle  $\frac12 \chi_{\rm cons}$.

The conservative scattering of the particle 1 corresponds to the change $p_1^-\to p_1^{+ \rm cons}$ of its linear momentum 
\beq
p_1^-=E_1U+p_-  {\mathbf n}_-\,,\qquad p_1^{+  \rm cons}=E_1U+p_-{\mathbf n}_+ ^{\rm cons}\,,
\eeq
such that
\beq
\Delta p_1^{\rm cons}=p_1^{+\rm cons}-p_1^-=p_-({\mathbf n}_+ ^{\rm cons}-{\mathbf n}_-)\,.
\eeq
The following representation
\beq
\Delta p_1^{\rm cons}=c_b^{1\rm cons} \hat {\mathbf b}+c_{u_1}^{1\rm cons}u_1+c_{u_2}^{1\rm cons}u_2 \,,
\eeq
with
\bea
c_b^{1\rm cons}&=&-p_- \sin\chi_{\rm cons}\,,\nonumber\\
c_{u_1}^{1\rm cons}&=&\frac{m_1E_2}{E}(\cos\chi_{\rm cons}-1)\,, \nonumber\\
c_{u_2}^{1\rm cons}&=&-\frac{m_2E_1}{E}(\cos\chi_{\rm cons}-1)\,, 
\eea
is also used.

For  particle 2 we have instead
\beq
p_2^-=E_1U-p_-  {\mathbf n}_-\,,\qquad p_2^{+\rm cons}=E_1U-p_-{\mathbf n}_+ ^{\rm cons}
\eeq
with
\beq
\Delta p_2^{\rm cons}=p_2^{+\rm cons}-p_2^-=-p_-({\mathbf n}_+ ^{\rm cons}-{\mathbf n}_-)\,.
\eeq
Therefore
$\Delta p_1^{\rm cons}+\Delta p_2^{\rm cons}=0\,,$ and then
\beq
\frac{d}{d\chi^{\rm cons}}\Delta p_1^{\rm cons}=-p_- \hat {\mathbf B}=-
\frac{d}{d\chi^{\rm cons}}\Delta p_2^{\rm cons}\,.
\eeq

\section{Relating hyperbolic-motion results to elliptic-motion ones by analytic continuation} \label{comparingDGI2004}

As a check on our computation, in Section \ref{2.5PNmotion}, of the 
2.5PN, radiation-reaction correction to the quasi-Keplerian parametrization of hyperboliclike motions,
we have (successfully) related it  to the corresponding 
2.5PN, radiation-reaction correction to the quasi-Keplerian parametrization of {\it ellipticlike} motions
derived in Ref. \cite{Damour:2004bz} (by using the elliptic version of  Lagrange's method of variation of constants).
As already mentioned in the text, this comparison used  two different ingredients: (i) analytic
continuation between elliptic and hyperbolic quasi-Keplerian parametrizations (at the Newtonian order); 
and (ii) the use of a different expression
for the radiation-reaction force, because of a difference in coordinates (ADM versus harmonic).

Let us only mention a few technical steps of this comparison.
The analytic continuation relating the elliptic eccentric anomaly, $u$, to the hyperbolic one, $v$ is simply
$u\to iv$. This has to be taken together with the replacement $a_r\to -\bar a_r$.
Concerning the gauge dependence of the radiation-reaction force, let us recall that, in a general
coordinate system, the 2.5PN-level radiation-reaction acceleration depends on two gauge parameters, $\alpha$
and $\beta$, and reads \cite{Iyer:1993xi,Iyer:1995rn}
\beq
\label{Arr}
{\mathbf A}^{\rm rr}=-\frac85 \nu \frac{G^2}{c^5} \frac{M^2}{r^3}\left[-A_{2.5\rm PN}\dot r {\mathbf n}
+B_{2.5\rm PN}{\mathbf v}\right]\,,
\eeq
where
\bea
\label{AB_alpha_beta}
A_{2.5\rm PN}&=&3(1+\beta)v^2+\frac13 (23+6\alpha-9\beta)\frac{GM}{r}-5\beta \dot r^2 
\,,\nonumber\\
B_{2.5\rm PN}&=&(2+\alpha)v^2+(2-\alpha)\frac{GM}{r}-3(1+\alpha)\dot r^2 \,.
\eea
For example, in harmonic coordinates $\alpha=-1$ and $\beta=0$,
\bea
A_{2.5\rm PN,h}&=&3v^2+\frac{17}3 \frac{GM}{r}\,, \nonumber\\
B_{2.5\rm PN,h}&=&v^2+3\frac{GM}{r} \,.
\eea
Other useful gauge-choices correspond to the Burke-Thorne reactive potential ($\alpha=4$, $\beta=5$), 
and to ADM coordinates ($\alpha=\frac53$, $\beta=3$). 

One can then easily derive the variation of constants in a general gauge. 
For example the $(\alpha,\beta)$-dependent equation for $\delta^{\rm rr} e_t$ reads
\bea
\frac{d\,\delta^{\rm rr} e_t}{dt} &=&
 \frac{8\nu(1-e_t^2)}{15a_r^4\,e_t}
 \biggl \{
\frac{12\alpha - 6\beta  + 15}{\chi^3}\nonumber\\ 
&+& 
\frac{-48\alpha + 33\beta  - 65}{\chi^4}\nonumber\\ 
&+& 
\frac{ 21 (e_t^2 - 3)\beta  - 9(2\alpha + 3) e_t^2 + 60\alpha + 109 }{\chi^5} \nonumber\\
&+& \frac{24(e_t^2 - 1) (\alpha - \frac{17}{8}\beta + \frac{59}{24})}{\chi^6} 
- \frac{15\beta (e_t^2 - 1)^2}{\chi^7}\biggr\}
\,.\nonumber\\
\eea
In the ADM case this equation becomes
\bea
\frac{d\,\delta^{\rm rr} e_t}{dt} &=&
 \frac{8\nu(1-e_t^2)}{15a_r^4\,e_t} 
 \biggl \{
 \frac{17}{\chi^3} - \frac{46}{ \chi^4} + \frac{6e_t^2 + 20}{\chi^5}\nonumber\\
&-&
 54 \frac{e_t^2 -1}{\chi^6} - \frac{45 (e_t^2 - 1)^2}{\chi^7}
\biggr\}\,,
\eea
as in Eq. (56.b) of Ref. \cite{Damour:2004bz}, 
while in the harmonic case we find
\bea
\frac{d\,\delta^{\rm rr} e_t}{dt} &=&
 \frac{8\nu(1-e_t^2)}{15a_r^4\,e_t}
 \left[
 \frac3{\chi^3} - \frac{17}{\chi^4} - \frac{9e_t^2 - 49 }{\chi^5}\right.\nonumber\\
&+&\left.
\frac{35 (e_t^2 - 1)}{\chi^6}
\right]\,.
\eea

\section{Radiation-reaction contribution to the relative scattering angle up to the 4.5 PN accuracy}
\label{chirel}

Ref. \cite{Bini:2012ji}  (see Eq. (5.99) there) has shown that, to linear order in radiation-reaction, and under the assumption
of a time-odd radiation-reaction force, the radiation-reaction contribution to the relative
scattering angle (in the c.m. frame), $\chi_{\rm rr, \, rel}$ can be 
 computed through a linear-response formula involving the radiative losses of  energy and angular momentum.
 We have generalized this linear-response formula above, see Eq. \eqref{chirrrel}, by including the term $\Delta c_\phi$
 that is non-zero when the radiation-reaction force contains a time-even piece. As discussed above, such a correction in  $\chi_{\rm rr, \, rel}$ starts to contribute  only at the 5PN (and 4PM) level. In other words, the first two terms on the right-hand side of Eq. \eqref{chirrrel}
 suffice to evaluate  $\chi_{\rm rr, \, rel}$ up to the 4.5PN level, by using the known radiative losses at the 4.5PN accuracy
 (as the radiative losses start at the 2.5PN level, this corresponds to a fractional 2PN accuracy).
 
 At the leading-order, 2.5PN level, we have given in the text a direct rederivation of the value of $\chi_{\rm rr, \, rel}$,
 see Eq. \eqref{deltaphivar}. The explicit expression of 
 $\chi_{\rm rr, rel}^{\rm 2.5PN}= [\delta^{\rm rr}\phi]^{\rm 2.5PN}$ in terms of $a_r$ and $e_r$ reads
\bea
\label{chi25a_re_r}
\chi^{\rm 2.5PN}_{\rm rr, rel}(a_r,e_r)&=&\frac{2\nu}{15\bar a_r^{5/2}(e_r^2-1)^{5/2}}\times  \nonumber\\
&&
\left[\frac{72e_r^4+1069e_r^2+134}{3e_r^2}\right.  \nonumber\\
&&\left.
+\frac{121e_r ^2+304}{\sqrt{e_r^2-1}} {\rm arccos}\left(-\frac{1}{e_r}\right) \right]\,,
\qquad 
\eea
which, when expressed in terms of the conserved energy and angular momentum, becomes
\bea
\chi^{\rm 2.5PN}_{\rm rr}(p_\infty,j)&=&\frac{2\nu}{15j^5}\times \nonumber\\
&\biggl[&
\frac{72p_\infty^4j^4+1213p_\infty^2j^2+1275}{3(1+p_\infty^2j^2)} \nonumber\\
&+&\left.
\frac{121p_\infty^2j^2+425}{p_\infty j}{\mathcal A}(p_\infty,j) \right]
\,,\nonumber\\
\eea
where 
\beq
{\mathcal A}(p_\infty,j)\equiv {\rm arccos}\left(-\frac{1}{\sqrt{1+p_\infty^2j^2}}\right)\,.
\eeq
The large-$j$ expansion of the latter expression reproduces the leading PN order of the PM expansion of $\chi_{\rm rr}$, the first terms of which (up to $O(G^7)$) are listed in Table XI of Ref. \cite{Bini:2021gat}.

When going to higher PN levels in the radiative losses (still keeping below the absolute 5PN level)
we must take into account that the radiative losses contain fractional corrections at the following levels:
1PN, 1.5PN and 2PN. The 1.5PN correction to the losses is the leading-order tail effect (which is still described
by a time-odd radiation reaction). Let us first discuss the 1PN and 2PN fractional corrections, leading to
contributions to $\chi_{\rm rr, rel}$ at the 3.5PN and 4.5PN levels.

The expressions of $\chi_{\rm rr, rel}$ at the $(n+\frac12)$-PN levels (for  $n=3,4$)
 have the general structure  
\bea
\chi_{\rm rr}(p_\infty,j)^{\rm n.5PN}&=&A_2^{\rm n.5PN}(p_\infty,j;\nu){\mathcal A}^2(p_\infty,j)
\nonumber\\
&+&
A_1^{\rm n.5PN}(p_\infty,j;\nu){\mathcal A}(p_\infty,j)
\nonumber\\
&+&
A_0^{\rm n.5PN}(p_\infty,j;\nu)\,.
\eea
Using the 2PN conservative scattering angle, Eq. (45) of Ref. \cite{Bini:2017wfr},
\bea
\label{chi_cons_2PN_formal}
\frac{\chi_{\rm cons}}{2}&=&\frac{\chi_{\rm cons}^{\rm N}}{2}
+\frac{\chi_{\rm cons}^{\rm 1PN}}{2}\eta^2 +\frac{\chi_{\rm cons}^{\rm 2PN}}{2}\eta^4+O(\eta^6)\,,
\eea 
where
\bea
\label{chi_cons_2PN_coeffs}
\frac{\chi_{\rm cons}^{\rm N}}{2}&=& {\mathcal A}(p_\infty,j)
-\frac{\pi}{2}
\,, \nonumber\\
\frac{\chi_{\rm cons}^{\rm 1PN}}{2}&=& \frac{3}{j^2}{\mathcal A}(p_\infty,j)
+\frac{ p_\infty (3+2 j^2 p_\infty^2)}{j (1+j^2 p_\infty^2)}
\,,\nonumber\\
\frac{\chi_{\rm cons}^{\rm 2PN}}{2}&=& -\frac{3[j^2 p_\infty^2 (2\nu-5)-35+10\nu]}{4 j^4} {\mathcal A}(p_\infty,j)
\nonumber\\
&-&\frac{p_\infty }{ 4 j^3 (1+j^2 p_\infty^2)^2}[j^4 p_\infty^4 (-81+26\nu)\nonumber\\
&+& 2 j^2 p_\infty^2 (-95+28\nu)+30\nu-105] \,,
\eea
and the fractionally 2PN-accurate expressions  (when excluding tails)  for the radiated energy and angular momentum given in Ref. \cite{Bini:2021gat}, Eqs. (C10)--(C13) and (E4)--(E10), we get the following explicit results:
\begin{widetext}
\bea
\chi^{\rm 3.5PN}_{\rm rr}(p_\infty,j)&=&\frac{2\nu}{j^7}\left\{
\left(\frac{168 (p_\infty j)^2}{5}+72\right)
{\mathcal A}^2(p_\infty,j)
\right.\nonumber\\
&+&
\left[(p_\infty j)^3 \left(\frac{23111}{840}-\frac{437 \nu }{30}\right)+(p_\infty j) \left(\frac{11647}{60}-\frac{424 \nu }{3}\right)+\frac{\frac{13447}{40}-\frac{1127 \nu}{6}}{(p_\infty j)}\right]
{\mathcal A}(p_\infty,j)
\nonumber\\
&+&
\frac1{\left((p_\infty j)^2+1\right)^2}\left[
(p_\infty j)^8 \left(\frac{40}{7}-\frac{8 \nu }{5}\right)+(p_\infty j)^6 \left(\frac{92639}{1400}-\frac{7681 \nu }{90}\right)+(p_\infty j)^4\left(\frac{5049251}{12600}-\frac{3503 \nu }{10}\right)\right.\nonumber\\
&+&\left.\left.
(p_\infty j)^2 \left(\frac{81889}{120}-\frac{8179 \nu }{18}\right)-\frac{1127 \nu}{6}+\frac{13447}{40}
\right]
\right\}
\,,\nonumber\\
\chi^{\rm 4.5PN}_{\rm rr}(p_\infty,j)&=&
\frac{2\nu}{j^9}\left\{
\left[(p_\infty j)^4 \left(\frac{534}{7}-\frac{373 \nu }{5}\right)+(p_\infty j)^2 \left(816-\frac{2898 \nu }{5}\right)-745 \nu +1586\right]
{\mathcal A}^2(p_\infty,j)
\right.\nonumber\\
&+&
\frac{1}{(p_\infty j) ((p_\infty j)^2+1)}\left[(p_\infty j)^8 \left(\frac{511 \nu ^2}{24}-\frac{75253 \nu }{1680}+\frac{44759}{1120}\right)+(p_\infty j)^6 \left(366 \nu ^2-\frac{136789 \nu}{168}+\frac{1020745}{1512}\right)\right.\nonumber\\
&+&
(p_\infty j)^4 \left(\frac{5237 \nu ^2}{4}-\frac{1579549 \nu }{420}+\frac{16375901}{5040}\right)+(p_\infty j)^2 \left(\frac{4949\nu ^2}{3}-\frac{149209 \nu }{24}+\frac{6034507}{1080}\right)\nonumber\\
&+&\left.
\frac{5481 \nu ^2}{8}-\frac{258051 \nu }{80}+\frac{5839651}{2016}\right]
{\mathcal A}(p_\infty,j)
\nonumber\\
&+&
\frac{1}{((p_\infty j)^2+1)^3}\left[
(p_\infty j)^{12} \left(\frac{8 \nu ^2}{5}-\frac{186 \nu }{35}+\frac{256}{63}\right)+(p_\infty j)^{10} \left(\frac{19781 \nu ^2}{120}-\frac{105913 \nu}{560}+\frac{45934963}{352800}\right)\right.\nonumber\\
&+&
(p_\infty j)^8 \left(\frac{439657 \nu ^2}{360}-\frac{47396053 \nu }{25200}+\frac{3027711913}{3175200}\right)+(p_\infty j)^6\left(\frac{607627 \nu ^2}{180}-\frac{95753533 \nu }{12600}+\frac{7101025663}{1587600}\right)\nonumber\\
&+&
(p_\infty j)^4 \left(\frac{796337 \nu^2}{180}-\frac{170414669 \nu }{12600}+\frac{434998411}{45360}\right)+(p_\infty j)^2 \left(\frac{66997 \nu ^2}{24}-\frac{520709 \nu}{48}+\frac{266996831}{30240}\right)\nonumber\\
&+&\left.\left.
\frac{5481 \nu ^2}{8}-\frac{258051 \nu }{80}+\frac{5839651}{2016}
\right]
\right\}
\,,
\eea
For completeness, the corresponding PN-expansion coefficients when considering  $\chi_{\rm rr}$ as a function
of $\bar a_r$ and $e_r$  are Eq. \eqref{chi25a_re_r} (at the 2.5PN accuracy) together with
\bea
\chi^{3.5\rm PN}_{\rm rr}(a_r,e_r)&=& \frac{\nu}{\bar a_r^{7/2}(e_r^2-1)^{7/2}}\left[C_2^{3.5\rm PN}{\rm arccos}^2\left(-\frac{1}{e_r}\right)+
\frac{C_1^{3.5\rm PN}}{\sqrt{e_r^2-1}}{\rm arccos}\left(-\frac{1}{e_r}\right)+C_0^{3.5\rm PN}\right]
\,,\nonumber\\
\chi^{4.5\rm PN}_{\rm rr}(a_r,e_r)&=& \frac{\nu}{\bar a_r^{9/2}(e_r^2-1)^{9/2}}\left[C_2^{4.5\rm PN}{\rm arccos}^2\left(-\frac{1}{e_r}\right)+
\frac{C_1^{4.5\rm PN}}{\sqrt{e_r^2-1}}{\rm arccos}\left(-\frac{1}{e_r}\right)+C_0^{4.5\rm PN}\right]\,,
\eea
where
\bea
C_2^{3.5\rm PN}&=& \frac{336}{5}e_r^2+\frac{384}{5}
\,,\nonumber\\
C_1^{3.5\rm PN}&=& \left(\frac{2783}{420}+\frac{47}{15}\nu\right) e_r^4+\left(-\frac{260}{3}\nu-\frac{1507}{7}\right) e_r^2
-\frac{1832}{15}\nu-\frac{14594}{105}
\,,\nonumber\\
C_0^{3.5\rm PN}&=& \left(\frac{8}{5}\nu+\frac{288}{35}\right) e_r^4+\left(-\frac{1253}{45}\nu-\frac{1396049}{6300}\right) e_r^2
-\frac{7498}{45}\nu-\frac{71683}{450}+\left(-\frac{64}{5}\nu+\frac{39394}{1575}\right)\frac{1}{e_r^2}
\,,\nonumber\\
\eea
and
\bea
C_2^{4.5\rm PN}&=& \left(-\frac{1716}{35}+\frac{94}{5}\nu\right) e_r^4+\left(-\frac{10008}{35}-\frac{2624}{5}\nu\right) e_r^2-480\nu+\frac{16904}{35}
\,,\nonumber\\
C_1^{4.5\rm PN}&=& \left(\frac{9}{20}\nu^2+\frac{7783}{840}\nu+\frac{82489}{1680}\right) e_r^6
+\left(\frac{49}{3}\nu^2+\frac{48821}{84}\nu-\frac{417001}{3780}\right) e_r^4
+\left(\frac{514}{5}\nu^2+\frac{427622}{105}\nu-\frac{1607}{63}\right) e_r^2\nonumber\\
&+& 88\nu^2+\frac{19066}{15}\nu-\frac{19882}{27}
\,,\nonumber\\
C_0^{4.5\rm PN}&=&\left(-\frac{2}{5}\nu^2+\frac{242}{35}\nu+\frac{808}{45}\right) e_r^6
+\left(\frac{1367}{180}\nu^2+\frac{72587}{2520}\nu+\frac{28987039}{176400}\right) e_r^4
+\left(\frac{365}{6}\nu^2+\frac{72257}{18}\nu-\frac{147017953}{793800}\right) e_r^2\nonumber\\
&+&\frac{5956}{45}\nu^2-\frac{98228321}{99225}+\frac{1299217}{630}\nu
+\left(\frac{36}{5}\nu^2-\frac{56108}{315}\nu+\frac{16847071}{99225}\right)\frac{1}{e_r^2}\,.
\eea

Let us finally discuss the tail-related contribution to $\chi_{\rm rr, rel}$. 
The leading-order, 4PN tail contribution
is obtained by inserting in  the linear-response formula the ($j$-expanded) Eqs. (D26) and (F2) of 
Ref. \cite{Bini:2021gat}. 
The result is the following
\bea
\chi^{4\rm PN}_{\rm rr, rel}(p_\infty,j)&=& 
\nu \left[\frac{7168}{45} \frac{p_\infty^3}{j^5}+\frac{573}{20} \pi^3\frac{p_\infty^2}{j^6}
+\left(\frac{512}{9}+\frac{153856}{675}\pi^2\right) \frac{p_\infty}{j^7}
+O\left(\frac{1}{j^8}\right)\right]
\,.
\eea
If we formally insert also the fractional 1PN correction to the linear tail, we get (by using the
2.5PN accurate expressions for $E^{\rm rad}$ and $J^{\rm rad}$ derived above in Eqs. \eqref{Erad2p5pn} and \eqref{Jrad2p5pn}, respectively) the following
5PN-level contribution to $\chi_{\rm rr, rel}$:
 \bea
 \chi^{5\rm PN \, from \, tail \, in \, losses}_{\rm rr,rel}(p_\infty,j)&=& 
 \nu \left[
\left(\frac{4992}{35}-\frac{676096}{1575}\nu\right)\frac{p_\infty^5}{j^5}\right.\nonumber\\
&+&\left( -\frac{32079}{1120}\pi^2+\frac{145536}{175}-\frac{7767}{70}\nu\pi^2+\frac{14032}{525}\nu\right)\pi \frac{p_\infty^4}{j^6}\nonumber\\
&+&
\left(\frac{7014}{5}\zeta(3)-\frac{515456}{33075}\pi^2+\frac{206188}{105}+\frac{207}{5}\pi^4-\frac{89216}{105}\nu-\frac{18853168}{33075}\nu\pi^2\right)\frac{p_\infty^3}{j^7}\nonumber\\
&+&\left.
O\left(\frac{p_\infty^2}{j^8}\right)\right]\,.
 \eea
Note, however, that, at this level, there are several other contributions that should be added to this result.

\section{3PN-accurate quasi-Keplerian parametrization of the hyperbolic motion}
\label{3PN_orb_par}

The 3PN-accurate quasi-Keplerian parametrization of the hyperboliclike motion is 
\begin{eqnarray} 
r&=& \bar a_r (e_r \cosh v-1)\,,\nonumber\\
\bar n\,  t&=&e_t \sinh v-v + f_t V+g_t \sin V+h_t \sin 2V+i_t \sin 3V\,,\nonumber\\
\phi &=&K[V+f_\phi \sin 2V+g_\phi \sin 3V+h_\phi \sin 4V+i_\phi \sin 5V]\,,
\end{eqnarray}
with
\beq
\label{Vdef}
V(v)=2\, {\rm arctan}\left[\sqrt{\frac{e_\phi+1}{e_\phi-1}}\tanh \frac{v}{2}  \right]\,.
\eeq
The 3PN orbital parameters in modified harmonic coordinates along hyperboliclike orbits were obtained in Ref. \cite{Cho:2018upo}. However, their expressions are affected by typos, which we discovered when rederiving the
3PN-accurate quasi-Keplerian parametrization of hyperboliclike motions.  We list below these typos.
\begin{enumerate}
\item Eq. 2.36b. Third line: the term $4\eta^3 )$ should be replaced by
$$
2\bar E j^2 \frac{4\eta^3 -195 \eta^2+1120 \eta -1488}{430080}.
$$
 
\item Eq. 2.36c. The third term in parenthesis  should have an overall factor of $15$ in front, and   one should replace the $+3\eta^3$ by $-3\eta^3$. 

\item Eq. 2.36j. The prefactor $\eta^3$ should instead be $\eta$.
\item Eq. 2.36k. The $+ -$ sign of the third term in parenthesis is a $-$.
\item Eq. 2.36m. Second line: the term $-30135\eta^2$ is $-30135\pi^2$.
\item Eq. 2.36o. There is a missing overall $3/35$ in front.
\end{enumerate}

It is convenient to express the orbital parameters is terms of $\bar a_r$ and $e_r$ through the relations 
\bea
\bar E&=& \frac{1}{2\bar a_r}+\left(\frac{7}{8}-\frac{1}{8}\nu\right)\frac{\eta^2}{\bar a_r^2}
+\left[\frac{25}{16}-\frac{7}{16}\nu+\frac{1}{16}\nu^2+\frac{(2-\frac72 \nu)}{e_r^2-1}\right]\frac{\eta^4}{\bar a_r^3}\nonumber\\
&+& \left[\frac{363}{128}-\frac{149}{128}\nu+\frac{21}{64}\nu^2-\frac{5}{128}\nu^3
+\frac{(5+(\frac{41}{128}\pi^2-\frac{17033}{840})\nu+\frac{7}{4}\nu^2)}{(e_r^2-1)}
+\frac{(4+(-\frac{12343}{420}+\frac{41}{32}\pi^2)\nu+\nu^2)}{(e_r^2-1)^2}\right]\frac{\eta^6}{\bar a_r^4}
\,,\nonumber\\
j&=& \sqrt{\bar a_r} \sqrt{e_r^2-1}+\left[(1-\frac12 \nu)\sqrt{e_r^2-1}+\frac{(3-\frac12 \nu)}{\sqrt{e_r^2-1}} \right]\frac{\eta^2}{\sqrt{\bar a_r}}\nonumber\\
&+& \left[\left(\frac{3}{2}-\frac{11}{8}\nu+\frac{3}{8}\nu^2\right)\sqrt{e_r^2-1}
+\frac{(\frac{5}{2}-\frac{75}{8}\nu+\frac{1}{4}\nu^2)}{\sqrt{e_r^2-1}}
+\frac{(\frac{7}{2}-\frac{25}{2}\nu-\frac{1}{8}\nu^2)}{(e_r^2-1)^{3/2}}\right]\frac{\eta^4}{\bar a_r^{3/2}}\nonumber\\
&+&\left[ \left(\frac{5}{2}-\frac{51}{16}\nu+\frac{27}{16}\nu^2-\frac{5}{16}\nu^3\right)\sqrt{e_r^2-1}
+\frac{(6+(\frac{41}{128}\pi^2-\frac{19697}{560})\nu+9\nu^2-\frac{3}{16}\nu^3)}{\sqrt{e_r^2-1}}\right.\nonumber\\
&+&\left. \frac{(5+(\frac{123}{32}\pi^2-\frac{22193}{280})\nu+\frac{71}{16}\nu^2)+\frac{1}{16}\nu^3}{(e_r^2-1)^{3/2}}
+\frac{(\frac{11}{2}+(\frac{41}{8}\pi^2-\frac{32887}{420})\nu-\frac{15}{8}\nu^2-\frac{1}{16}\nu^3)}{(e_r^2-1)^{5/2}} \right]\frac{\eta^6}{\bar a_r^{5/2}}\,.
\eea
We find
\bea
\bar n &=& \frac{1}{\bar a_r^{3/2}}
-\frac12 (-9+\nu)\frac{\eta^2}{\bar a_r^{5/2}}
+\left[\frac{3}{8}\nu^2-\frac{25}{8}\nu+\frac{147}{8}+\frac{(-\frac{21}{2}\nu+6)}{\sqrt{e_r^2-1}}\right]\frac{\eta^4}{\bar a_r^{7/2}}\nonumber\\
&+&\left[\frac{1181}{16}-\frac{235}{16}\nu+3\nu^2-\frac{5}{16}\nu^3
+\frac{(39+(\frac{123}{128}\pi^2-\frac{29353}{280})\nu+\frac{35}{4}\nu^2)}{\sqrt{e_r^2-1}}
+\frac{(12+(-\frac{12343}{140}+\frac{123}{32}\pi^2)\nu+3\nu^2)}{(e_r^2-1)^2}\right]\frac{\eta^6}{\bar a_r^{9/2}}
\,,\nonumber\\
K&=&1+\frac{3}{(e_r^2-1)}  \frac{\eta^2}{\bar a_r}+  \left[\frac{(\frac32 \nu-\frac94)}{e_r^2-1}+\frac{(-\frac92 \nu+\frac{33}{4})}{(e_r^2-1)^2}\right]\frac{\eta^4}{\bar a_r^2}\nonumber\\
&+&\left[\frac{(\frac{3}{2}\nu-\frac{3}{8}\nu^2)}{e_r^2-1}+\frac{(-\frac{39}{4}+(\frac{13}{2}+\frac{123}{128}\pi^2)\nu-\frac{9}{4}\nu^2)}{(e_r^2-1)^2}+\frac{(\frac{135}{4}+(-122+\frac{615}{128}\pi^2)\nu+\frac{9}{8}\nu^2)}{(e_r^2-1)^3}\right]   \frac{\eta^6}{\bar a_r^3}\,,\nonumber\\
\frac{e_t}{e_r}&=& 1+\left(4-\frac32 \nu\right) \frac{\eta^2}{\bar a_r}+\left(16-\frac{67}{8}\nu+\frac{15}{8}\nu^2+\frac{4-7\nu}{e_r^2-1}  \right)\frac{\eta^4}{\bar a_r^2}\nonumber\\
&+&\left( 64-\frac{599}{16}\nu+\frac{219}{16}\nu^2-\frac{35}{16}\nu^3+\frac{(28+(\frac{41}{64}\pi^2-\frac{34463}{420})\nu+21\nu^2)}{e_r^2-1}
+\frac{(8+(-\frac{12343}{210}+\frac{41}{16}\pi^2)\nu+2\nu^2)}{(e_r^2-1)^2} \right)\frac{\eta^6}{\bar a_r^3}
\,,\nonumber\\
\frac{e_\phi}{e_r}&=& 1-\frac{\nu}{2} \frac{\eta^2}{\bar a_r}
+\left(-\frac{29}{32}\nu +\frac{15}{32}\nu^2+\frac{(-5-\frac{357}{32}\nu+\frac{15}{32}\nu^2)}{e_r^2-1} \right)\frac{\eta^4}{\bar a_r^2}+\left( -\frac{213}{128}\nu+\frac{213}{128}\nu^2-\frac{61}{128}\nu^3\right.\nonumber\\
&+&\left. \frac{(4+(\frac{205}{256}\pi^2-\frac{10463}{448})\nu+\frac{799}{64}\nu^2-\frac{15}{64}\nu^3)}{e_r^2-1}
+\frac{(-16+(\frac{533}{256}\pi^2+\frac{276553}{4480})\nu+\frac{585}{128}\nu^2+\frac{95}{128}\nu^3)}{(e_r^2-1)^2} \right)\frac{\eta^6}{\bar a_r^3}\,.
\eea
The remaining 3PN orbital parameters still expressed as functions of $\bar a_r$ and $e_r$ are listed in Table \ref{tab:orbpar_3PN}.


\begin{table*}  
\caption{\label{tab:orbpar_3PN}  The orbital parameters of the 3PN  quasi-Keplerian hyperbolic representation
in modified harmonic coordinates,  expressed as functions of $\bar a_r$ and $e_r$. The corresponding (equivalent) expressions in terms of $\bar E$ and $j$ have been given in Ref. \cite{Cho:2018upo}.
}
\begin{ruledtabular}
\begin{tabular}{l|l}
$f_t $& $ \frac{3 (5 - 2 \nu)}{ 2 {\bar a}_r^2\sqrt{e_r^2 - 1} }\eta^4 
+ \frac{144(4\nu^2-19\nu+40)e_r^2+\nu(-8768+576\nu+123\pi^2)}{ 192 (e_r^2 - 1)^{3/2}{\bar a}_r^3 }\eta^6 $\\
\hline
$f_\phi$& $\frac{e_r^2 (1+19 \nu-3\nu^2)}{ 8 {\bar a}_r^2 (e_r^2 - 1)^2 }\eta^4 
- \frac{e_r^2 [-280\nu(9\nu^2-177\nu+458)e_r^2-26880+26880\nu^3+36960\nu^2+(-107104+30135\pi^2)\nu]}{ 26880 {\bar a}_r^3 (e_r^2 - 1)^3 }\eta^6$\\
\hline
$g_t $& $\frac{e_r \nu (15 - \nu)}{ 8 {\bar a}_r^2\sqrt{e_r^2 - 1} }\eta^4 
- \frac{e_r [-35\nu(27\nu^2-263\nu+717)e_r^2-22400+700\nu^3+8820\nu^2+(-5956+1435\pi^2)\nu]}{2240 (e_r^2 - 1)^{3/2}{\bar a}_r^3}\eta^6$\\
\hline
$g_\phi$& $\frac{ (1-3 \nu)\nu e_r^3}{ 32 {\bar a}_r^2 (e_r^2 - 1)^2}\eta^4  
- \frac{\nu e_r^3 [-35(23\nu^2-87\nu+27)e_r^2+1960\nu^2+14840\nu+1435\pi^2-31856]}{ 8960 {\bar a}_r^3 (e_r^2 - 1)^3 }\eta^6$\\
\hline
$h_t $& $ \frac{(3 \nu^2 - 49 \nu + 116) \nu e_r^2 }{16 (e_r^2 - 1)^{3/2} {\bar a}_r^3}\eta^6$\\
\hline
$h_\phi$& $ \frac{ (15 \nu^2 - 57 \nu + 82) e_r^4 \nu}{ 192 {\bar a}_r^3 (e_r^2 - 1)^3 }\eta^6$ \\
\hline
$i_t$& $ \frac{ (13 \nu^2 - 73 \nu + 23) \nu e_r^3}{ 192 (e_r^2 - 1)^{3/2}{\bar a}_r^3 }\eta^6 $\\
\hline
$i_\phi$& $ \frac{ (5 \nu^2 - 5 \nu + 1) \nu e_r^5}{ 256 {\bar a}_r^3 (e_r^2 - 1)^3 }\eta^6 $\\
\end{tabular}
\end{ruledtabular}
\end{table*}

\end{widetext}

\end{document}